\documentclass[linenumbers,letterpaper,twocolumn,english,aps,showpacs,showkeys,preprintnumbers,prd,floatfix,nofootinbib,superscriptaddress]{revtex4-1}

\makeatletter
\def\@fnsymbol#1{%
 \ensuremath{%
  \ifcase#1\or
  *\or                        \dagger                   \or
  \ddagger                \or \mathsection              \or
  \mathparagraph\or
  **\or                       \dagger\dagger            \or
  \ddagger\ddagger        \or \mathsection \mathsection \or
  \mathparagraph\mathparagraph\or
  *{*}*\ignorespaces      \or \dagger\dagger\dagger     \or
  \ddagger\ddagger\ddagger\or \mathsection \mathsection \mathsection \or
  \mathparagraph\mathparagraph\mathparagraph\or
  \else
  \@ctrerr
  \fi
 }%
}%
\makeatother

\usepackage[colorlinks = true,linkcolor = blue, urlcolor  = blue,citecolor = red,anchorcolor = blue]{hyperref}

\usepackage[T1]{fontenc}
\usepackage{pdfpages}
\usepackage{graphicx}

\usepackage[figuresright]{rotating}
\usepackage{amsmath}
\usepackage{amsfonts}
\usepackage{hyperref}
\usepackage{color}
\usepackage{float}
\usepackage{verbatim}
\usepackage{booktabs}
\usepackage{multirow}
\usepackage{subfigure}

\DeclareMathSizes{10}{9}{7}{7}
\makeatletter
\AtBeginDocument{\let\LS@rot\@undefined}
\makeatother

\newcommand{\orcid}[1]{\,\href{https://orcid.org/#1}{\includegraphics[width=9pt]{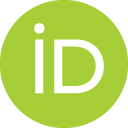}}\,}

\newcommand{\orcidTJ}{0000-0002-1334-7607} 
\newcommand{\orcidPD}{0000-0001-7960-7953} 
\newcommand{\orcidMK}{0000-0002-4665-3088} 
\newcommand{\orcidKK}{0000-0003-1412-447X} 
\newcommand{\orcidAK}{0000-0002-4090-0084} 
\newcommand{\orcidJM}{0000-0001-9343-9351} 
\newcommand{\orcidFM}{0000-0002-3888-1697} 
\newcommand{\orcidFO}{0000-0001-6799-2436} 
\newcommand{\orcidIS}{0000-0003-0373-474X} 
\newcommand{\orcidRR}{0000-0002-3316-2175} 

\begin{document}

\nolinenumbers  


	\title{Compatibility of neutrino DIS data and its impact on \\ nuclear parton distribution functions}
	
	\author{K.F.~Muzakka\orcid{\orcidFM}}
	\email{khoirul.muzakka@uni-muenster.de}
	\affiliation{Institut f{ü}r Theoretische Physik, Westf{ä}lische Wilhelms-Universit{ä}t
		M{ü}nster, Wilhelm-Klemm-Stra{ß}e 9, D-48149 M{ü}nster, Germany}
		
	\author{P.~Duwent\"aster\orcid{\orcidPD}}
	\affiliation{Institut f{ü}r Theoretische Physik, Westf{ä}lische Wilhelms-Universit{ä}t
		M{ü}nster, Wilhelm-Klemm-Stra{ß}e 9, D-48149 M{ü}nster, Germany}
		
	\author{T.J.~Hobbs}
	\affiliation{Fermi National Accelerator Laboratory, Batavia, IL 60510, USA}
	\affiliation{Department of Physics, Illinois Institute of Technology, Chicago, Illinois 60616, USA}

	\author{T.~Je\v{z}o\orcid{\orcidTJ}}
	\affiliation{Institut f{ü}r Theoretische Physik, Westf{ä}lische Wilhelms-Universit{ä}t
		M{ü}nster, Wilhelm-Klemm-Stra{ß}e 9, D-48149 M{ü}nster, Germany}
		
	\author{M.~Klasen\orcid{\orcidMK}}
	\affiliation{Institut f{ü}r Theoretische Physik, Westf{ä}lische Wilhelms-Universit{ä}t
		M{ü}nster, Wilhelm-Klemm-Stra{ß}e 9, D-48149 M{ü}nster, Germany}
	
	\author{K.~Kova\v{r}\'{\i}k\orcid{\orcidKK}}
	\email{karol.kovarik@uni-muenster.de}
	\affiliation{Institut f{ü}r Theoretische Physik, Westf{ä}lische Wilhelms-Universit{ä}t
		M{ü}nster, Wilhelm-Klemm-Stra{ß}e 9, D-48149 M{ü}nster, Germany}
	
	\author{A.~Kusina\orcid{\orcidAK}}
	\affiliation{Institute of Nuclear Physics, Polish Academy of Sciences, ul. Radzikowskiego, Cracow 31-342, Poland}

	\author{J.G.~Morf\'{i}n\orcid{\orcidJM}} 
    \affiliation{Fermi National Accelerator Laboratory, Batavia, IL 60510, USA}
	
	\author{F.~I.~Olness\orcid{\orcidFO}}
	\affiliation{Southern Methodist University, Dallas, TX 75275, USA }
	
	\author{R.~Ruiz\orcid{\orcidRR}}
	\affiliation{Institute of Nuclear Physics, Polish Academy of Sciences, ul. Radzikowskiego, Cracow 31-342, Poland}
	
	\author{I.~Schienbein\orcid{\orcidIS}}
	\affiliation{Laboratoire de Physique Subatomique et de Cosmologie, Université
		Grenoble-Alpes, CNRS/IN2P3, 53 avenue des Martyrs, 38026 Grenoble, France}
	
	\author{J.~Y.~Yu}
	\affiliation{Southern Methodist University, Dallas, TX 75275, USA }
	
	\preprint{MS-TP-22-06}
	\preprint{SMU-HEP-22-04}
	\preprint{IFJPAN-IV-2022-2}
	\preprint{FERMILAB-PUB-22-119-ND-SCD-T}
	\date{\today}
	
	\begin{abstract}
In global analyses of nuclear parton distribution functions (nPDFs), neutrino deep-inelastic scattering (DIS) data have been argued to exhibit tensions with the data from charged-lepton DIS. Using the nCTEQ framework, we investigate these possible tensions both internally and with the data sets used in our recent nPDF analysis nCTEQ15WZSIH. We take into account nuclear effects in the calculation of the deuteron structure function $F_2^D$ using the CJ15 analysis. The resulting nPDF fit, nCTEQ15WZSIHdeut, serves as the basis for our comparison with inclusive neutrino DIS and charm dimuon production data. Using $\chi^2$ hypothesis testing, we confirm evidence of tensions with these data and study the impact of the proton PDF baseline as well as the treatment of data correlation and normalization uncertainties. We identify the experimental data and kinematic regions that generate the tensions and present several possible approaches how a consistent global analysis with neutrino data can be performed. We show that the tension can be relieved using a kinematic cut at low $x$ ($x>0.1$) and also investigate a possibility of managing the tensions by using uncorrelated systematic errors. Finally, we present a different approach identifying a subset of neutrino data which leads to a consistent global analysis without any additional cuts. Understanding these tensions between the neutrino and charged-lepton DIS data is important not only for a better flavor separation in global analyses of nuclear and proton PDFs, but also for neutrino physics and for searches for physics beyond the Standard Model.
	\end{abstract}
	\maketitle
	\tableofcontents{}
%
\section{Introduction and review of previous analyses}
Charged-current (CC) deep-inelastic scattering (DIS) of neutrinos off nuclei has long been recognized to have a significant impact on global analyses of proton~\cite{Hou:2019efy,Accardi:2016qay,Bailey:2020ooq,Abramowicz:2015mha,Ball:2017nwa,Alekhin:2017kpj} and nuclear~\cite{deFlorian:2011fp,Kovarik:2015cma,Kusina:2016fxy,Kusina:2020lyz,Duwentaster:2021ioo,Eskola:2016oht,AbdulKhalek:2020yuc,Walt:2019slu,Khanpour:2020zyu} parton distribution functions (PDFs), mainly due to its discriminating power in separating quark flavors~\cite{Kovarik:2019xvh,Ethier:2020way}. A good theoretical understanding of neutrino DIS is also an important ingredient for determinations of the weak mixing angle and for searches for physics beyond the Standard Model~\cite{Zyla:2020zbs}. Apart from inclusive neutrino DIS, the semi-inclusive charm dimuon production $\nu N \to \mu D +X$ with $D \to \mu+X'$ plays a crucial role in determining the strange quark content of the nucleon \cite{Goncharov:2001qe, Kusina:2012vh, Faura:2020oom}. 

Due to the weak nature of the neutrino-nucleus interaction, heavy nuclei such as iron or lead have been usually used as targets in neutrino scattering experiments in order to obtain data with sufficiently high statistics. Therefore, if one were to use the neutrino DIS data in an analysis of the structure of the proton, a nuclear correction factor would be required. Indeed it is much more natural to analyze neutrino DIS in the framework of nuclear PDFs (nPDFs). Out of all available up-to-date global analyses of nPDFs, most include a small selection of neutrino inclusive or semi-inclusive DIS data. The reason why nPDF analyses do not include the totality of neutrino DIS data can be traced back to concerns about possible tensions between neutrino DIS data and the charged-lepton data fitted in nPDF frameworks.

In the past decades, there have been several dedicated analyses of neutrino DIS data using the framework of nPDFs. They started with Ref.~\cite{Schienbein:2007fs}, where it was shown by conducting an analysis of neutrino DIS cross-section data from NuTeV~\cite{Tzanov:2005kr} and dimuon data from NuTeV and CCFR~\cite{Goncharov:2001qe} that the extracted iron PDFs in the nCTEQ framework led to a nuclear ratio of the charged-current structure function $F_2$ that is flatter and significantly different from the similar ratio extracted directly from the charged-lepton DIS data, as described, e.g., by the Kulagin-Petti model \cite{Kulagin:2004ie} or the SLAC/NMC parametrization \cite{Abramowicz:1991xz}. In particular, the lack of shadowing of the charged-current structure function ratio in the low-$x$ ($x\leq 0.1$) region is quite atypical. Another peculiarity can also be observed: the typical antishadowing which is present in the neutral current data at moderate $x$ ($0.06< x<0.3$) is shifted to much smaller $x$. The stark difference in the nuclear correction factor triggered a follow-up study \cite{Kovarik:2010uv}, where a global analysis that included charged-lepton and Drell-Yan (DY) data as well as neutrino DIS from NuTeV~\cite{Tzanov:2005kr} and Chorus~\cite{Onengut:2005kv} was performed. It concluded that the neutrino DIS data is incompatible with the charged-lepton data citing the high precision of the NuTeV cross-section data and especially the correlated systematic uncertainties as the main reason for the conclusion.

Some time later two related studies \cite{Paukkunen:2010hb,Paukkunen:2013grz} were carried out in the EPS nPDF framework. The authors found only a mild tension between the neutrino DIS data and the charged-lepton DIS data. They further suggested~\cite{Paukkunen:2013grz} that data normalization might be the reason of the apparent incompatibility. By normalizing cross-section data with the integrated cross-section in each energy bin and using a Hessian reweighting analysis based on linearization of theory predictions near the minimum, it was shown that the neutrino DIS data, in particular those from NuTeV, could be included in a global analysis with charged-lepton DIS data without causing significant tensions. It is worth noting that the NuTeV data used in Ref.~\cite{Paukkunen:2013grz} were without point-to-point correlations, which as it was also shown in the previous nCTEQ analysis \cite{Kovarik:2010uv} makes a large difference. With uncorrelated systematic errors the NuTeV data can be described with a very good $\chi^2$ even in Ref.~\cite{Kovarik:2010uv}. Nevertheless, even if NuTeV data is described well, some charged-lepton DIS data, especially those taken on a nucleus close to iron in the mass number, have $\chi^2$/pt significantly larger than unity. Furthermore, without a proper global analysis, the linearization method employed in Ref.~\cite{Paukkunen:2013grz} might not be sufficient to capture the true minimum, considering the fact that there are almost four times as many neutrino DIS data points as there are charged-lepton and DY data. 

Another intriguing study aiming at comparing the neutrino DIS data with the rest of the data was performed by Kalantarians \textit{et al.}~\cite{Kalantarians:2017mkj}. There, $F_2^{\mathrm{Fe}}/F_2^{\mathrm{D}}$ data from BCDMS and NMC were transformed into $F_2^{\mathrm{Fe}}$ by multiplying the data with $F_2^{\mathrm{D}}$ from the NMC parametrization \cite{Abramowicz:1991xz}. This neutral current $F_2^{\mathrm{Fe}}$ data was then compared with charged current $F_2^{\mathrm{Fe}}$ data from the NuTeV, CCFR and CDHSW experiments, after correcting them using the well-known ``18/5-rule''. Agreement in the valence region ($x>0.3$) could be shown but around 15\% discrepancies at \hbox{$x<0.15$} were still visible. These still could be explained by a proper NLO treatment including also heavy-quark effects, which also lead to differences of similar size in the same kinematic region.  

Apart from the aforementioned dedicated analyses, the neutrino DIS data have been used in numerous global analyses of nPDFs. In the past, the analyses such as Ref.~\cite{deFlorian:2011fp} included $F_2$ and $F_3$ neutrino data from CDHSW, NuTeV, and Chorus. The downside of using the structure function data is that these data are not as precise and therefore much less sensitive to any tension. Currently all global analyses that use neutrino DIS data to aid in flavour decomposition, e.g. Refs.~\cite{AbdulKhalek:2020yuc,Walt:2019slu,Khanpour:2020zyu,Eskola:2021nhw,Khalek:2022zqe}, prefer to avoid the NuTeV cross-section data.%
    \footnote{One should also mention that the HKN group observed similar incompatibilities in the nuclear modifications extracted from charged lepton and neutrino DIS data. However, these results are still preliminary~\cite{Nakamura:2016cnn}.}
 
It is important to emphasize that the nuclear effects determined in global nPDF analyses are relatively small and that there is insufficient data to constrain all parton densities in the nuclear environment. The notion of compatibility or lack of compatibility of the neutrino DIS cross-section data depends on the specific nPDF fitting framework such as the parameterization, the choice of free parameters, data selection or even the proton PDF baseline. Moreover, compatibility criteria differ from analysis to analysis.
 
In this paper, we study the compatibility of the neutrino data by performing global analyses that include both charged-lepton data and neutrino DIS data. To extend the previous analyses, we include data sets that were not used in Ref.~\cite{Kovarik:2010uv}. Specifically, in addition to the charged-lepton DIS, DY, and neutrino DIS data from NuTeV and Chorus, we now include the $W$ and $Z$ boson production data from the LHC~\cite{AtlasWpPb,Aad:2015gta,Khachatryan:2015hha,Khachatryan:2015pzs,Sirunyan:2019dox,ALICE:2016rzo,Aaij:2014pvu}, single inclusive hadron production data from both RHIC~\cite{Adler:2006wg,PHENIX:2013kod,Abelev:2009hx,STAR:2006xud} and the LHC \cite{ALICE:2016dei,ALICE:2018vhm,ALICE:2021est}, charm-dimuon data from NuTeV and CCFR~\cite{Goncharov:2001qe}, and neutrino DIS data from CDHSW~\cite{Berge:1989hr} and CCFR~\cite{CCFRNuTeV:2000qwc,Yang:2001rm}. Furthermore, we improve on the treatment of the deuteron corrections which are applied to $F_2$ theory predictions. We also improve the treatment of normalization uncertainties by fitting their fluctuations to the data. To have maximal discriminatory power from the highly correlated data like NuTeV and Chorus, we take into account their correlated systematic uncertainties in all fits. We also allow the strange quark PDF parameters to vary, in contrast to our previous analysis \cite{Kovarik:2010uv} where we assumed that they are fixed by requiring $s+\bar{s}=\kappa (\bar{u}+\bar{d})$. As a result of all the aforementioned improvements and additions, the analysis presented in this paper is the most comprehensive analysis of the neutrino DIS data available so far. 

As a result of our compatibility study we also identify several approaches how neutrino DIS data can be used together with the charged lepton DIS data in global nPDF analyses while avoiding much of the tension. We also present the best approach which will be used in our future global release of nCTEQ nPDFs with neutrino data. In the meantime, we also publish the nPDFs obtained in the current analysis which are our most complete set of nPDFs until now.

The remaining part of the paper is organized as follows. The analysis framework that serves as the basis for this work is briefly reviewed in Sec.~\ref{sec:framework}. Section~\ref{sec:data} is dedicated to the neutrino data new to this analysis. This section also contains some preliminary checks of the internal consistency of the neutrino data among themselves. Section~\ref{sec:nuglobal} is the core of this paper and introduces the compatibility criteria used in reaching the conclusions. The main point is the discussion of the compatibility between the charged-lepton and neutrino data. We investigate the impact of data selection, treatment of errors and the kinematic cuts in Sec.~\ref{sec:nufinal}. The details of the combined fit with neutrino and other data are given in Section~\ref{sec:ncteqnu}. The whole study is then summarized in Section~\ref{sec:conclusion} which also provides an outlook and a possible interpretation of the results. In addition, we list the explicit results of all fits performed in the course of this analysis in Appendix~\ref{sec:fitresults} and we discuss normalization issues and our method to handle the d' Agostini bias in Appendix~\ref{sec:app_norm_unc}. 
%

\section{Analysis Framework}
\label{sec:framework}

\subsection{nPDF fitting framework}
The extraction of nuclear PDFs in this analysis is performed using the same framework already employed in the nCTEQ15 analysis~\cite{Kovarik:2015cma} and all our subsequent analyses~\cite{Kusina:2020lyz,Duwentaster:2021ioo}. Specifically, for a nucleus with mass number $A$ the full nPDF, $f_i^A$, is expressed in terms of effective bound-nucleon distributions:
\begin{equation}\label{fiA}
    f_i^A(x, Q) = \frac{Z}{A} f_i^{p/A}(x, Q) + \frac{N}{A} f_i^{n/A}(x, Q),
\end{equation}
where $i$ is a parton flavor, $Q$ is the factorization/evolution scale, $x$ is the fractional momentum of the parton with respect to the average momentum of the nucleons, $Z$ and $N=(A-Z)$ are respectively the number of protons and neutrons inside the nucleus, while $f_i^{p/A}$ and $f_i^{n/A}$ are the effective bound proton and neutron PDFs respectively. The momentum fraction $x$ in this case takes in principle the values $0\leq x\leq A$. However, we assume that $f_i^A(x, Q)=0$ for $x>1$ which is reasonable as long as we neglect the motion of bound nucleons inside the nucleus~\cite{Segarra:2020gtj}.

The bound neutron PDFs can be obtained from the bound proton ones by assuming isospin symmetry. The bound proton PDFs are parametrized at the input scale $Q_0=1.3$ GeV using the following parametrization \cite{Kovarik:2015cma}:
\begin{align}
xf_i^{p/A}(x, Q_0) = c_0 x^{c_1}(1-x)^{c_2}e^{c_3 x}\left(1+e^{c_4} x\right)^{c_5},\\
\frac{\bar{d}(x, Q_0)}{\bar{u}(x, Q_0)} = c_0 x^{c_1} (1-x)^{c_2}+ (1+c_3)(1-x)^{c_4},
\end{align}
where the flavor index $i$ runs over $i=u_v, d_v, g, \bar{u}+\bar{d}, s+\bar{s}, s-\bar{s}$. Here $u_v$ and $d_v$ are the up and down quark valence distributions, and $g,\bar{u},\bar{d}, s, \bar{s}$ are the gluon, anti-up, anti-down, strange, and anti-strange quark distributions, respectively. The free coefficients $c_i$ are assumed to be $A$-dependent and the general form of this dependence is given by
\begin{equation}\label{ck}
c_i(A,Z) = p_i+ a_i(1-A^{-b_i})\,.
\end{equation} 
Here, $p_i$ are the free-proton PDF parameters obtained in a dedicated proton PDF analysis of Ref.~\cite{Owens:2007kp}, which are close in value to the CTEQ6.1M parameters \cite{Stump:2003yu}. We have chosen the free-proton PDF parameters in order to avoid possible inconsistencies when proton PDF analyses use data taken on nuclei. The analysis \cite{Owens:2007kp} excludes all nuclear data such as the CCFR $F_2$ and $F_3$ neutrino DIS data~\cite{CCFRNuTeV:2000qwc}. The nPDFs for different nuclei are obtained by fitting the nuclear parameters $a_i$ and $b_i$ to the experimental data.

In total, there are about 40 $a_i$ and $b_i$ parameters each. Some of these parameters are constrained by the usual sum rules, but the rest remains to be constrained by the data. Given that in the case of nuclear PDFs the data are not so numerous and precise as in the proton case, many of the free parameters need to be fixed in any nPDF analysis. Comparing two different nPDF extractions can be made difficult if the analyses in question use vastly different numbers of free parameters. In such a case, parametrization bias becomes an issue which is difficult to overcome. In this analysis we have succeeded to perform every relevant fit containing a sufficient number of data points with the same large number of free parameters. Only for special fits to a very small subset of data, we were forced to use a smaller number of free parameters to reliably estimate the uncertainties of these analyses within the Hessian approach.

In general, even though the $A$-dependence of the parton distribution functions given in Eq.~(\ref{ck}) allows for great flexibility, there is insufficient data to constrain the whole functional form. Therefore, we opt to fix most of the $b_i$ coefficients and let them vary only in cases where we expect precise data taken on multiple nuclei can constrain them.
\subsection{nCTEQ15WZSIHdeut}\label{sec:nCTEQ15}
\begin{figure}[t]
    \centering
 \includegraphics[width=0.9\columnwidth]{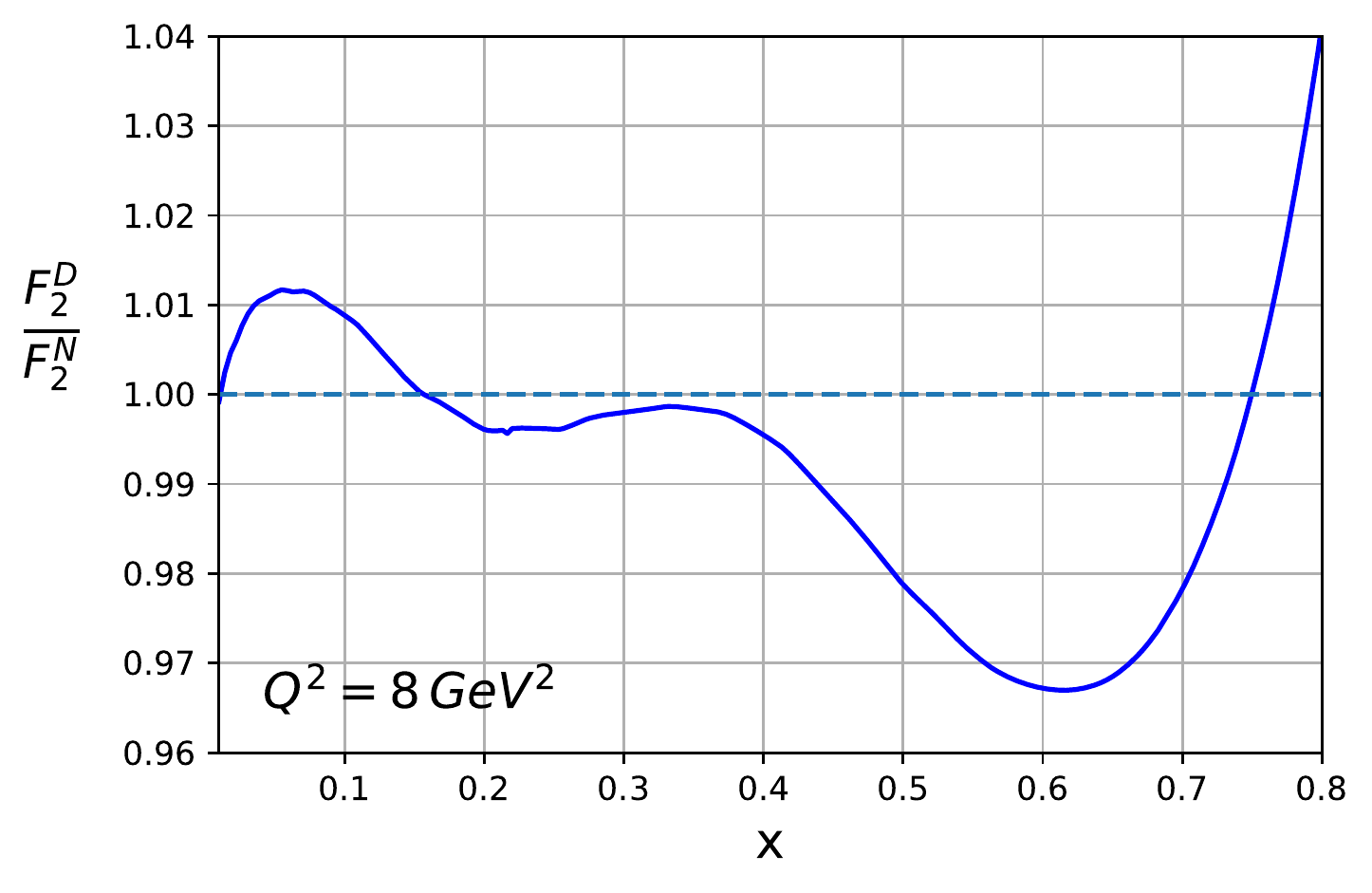}
    \caption{The ratio $F_2^{\mathrm{D}}/F_2^N$ of deuteron to isoscalar structure functions at $Q^2=8$ GeV$^2$, where $F_2^{\mathrm{D}}$ is computed using Eq.~\eqref{F2D}. }
    \label{fig:F2D/F2N}
\end{figure}
\begin{table*}[h!]
\renewcommand{\arraystretch}{1.2}  
\caption{Comparison of the $\chi^2$/pt for the nCTEQ15, nCTEQWZSIH and nCTEQ15WZSIHdeut analyses for selected data sets. Numbers appearing inside brackets show the $\chi^2$/pt values for data sets that are not used in the corresponding fits.\label{tab:Chi}}
\centering
\begin{tabular}{|c|c|c|c|c|c|c|c|c|c|c|c|c|c|c|c|c|c|}
\hline 
 & \multicolumn{3}{c|}{ATLAS Run I} & \multicolumn{3}{c|}{CMS Run I} & \multicolumn{2}{c|}{CMS Run II} & \multicolumn{2}{c|}{ALICE} & LHCb &
 & DIS & DY & SIH & $W$,$Z$ & {\bf~Total~}    \tabularnewline
\cline{1-12} \cline{2-12} \cline{3-12} \cline{4-12} \cline{5-12} \cline{6-12} \cline{7-12} \cline{8-12} \cline{9-12} \cline{10-12} \cline{11-12} \cline{12-12} \cline{12-12} 
 & $W^{-}$ & $W^{+}$ & $Z$ & $W^{-}$ & $W^{+}$ & $Z$ & $W^{-}$ & $W^{+}$ & $W^{-}$ & $W^{+}$ & $Z$ & &  &  &  & LHC  & \tabularnewline
\hline \hline
 nCTEQ15 & (1.38) & (0.71) & (2.88) & (6.13) & (6.38) & (0.05) & (9.65) & (13.20) & (2.30) & (1.46) & (0.70) & & 0.91 & 0.73 & (0.25) & (6.20) & {\bf 1.66} 
 \tabularnewline
\hline \hline
nCTEQ15WZSIH & 0.64 & 0.26 & 1.76 & 1.31 & 1.16 & 0.11 & 0.74 & 1.14 & 0.76 & 0.04 & 0.56 & & 0.91 & 0.78 & 0.41 & 0.91  &  {\bf 0.83} \tabularnewline
\hline \hline
nCTEQ15WZSIHdeut & 0.56 & 0.37 & 1.33 & 1.01 & 1.13 & 0.13 & 0.70 & 0.90 & 0.75 &  0.05 &  0.63 & &0.85 & 0.79 & 0.45 & 0.77   &  {\bf 0.78} \tabularnewline
\hline
\end{tabular}
\end{table*}
\begin{figure*}[h!]
    \centering
   \includegraphics[width=0.95\textwidth]{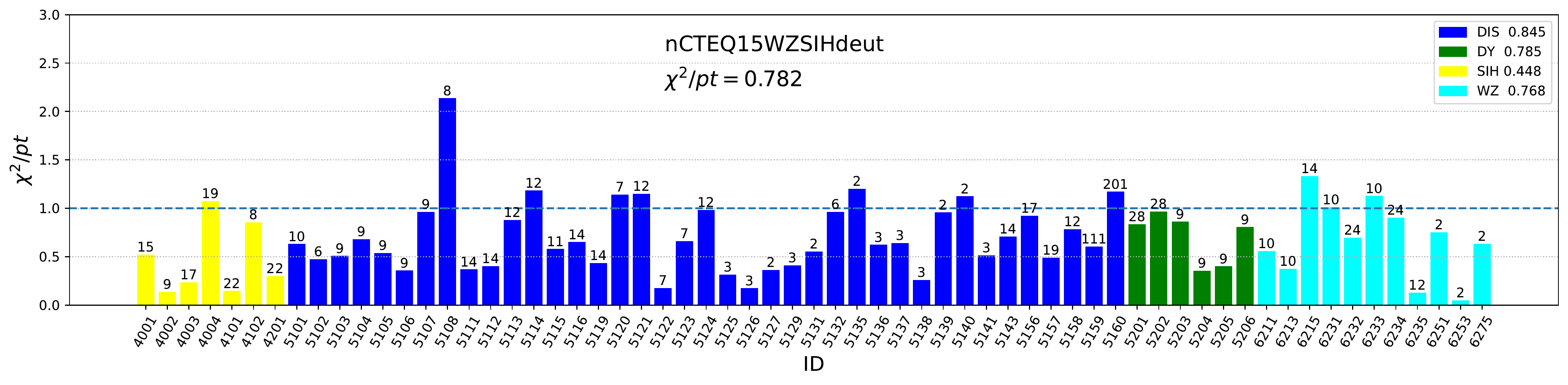}
    \caption{Values of $\chi^2$/pt for the nCTEQ15WZSIHdeut fit for individual experiments.\footnote{We find the DIS experiment 5108 (Sn/D EMC-1998) to be an outlier and our result is consistent with other results from literature.} The IDs of the experiments can be found in Tabs.~I-IV of Ref.~\cite{Kovarik:2015cma}, Tab.~II of Ref.~\cite{Kusina:2020lyz} and Tab.~I of Ref.~\cite{Duwentaster:2021ioo}.}
    \label{fig:ncteq15wzchi2}
\end{figure*}
\begin{figure*}[h!]
    \centering
   \includegraphics[width=0.95\textwidth]{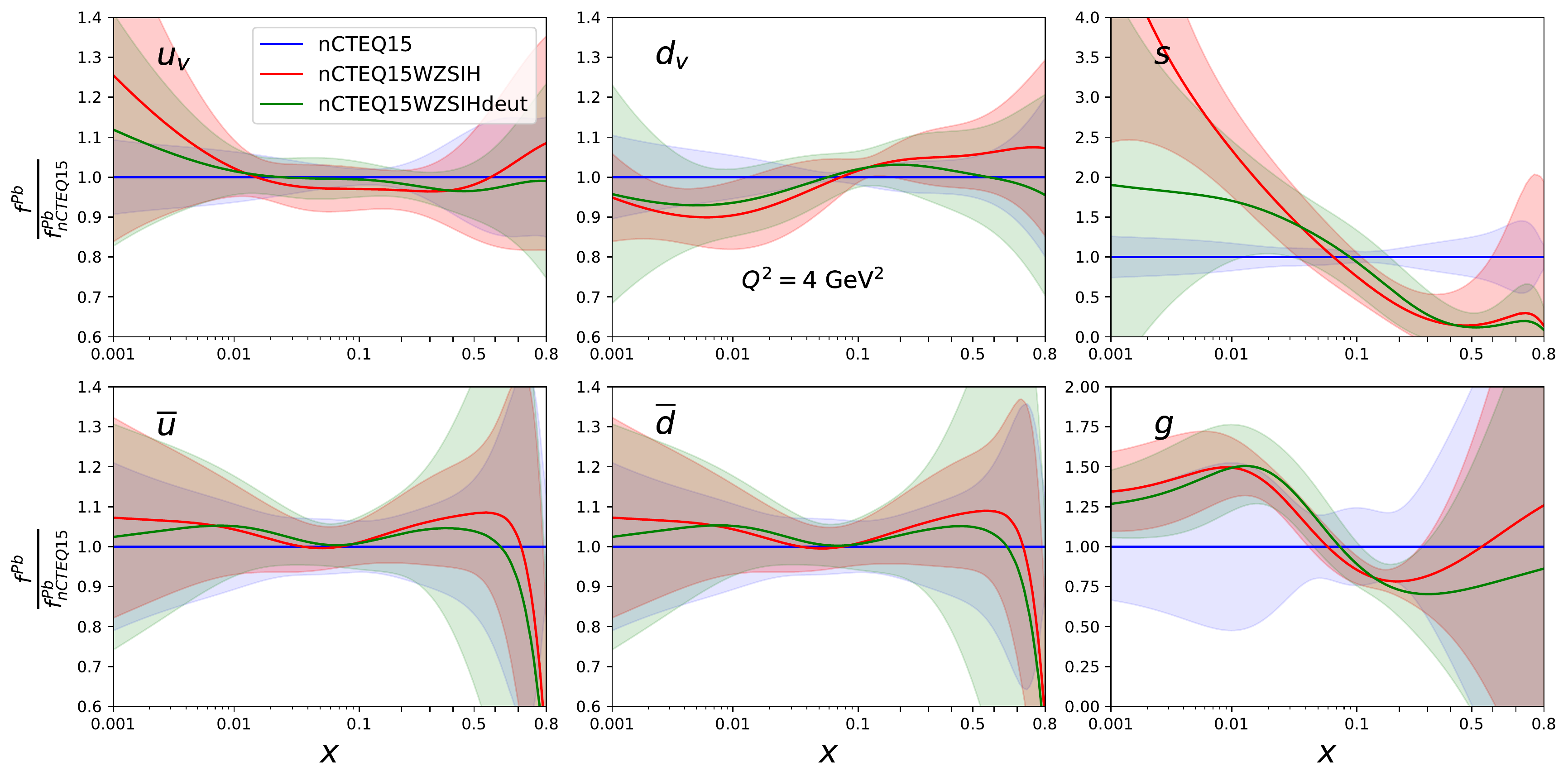}
    \caption{The ratio of nuclear parton distribution functions of the nCTEQ15WZSIH and nCTEQ15WZSIHdeut analyses with respect to the nCTEQ15 analysis for lead at the scale $Q^2=4\ {\rm GeV}^2$.}
    \label{fig:pdfncteq15wz}
\end{figure*}
Before discussing the neutrino data, we need to carefully specify the nPDFs we will compare our results against. The global analysis that we use as a reference here is based on the recent nCTEQ15WZSIH analysis~\cite{Duwentaster:2021ioo} which uses charged lepton DIS, DY, LHC $W$ and $Z$ boson production data and single inclusive hadron production data from both RHIC and LHC to determine the nPDFs. 

However, we improve upon the nCTEQ15WZSIH analysis in several respects. First, we remove the isoscalar corrections that were applied when the data were published using the same method as used in Ref.~\cite{Segarra:2020gtj}, to improve the up- and down-quark PDF separation. Moreover, in order to take into account the nuclear corrections in deuteron data, we correct the deuteron $F_2$ structure function predictions using the method discussed in Ref.~\cite{Segarra:2020gtj}. Specifically, the deuteron $F_2^{\mathrm{D}}$ is computed as
\begin{equation}\label{F2D}
F_2^{\mathrm{D}} = F_2^{p, nCTEQ15} \times \frac{F_2^{\mathrm{D}, CJ}}{F_2^{p,CJ}}
\end{equation}
where $F_2^{\mathrm{D}, CJ}$ and $F_2^{p, CJ}$ are the fitted deuteron and proton structure functions from the CJ15 analysis~\cite{Accardi:2016qay} and $F_2^{p, nCTEQ}$ is the computed proton structure function using our base proton PDFs. Without this method, the deuteron $F_2$ is traditionally computed as a simple isoscalar combination, $F^N_2 \equiv F^p_2+F_2^n$ \cite{Kovarik:2015cma, Eskola:2016oht}. In Fig.~\ref{fig:F2D/F2N}, we show the ratio $F_2^{\mathrm{D}}/F_2^N$ at $Q^2=8$ GeV$^2$. 
We can see that our treatment for the deuteron structure function modifies $F_2^N$ by $\sim 1\%$ at $x\leq 0.1$ and $\sim 3.5\%$ at $x\approx 0.65$. The different treatment of the deuteron structure function influences the description of all the charged-lepton DIS data which are published as ratios $F_2^A/F_2^{\mathrm{D}}$. This set of data includes data taken on a wide range of nuclear targets and it constitutes about a half of the data in the nCTEQ15WZSIH analysis.

For DIS data, we apply our standard kinematic cuts namely we only keep data with $Q^2>4$ GeV$^2$ and $W^2= M_p^2+Q^2(1-x)/x >12.25$ GeV$^2$, where $M_p$ is the nucleon mass.%
    \footnote{We refrain from using less restrictive kinematic cuts like the ones in our recent analysis of JLab data~\cite{Segarra:2020gtj} as we want to stay in the purely perturbative regime and we do not want to complicate the picture by additional effects like the higher twist or the target mass corrections.}
As in \cite{Duwentaster:2021ioo}, we use the same strict $p_T\geq 3$ GeV cut for all single inclusive hadron data (compared to $p_T\geq 1.7$ GeV in nCTEQ15 and EPPS16). We have repeated the nCTEQ15WZSIH analysis with all corrections and cuts mentioned above and enlarged the set of free parameters from 19 to 27. Specifically we fit: 
\begin{align*}
    & a_1^{u_v},\; a_2^{u_v},\; a_4^{u_v},\; a_5^{u_v},\; b_1^{u_v},\; b_2^{u_v}, \\
    & a_1^{d_v},\; a_2^{d_v},\; a_4^{d_v},\; a_5^{d_v},\; b_1^{d_v},\; b_2^{d_v}, \\
    & a_1^{\bar{u}+\bar{d}},\; a_2^{\bar{u}+\bar{d}},\; a_5^{\bar{u}+\bar{d}}, \\ 
    & a_1^{g},\; a_4^{g},\; a_5^{g},\; b_0^{g},\; b_1^{g},\; b_4^{g},\; b_5^{g}, \\ 
    & a_0^{s+\bar{s}},\; a_1^{s+\bar{s}},\; a_2^{s+\bar{s}},\; b_0^{s+\bar{s}},\; b_2^{s+\bar{s}} \; .
\end{align*}
On top of these free parameters, there are 10 additional free normalisation parameters which are also determined in the fit using the approach highlighted in App.~\ref{sec:app_norm_unc}. Similar to the analysis presented in \cite{Duwentaster:2021ioo}, 7 normalisation parameters are used to describe the single inclusive hadron experimental data and 3 normalisations are used for the description of the $W$- and $Z$-boson production measurements from the LHC. After fitting 940 data points from the same experiments that were also used in the nCTEQ15WZSIH analysis \cite{Duwentaster:2021ioo}, we obtain a $\chi^2=735$ corresponding to $\chi^2$/pt = 0.782.

The list of values of all parameters obtained in this analysis is given in App.~\ref{sec:fitresults}. In the following text we refer to this new analysis as nCTEQ15WZSIHdeut. For completeness, in Tab.~\ref{tab:Chi}, we compare the quality of the new nCTEQ15WZSIHdeut fit with the previous nCTEQ15WZSIH and the nCTEQ15 analyses. The values of $\chi^2$/pt for each experiment are displayed in Fig.~\ref{fig:ncteq15wzchi2}. The resulting PDFs are then compared for all relevant flavours at the scale $Q^2= 4\ {\rm GeV}^2$ in Fig.~\ref{fig:pdfncteq15wz}. For comparison, we use the same $\Delta\chi^2 =45$ tolerance to define the uncertainties for all three analyses. There are several differences which can be observed between the original nCTEQ15WZSIH and the nCTEQ15WZSIHdeut analyses. In all parton flavors, we observe larger uncertainties compared to the nCTEQ15WZSIH analysis. This is connected to the enlarged number of free parameters which now can more realistically describe the true uncertainty. The differences in the central values for the up- and down-quark parton distributions are the expected consequences of removing the isoscalar corrections and of the different treatment of the deuterium in DIS data together with a slightly larger number of free parameters. The differences seen in the gluon distribution can be attributed to different free parameters used to describe the gluon PDF as well as secondary effects on the gluon from altered scaling violations coming from the modified deuteron data. In the case of the strange quark, the only constraint comes from the $W$ and $Z$ boson data from the LHC as well as the sum rules linking all PDFs together. Given the lack of data constraining the strange quark, we conclude that what is displayed in Fig.~\ref{fig:pdfncteq15wz} is just the parametrization bias where even our parametrization with a large number of free parameters cannot reproduce the true uncertainty in the determination of the strange quark PDF, which should be regarded as much wider than the plotted bands in Fig.~\ref{fig:pdfncteq15wz}. It is here where neutrino DIS could play a major role in a global PDF analysis, providing additional sensitivity to the strange quark PDF. 
\section{Neutrino DIS data}
\label{sec:data}
\subsection{Neutrino data and observables}
As in any global analysis, data selection is an important factor which, as previous analyses of neutrino data show, can largely influence the obtained results. Given that we investigate the compatibility of neutrino DIS data with the rest of nuclear data, we aim at including all available neutrino DIS data. The experimental collaborations usually publish their results for different observables as differential cross-sections or structure functions. Given that the structure functions are extracted from the cross-section data and that this extraction often requires certain assumptions or input from theory, we prefer to use the differential cross-section data whenever possible.

There are two kinds of neutrino data included in the current analysis. All the new data with a breakdown of the number of neutrino and anti-neutrino DIS cross-section data points that satisfy the kinematic cuts $Q^2>4$ GeV$^2$ and $W^2>12.25$ GeV$^2$ applied in our analysis are listed in Tab.~\ref{tab:nudata}. We also give the range of (anti-)neutrino energy bins for each data set.

The largest and the most important contribution comes from the measurements of the inclusive double-differential cross section for the scattering of neutrinos and anti-neutrinos on iron or lead nuclei. The data taken on iron targets come from the CDHSW \cite{Berge:1989hr}, CCFR \cite{CCFRNuTeV:2000qwc,Yang:2001rm} and NuTeV \cite{Tzanov:2005kr} collaborations whereas Chorus \cite{Onengut:2005kv} data are taken on lead. For Chorus, CCFR and NuTeV data the electroweak corrections were applied directly to the experimental data. The Chorus and NuTeV data provide point-by-point correlated systematic uncertainties which we include in our analysis.\footnote{The correlated systematic uncertainties for NuTeV data have been used but not given explicitly in the official publication \cite{Tzanov:2005kr}. They can be found in the supplemental material of the corresponding arXiv submission.} There is one issue that needs to be mentioned here. Given that the NuTeV experiment was conceived as a follow-up experiment to the older CCFR experiment and given that in \cite{Tzanov:2005kr} it was claimed that the CCFR experiment had issues such as with mapping of the magnetic field affecting the measurements at large $x$, we apply a cut excluding all CCFR data with $x>0.4$. Apart from the data mentioned before, there have been measurements of neutrino DIS reported by the NOMAD \cite{NOMAD:2007krq,Petti:2006tu}, IceCube \cite{IceCube:2017roe} and Minerva \cite{PhysRevD.93.071101} collaborations which we do not consider in this analysis for different reasons. The NOMAD cross-section data would be the most promising given the high statistics and given that the data were taken on multiple nuclear targets. Unfortunately, the inclusive differential cross-section data have never been publicly released. The IceCube data are measured at extremely small $x\sim 10^{-6}$ where a possibly different theoretical treatment might be required and come with large uncertainties. Finally, the latest results come from the MINER$\nu$A neutrino scattering experiment on polystyrene, graphite, iron and lead targets. The collaboration published the ratio of the neutrino scattering single-differential cross section, $d\sigma/dx$, as function of $x$ and neutrino energy $E_\nu$. Unfortunately the average virtuality $\langle Q^2\rangle$ is below the $Q^2=4$ GeV$^2$ threshold and so the data are excluded from the analysis by our kinematic cuts.

The second class of data we consider is the semi-inclusive production of di-muons in (anti-)neutrino DIS measured by the NuTeV and CCFR experiments \cite{Goncharov:2001qe}. There are additional numerous data from the CDHS \cite{Abramowicz:1982zr}, Chorus \cite{CHORUS:2008vjb} and NOMAD \cite{NOMAD:2013hbk} collaborations which we do not include in our analysis. The older data from CDHS and Chorus experiments provide no additional constraint compared to the di-muon data we include. The NOMAD data are more precise but due to technical difficulties we were unable to make use of them in this analysis. However, at the end of this paper, we compare the results of our analysis against the NOMAD data and show that the theoretical prediction from the final result of our analysis correctly describes the data. Still, precision of the NOMAD data suggests that further studies of their PDF constraints could be valuable.
%
\begin{table}[tb]
\label{numpoints}
	\caption{New neutrino data sets used in this analysis.}\label{tab:nudata}
\begin{center}
\resizebox{\columnwidth}{!}{
	\begin{tabular}{|lccccc|}
	\hline
		Data set &  Nucleus & $E_{\nu/\bar{\nu}}$(GeV) &  \#pts & Corr.sys. & Ref.\\
		\hline\hline
		CDHSW $\nu$ & \multirow{2}{*}{Fe} & \multirow{2}{*}{23 - 188}  & 465 & \multirow{2}{*}{No} & \multirow{2}{*}{\cite{Berge:1989hr}}\\
		CDHSW $\bar{\nu}$ & & & 464 & &\\ \hline
		CCFR $\nu$ & \multirow{2}{*}{Fe} & \multirow{2}{*}{35 - 340} & 1109 & \multirow{2}{*}{No} & \multirow{2}{*}{\cite{Yang:2001rm}}\\
		CCFR $\bar{\nu}$ & & & 1098 & &\\ \hline
		NuTeV $\nu$ & \multirow{2}{*}{Fe} & \multirow{2}{*}{35 - 340} & 1170 & \multirow{2}{*}{Yes} &\multirow{2}{*}{\cite{Tzanov:2005kr}}\\
		NuTeV $\bar{\nu}$ & & & 966 & &\\ \hline
		Chorus $\nu$ &  \multirow{2}{*}{Pb} & \multirow{2}{*}{25 - 170} & 412 & \multirow{2}{*}{Yes} & \multirow{2}{*}{\cite{Onengut:2005kv}}\\
		Chorus $\bar{\nu}$ & & & 412 & &\\ \hline
		CCFR dimuon $\nu$ & \multirow{2}{*}{Fe} & 110 - 333 & 40 & \multirow{2}{*}{No} & \multirow{2}{*}{\cite{Goncharov:2001qe}}\\
		CCFR dimuon $\bar{\nu}$ & & 87 - 266 & 38 & &\\ \hline
		NuTeV dimuon $\nu$ & \multirow{2}{*}{Fe} & 90 - 245 & 38 & \multirow{2}{*}{No} & \multirow{2}{*}{\cite{Goncharov:2001qe}}\\
		NuTeV dimuon $\bar{\nu}$ & & 79 - 222 & 34 & &\\
		\hline
	\end{tabular}
}
\end{center}
\end{table}
%

It is not a simple task to compare the precision of different experimental measurements if the measurements extend over different kinematic regions or include correlated systematic uncertainties. However, we show the results of a simplified comparison of the measurements of inclusive (anti-)neutrino DIS double-differential cross-sections in Tab.~\ref{tab:dataunc}. We choose an incoming neutrino energy $E_\nu\sim 85$ GeV which is common and typical for each of the experiments and average over the uncertainties (statistical and systematical errors are added in quadrature) for the corresponding data at the given neutrino beam energy. Due to the oversimplifications contained in this comparison we cannot draw very detailed conclusions but we clearly see a general trend. The neutrino data are much more precise than their anti-neutrino counterparts. This conclusion is true also for the remaining data not considered in Tab.~\ref{tab:dataunc}. For neutrino data, we see that at this energy NuTeV and CCFR data are the most precise, followed by the data from Chorus and CDHSW. For anti-neutrino data, the order is somewhat different: NuTeV and CDHSW are comparable in precision, followed by CCFR and Chorus. This conclusion has to be taken with a grain of salt. The averaging procedure and most importantly discarding the correlations might change this simple picture. We will perform much more detailed studies in the following.
%
\begin{table}[tb]
	\caption{Relative experimental uncertainties (in percent) of various data sets at $E_\nu \sim 85$ GeV where all the data sets overlap.}   \label{tab:dataunc}
	\centering
	\begin{tabular}{|lcc|}
		\hline
		Experiment & \#pts & Relative Error($\%$)  \\ \hline \hline
        CDHSW $\nu$  & 59 & 8.36\\
        CDHSW $\bar{\nu}$ & 59& 10.75\\ \hline
        CCFR $\nu$ & 54 & 6.01 \\
        CCFR $\bar{\nu}$ & 54 & 16.90\\ \hline
        NuTeV $\nu$ &  55 & 5.88\\
        NuTeV $\bar{\nu}$  & 54  & 10.29\\ \hline
        Chorus $\nu$  & 65 & 7.70\\
        Chorus $\bar{\nu}$  & 65 & 18.32\\
        \hline
	\end{tabular}
\end{table}
%
\subsection{Nuclear corrections from neutrino cross-section data}\label{sec:weightedaverage}
%
\begin{figure*}[t]
    \centering
 \includegraphics[width=0.4\textwidth]{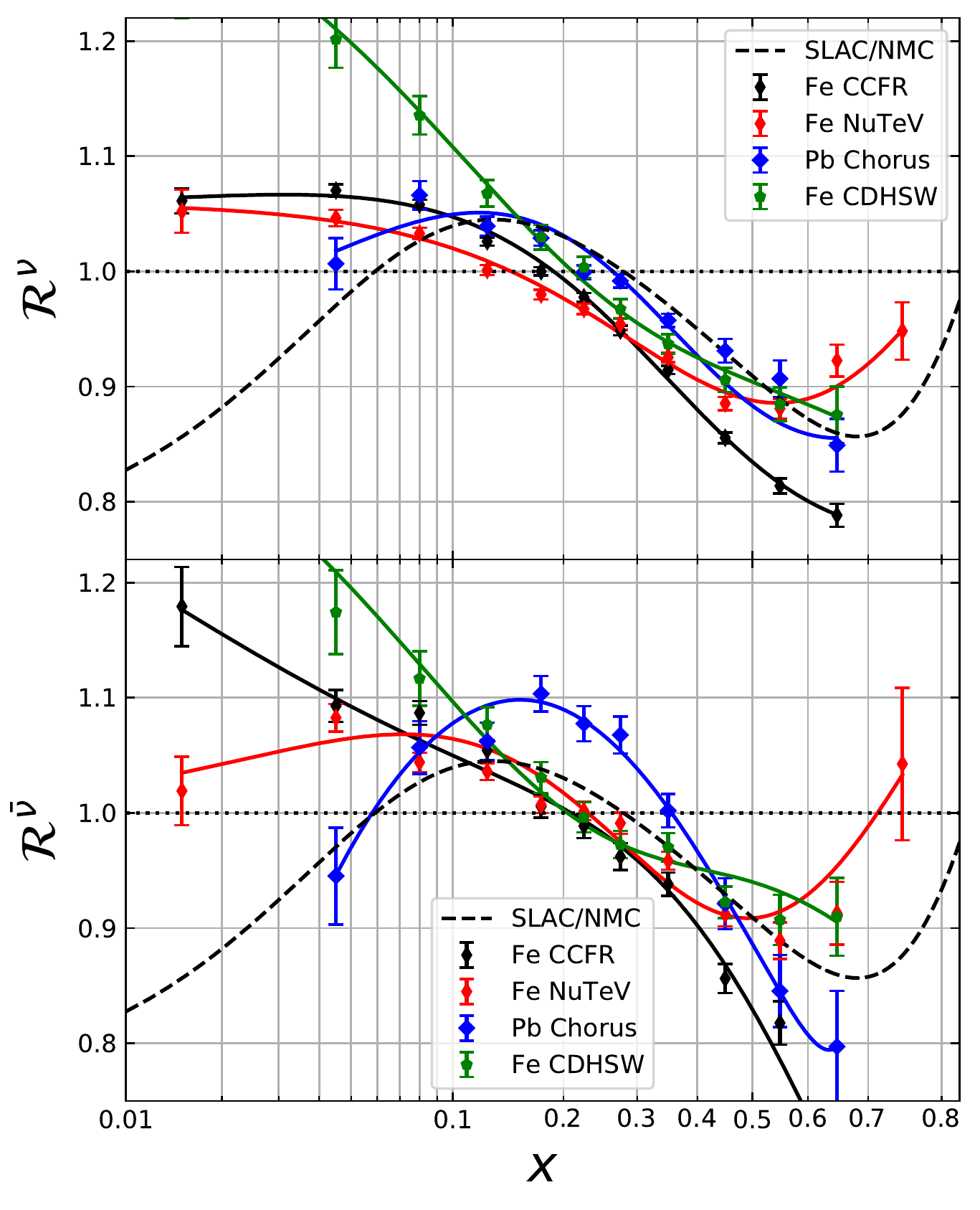}\hfil
  \includegraphics[width=0.4\textwidth]{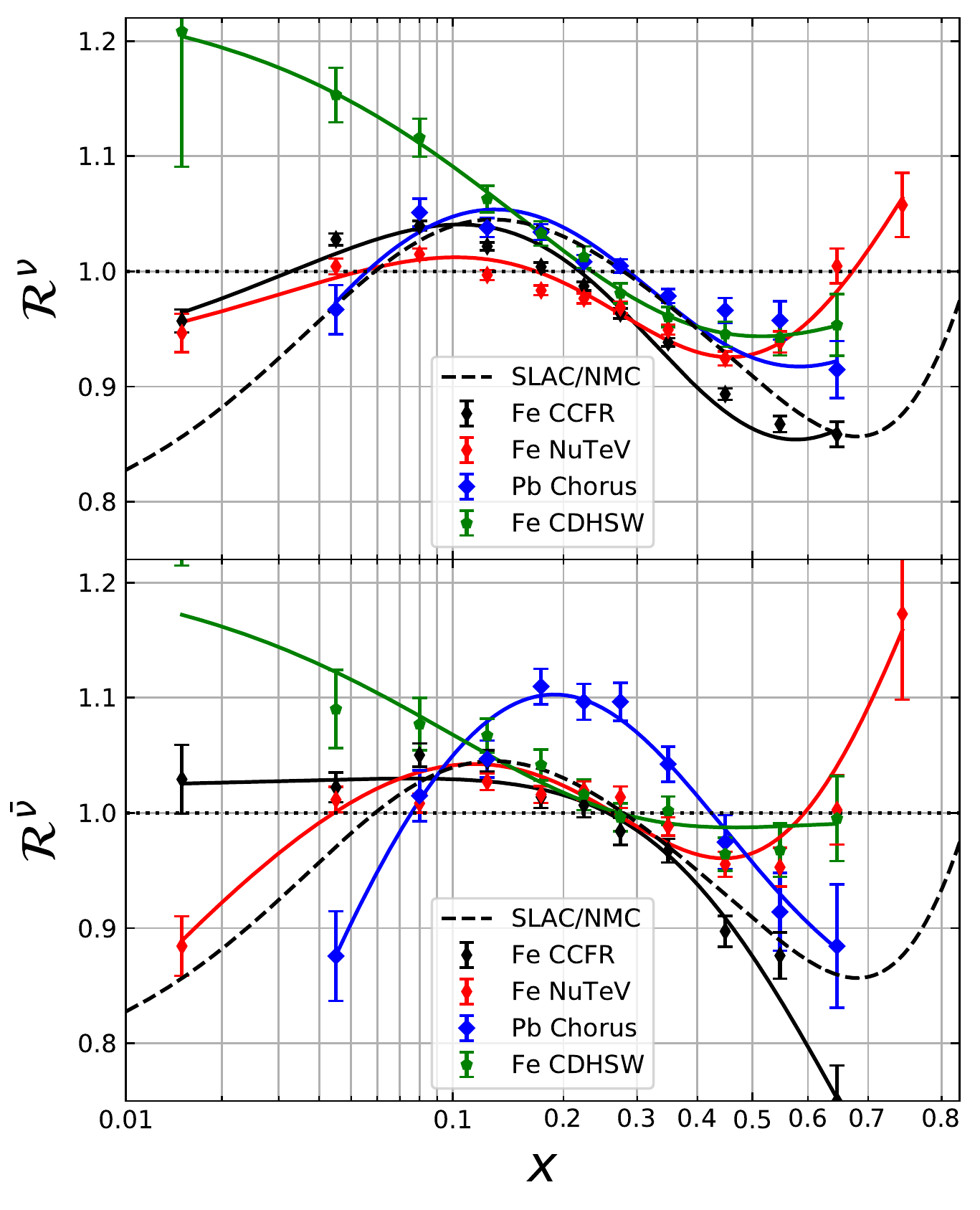}
    \caption{ The weighted average of the cross-section ratios for $Q^2>4$ GeV$^2$ and $W^2>12.25$ GeV$^2$ from CDHSW, CCFR, NuTeV, and Chorus data. The denominator ($\sigma_{free}$) is computed using nCTEQ15 proton baseline (left) and CT18 (no nu A) NLO proton PDFs without neutrino data of Ref.~\cite{Accardi:2021ysh} (right).
    }
    \label{fig:Rnunubar}
\end{figure*}
%
Before we perform a global analysis including the neutrino data in our nPDF framework, it is instructive to attempt to quantify a nuclear correction factor extracted purely from these data alone. Given that the neutrino double-differential cross-section data are reported as a function of the usual DIS variables $x, \,y,\,$and $E_{\nu}$, while the nuclear ratio is typically given only as a function of $x$ assuming the variation with changing $Q^2$ is small, an averaging procedure is necessary. We define the nuclear ratio of the cross-section and its uncertainty for each data point as
\begin{eqnarray}\label{Rsigma}
    R_i^\sigma(x) &=& \frac{\sigma(x, y_i, E_i)}{\sigma_{\rm free}(x, y_i, E_i)}\,,\\ \label{DRsigma}
    \Delta R_i^\sigma(x) &=&  \frac{\Delta\sigma(x, y_i, E_i)}{\sigma_{\rm free}(x, y_i, E_i)}\,,
\end{eqnarray}
where $\sigma_{\rm free}$ is the predicted differential cross section using ``free'' iron or lead PDFs, $f_i^{A,{\rm free}}$, defined by 
\begin{equation}\label{ffree}
f_i^{A,{\rm free}} = \frac{Z}{A}f_i^{p}+ \frac{A-Z}{A}f_i^{n} \; .
\end{equation}
Here, $f_i^{p (n)}$ are the free proton (neutron) PDFs, which in our case are taken from our proton baseline. The quantity $\Delta\sigma(x, y_i, E_i)$ is the total sum of statistical and systematic uncertainties for the data points added in quadrature, except for the normalization uncertainty. We construct a weighted average of the nuclear ratios, such that for a given $x$ the weighted-average ratio and its uncertainty are:
\begin{eqnarray}
    \mathcal{R}(x)  &=& \sum_{i} w_i R^\sigma_i,\\
    \Delta\mathcal{R}(x) &=& \left(\sum_{i} w_i^2(\Delta R^\sigma_i)^2\right)^{1/2} \; .
\end{eqnarray}
The weight $w_i$ is defined as
\begin{equation}
    w_i = \left( \sum_{j} \frac{1}{(\Delta R^\sigma_j)^2}\right)^{-1} \frac{1}{(\Delta R^\sigma_i)^2}\,,
\end{equation}
where the sum runs over data points with the same $x$. This averaging procedure is similar to the one used in Ref.~\cite{Paukkunen:2013grz}, although there are differences in the definition of the weight $w_i$ and of the uncertainty $\Delta\mathcal{R}(x)$. In such a procedure the dependence on the remaining variables is averaged out. This of course is only reasonable if there is just a mild dependence of the nuclear correction factor on the remaining variables.  We have checked that this assumption is reasonably valid for a wide range of $Q^2$ and $y$ within the kinematic range allowed by our cuts. Some deviations from this assumption can be observed below $x=0.015$ and above $x=0.75$, where $R$ can be spread around unity quite widely. Therefore, any inference based on this averaging procedure in these regions should be done with caution. 

In Fig.~\ref{fig:Rnunubar}, we show the nuclear correction factors $\mathcal{R}^\nu(x)$ and $\mathcal{R}^{\bar{\nu}}(x)$ obtained from the inclusive neutrino and anti-neutrino cross-section data from CDHSW, CCFR, NuTeV and Chorus. To better compare the shape of the nuclear corrections from different data sets, we also show an interpolation (solid lines), obtained from fits with the parametrization of the ratio~\cite{Tzanov:2005kr}
\begin{equation}
\mathcal{R}(x) = a_1+ a_2x+a_3e^{a_4x}+a_5x^{a_6}.
\end{equation}
For comparison, we also include the SLAC/NMC nuclear correction factor \cite{Abramowicz:1991xz} which approximately describes the nuclear effects in the charged lepton data.

    In the left panels of Fig.~\ref{fig:Rnunubar}, we show the shape of cross-section ratios where $\sigma_{free}$ is computed using our proton baseline PDFs. We observe that the CCFR and NuTeV ratios generally agree at low $x$, but the NuTeV ratio is consistently above the CCFR one for $x>0.4$. This is consistent with the observation in Ref.~\cite{Tzanov:2005kr} where issues with the CCFR experiment were cited which account for this discrepancy. In the following we will also apply a cut $x<0.4$ to the CCFR data. Overall, for the iron neutrino data (CDHSW, CCFR and NuTeV), there is no obvious shadowing, i.e. the appearance of $R<1$, at low $x$ ($x\leq 0.1$) as one expects from the SLAC/NMC model. This is even more so for CDHSW data. However, the bin center correction was not applied for the CDHSW data, which affects largely low- and high-$x$ data~\cite{Tzanov:2005kr}. In contrast to the data on iron, the nuclear ratio obtained from the Chorus data shows a shape more similar to the traditional SLAC/NMC ratio.

\begin{table*}[t!]
	\caption{$\chi^2$/pt value for each data set from the DimuNeu fit.}
	\centering
	\begin{tabular}{|c|c||c|c|c|c||c|c|c|c||c|c|c|c||c|c|c|c||c|c|}
	\hline
	\multicolumn{2}{|c||}{Dimuon} & \multicolumn{2}{c|}{NuTeV $\nu$}
    & \multicolumn{2}{c||}{NuTeV $\bar{\nu}$} & \multicolumn{2}{c|}{CCFR $\nu$} & \multicolumn{2}{c||}{CCFR $\bar{\nu}$} &
	\multicolumn{2}{c|}{Chorus $\nu$} & \multicolumn{2}{c||}{Chorus $\bar{\nu}$} & \multicolumn{2}{c|}{CDHSW $\nu$} & 
	\multicolumn{2}{c||}{CDHSW $\bar{\nu}$} & 
	\multicolumn{2}{c|}{Total} \\ \hline 
	$\chi^2\!$/pt & \#pts & $\chi^2\!$/pt & \#pts & $\chi^2\!$/pt & \#pts & $\chi^2\!$/pt & \#pts & $\chi^2\!$/pt & \#pts & $\chi^2\!$/pt & \#pts &
	$\chi^2\!$/pt & \#pts & $\chi^2\!$/pt & \#pts & $\chi^2\!$/pt & \#pts & $\chi^2\!$/pt & \#pts \\ \hline
	  1.06 & 150 & 1.51 & 1170 & 1.25 & 966 & 1.00 & 824 & 1.00 & 826 & 1.21 & 412 & 1.09 & 412 &
	  0.68 & 465 & 0.72 & 464 & 1.12 & 5689 \\ \hline 
	\end{tabular}
	\label{tab:chi2dimuneu}
\end{table*}
The nuclear ratio defined above obviously depends on the underlying proton PDFs used for the free proton cross-section in the denominator of Eq.~(\ref{Rsigma}). This dependence can be seen when we compare the left and the right panels in Fig.~\ref{fig:Rnunubar}. The right panels show the same nuclear ratios as the ones on the left, but the ratios are constructed using the more recent CT18 NLO PDFs. Here we have used a dedicated fit which does not include any neutrino data in the CT18 analysis to avoid inconsistencies \cite{Accardi:2021ysh}. Comparing the nuclear ratios coming from different underlying proton PDFs, we can clearly see differences in the $x$-shape of these ratios. The largest difference is apparent at low $x$. The ratios constructed from CT18 NLO PDFs show signs of shadowing at $x\leq 0.1$ in contrast to the ones where the nCTEQ15 proton baseline PDFs were used. This should serve as a warning to draw conclusions about the existence of shadowing in neutrino data from observables, which are not purely data driven and depend on some assumptions such as the proton parton distributions.
%
\subsection{Neutrino DIS Data Fit}\label{sec:neutrino}
\begin{figure*}[t!]
    \centering
    \includegraphics[width=0.95\textwidth]{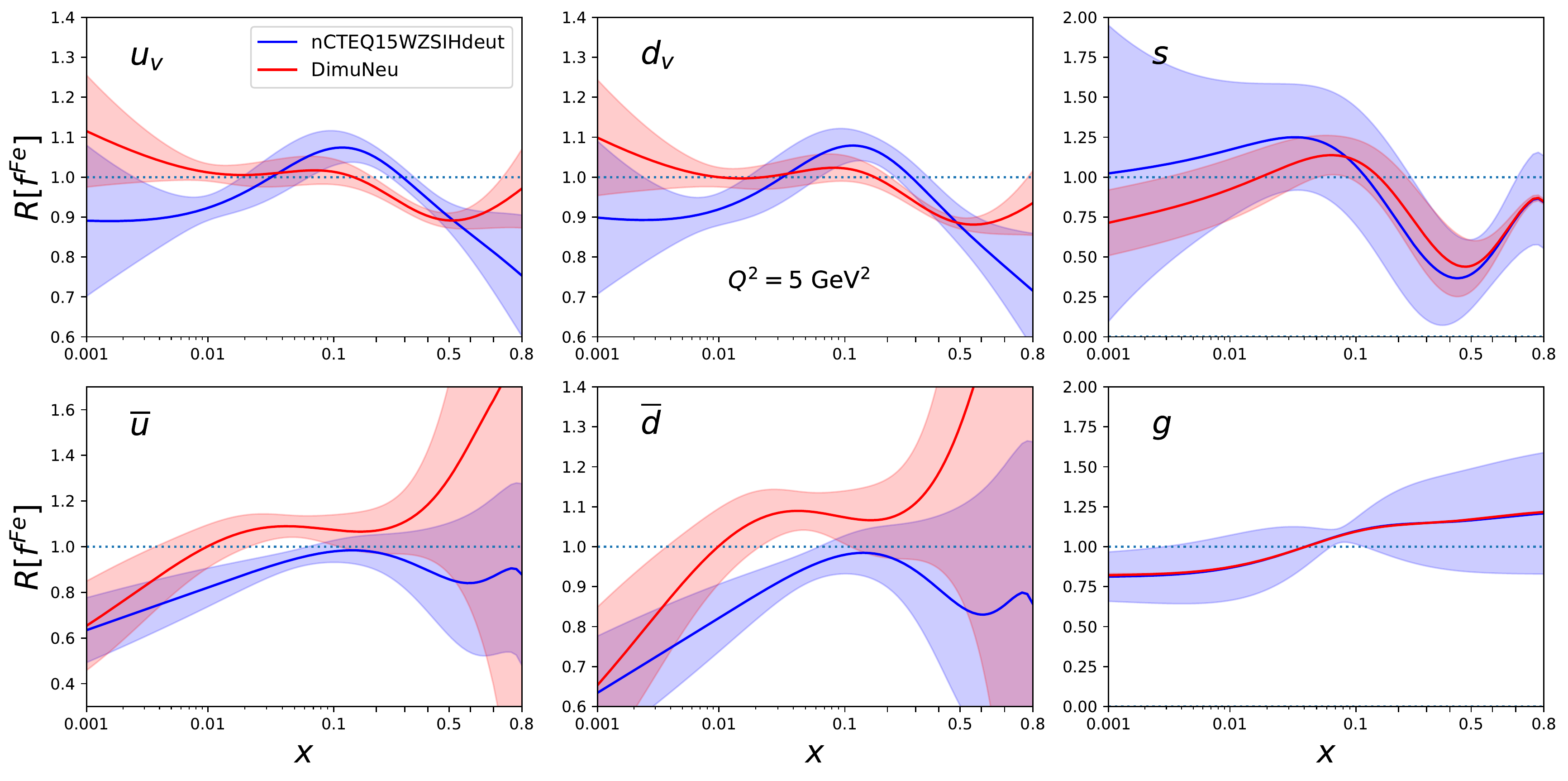}
    \includegraphics[width=0.95\textwidth]{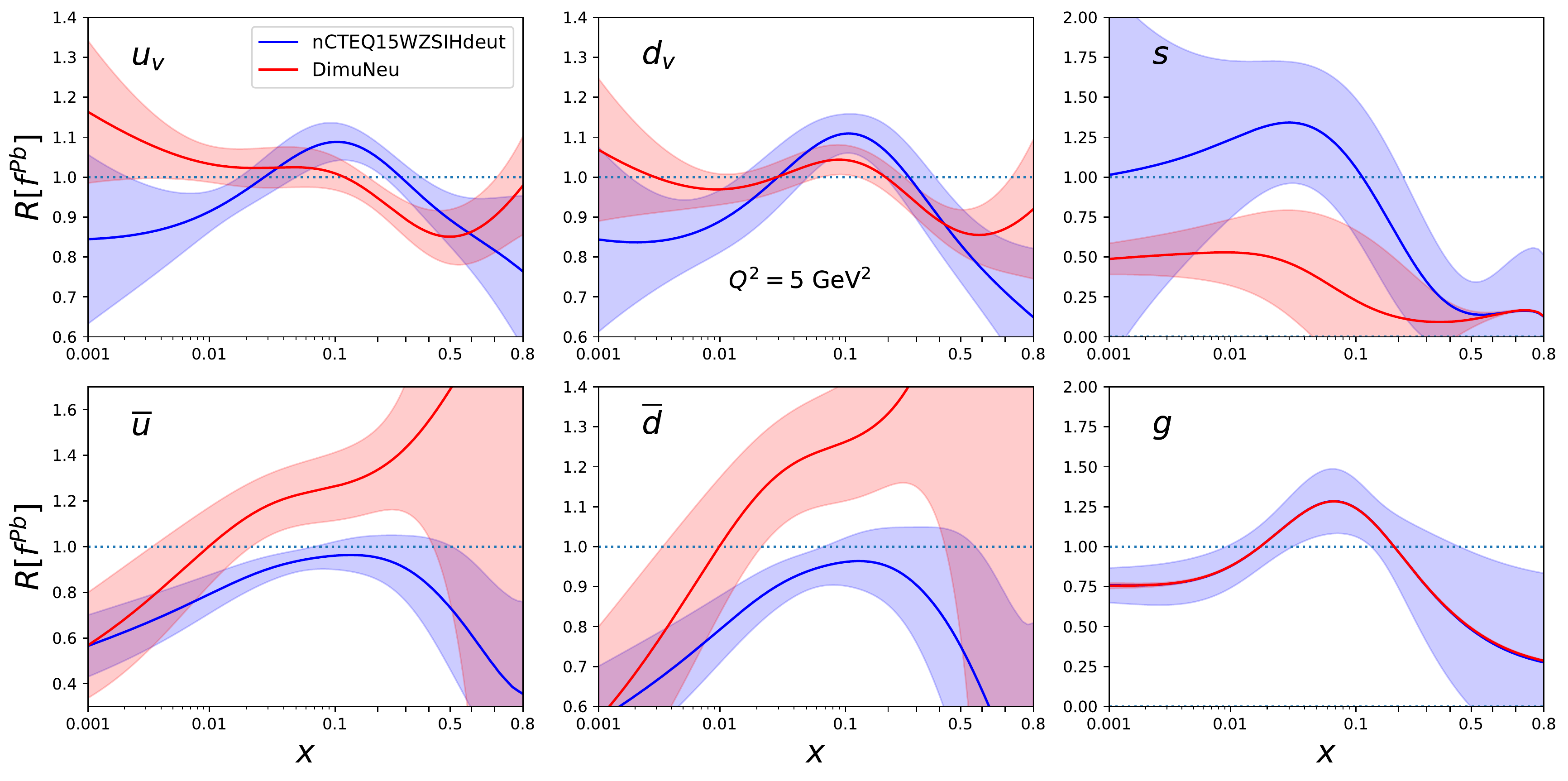}
    \caption{The ratio of nuclear parton distribution functions for the full nuclei - iron $(A=56,Z=26)$ (top) and lead $(A=208,Z=82)$ (bottom) - to the nPDF of full nuclei made up of free protons and neutrons both at the scale $Q^2=5\,{\rm GeV}^2$.}
    \label{fig:dimuneu_pdfs}
\end{figure*}
%
\begin{figure*}[t!]
    \centering
    \includegraphics[width=0.98\columnwidth]{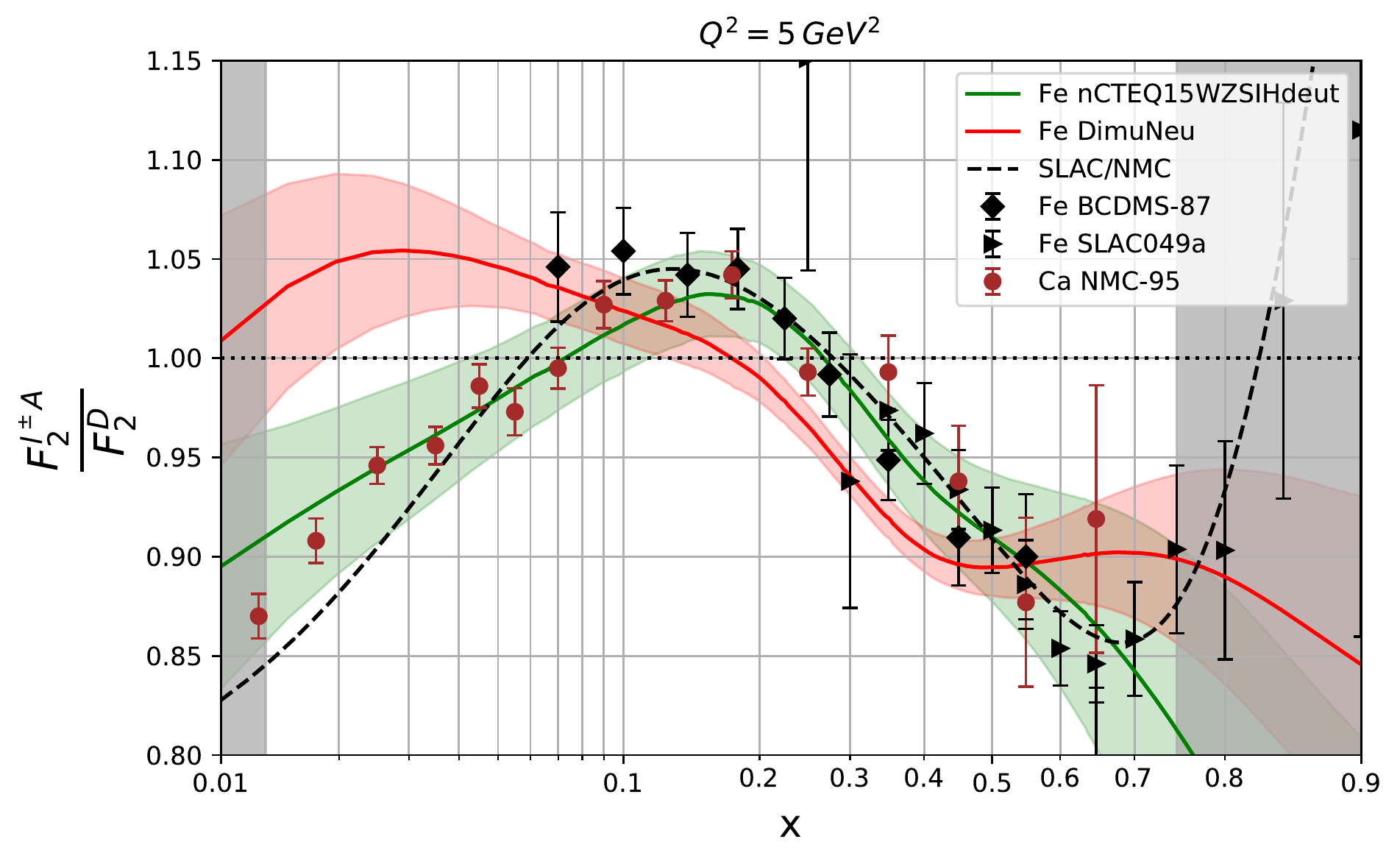}
    \hfil
    \includegraphics[width=0.94\columnwidth]{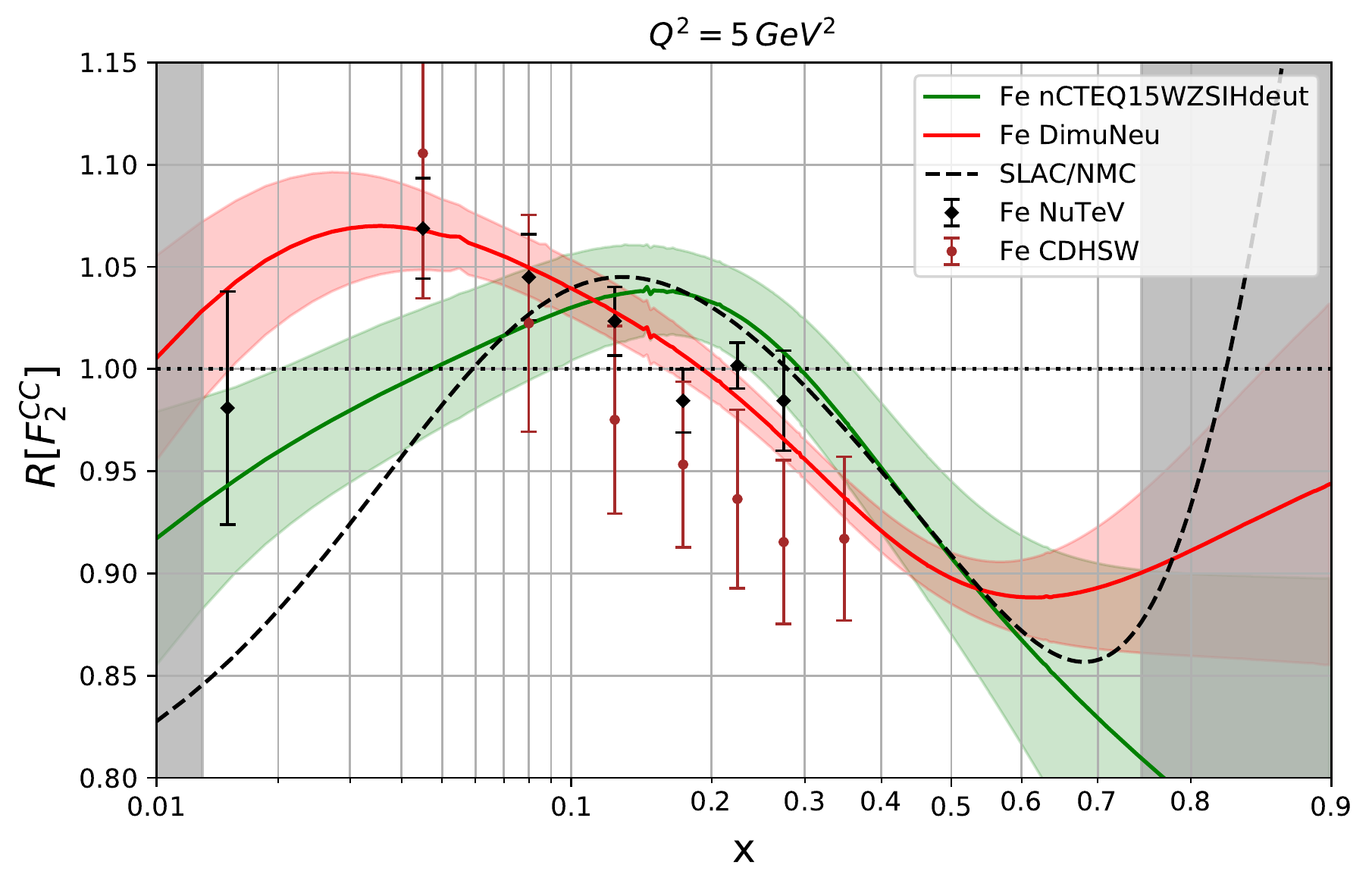}
    \caption{The structure function ratio predictions from DimuNeu and nCTEQ15WZSIHdeut fits. The grey bands on the left and on the right highlight the regions without any data points passing the kinematic cuts. 
    }
    \label{RF2_CC_dimuneu}
\end{figure*}
%
In the previous section, we have investigated the nuclear effects using just the data, constructing the weighted average of cross section ratios. We have observed in Fig.~\ref{fig:Rnunubar} that the resulting $x$-dependence varies between neutrino experiments and is different from the expected SLAC/NMC result. Here we will go one step further and perform a neutrino analysis using the nPDF framework detailed in Sec.~\ref{sec:framework}. In this analysis, which we will refer to as ``DimuNeu'', we include {\bf only} the inclusive and semi-inclusive neutrino data listed in Tab.~\ref{tab:nudata}. Compared to our previous analyses, we improve on the treatment of correlated errors and normalisation uncertainties. The details of this treatment are given in App.~\ref{sec:app_norm_unc}. Before going further, we note that extracting a reliable set of nPDFs from neutrino data alone is not possible without making some assumptions given that the neutrino data alone cannot constrain all possible parton distributions. In this global neutrino analysis, we set the gluon PDF parameters to be the same as those in the nCTEQ15WZSIHdeut fit. Furthermore, we set the $\bar{d}/\bar{u}$ ratio to be the same as in the free proton case, as we assume that the nuclear corrections to $\bar{u}$ and $\bar{d}$ are similar and cancel in the ratio~\cite{Schienbein:2007fs}. This fit therefore uses 20 free parameters. In addition, the normalizations of all data sets are also determined from the fit, which introduces 10 additional free parameters. The uncertainties of the parameters are determined using the Hessian method (for details see \cite{Kovarik:2015cma}) with the same $\Delta\chi^2 = 45$ tolerance criterion as the one used in the nCTEQ15WZSIHdeut analysis.

The results of the DimuNeu analysis are threefold. First, the list of final values of all parameters after the DimuNeu analysis can be found in App.~\ref{sec:fitresults}. Next, the $\chi^2$ values for all data and for each data set separately are given in Tab.~\ref{tab:chi2dimuneu}. Lastly, in Fig.~\ref{fig:dimuneu_pdfs} we show the ratio of nuclear PDFs for the whole nucleus to the PDFs for the whole nucleus obtained using the free proton PDFs. We compare the nuclear parton distribution functions extracted from the neutrino data to the ones extracted in the nCTEQ15WZSIHdeut analysis in Sec.~\ref{sec:nCTEQ15}. We observe that the results from the DimuNeu and nCTEQ15WZSIHdeut analyses are distinctly different for the valence quark PDFs as well as for the non-valence quark PDFs. The shapes are different even if we consider the PDF errors of both analyses. The strange quark nPDF also differs between the two analyses. In the case of iron PDFs the changes in the strange quark PDF are still within the uncertainties but for lead the strange quark PDF is distinctly different. The gluon PDF parameters were fixed and so the gluon PDF is the same in both analyses. 

\begin{figure*}[t!]
    \centering
    \includegraphics[width=0.90\columnwidth]{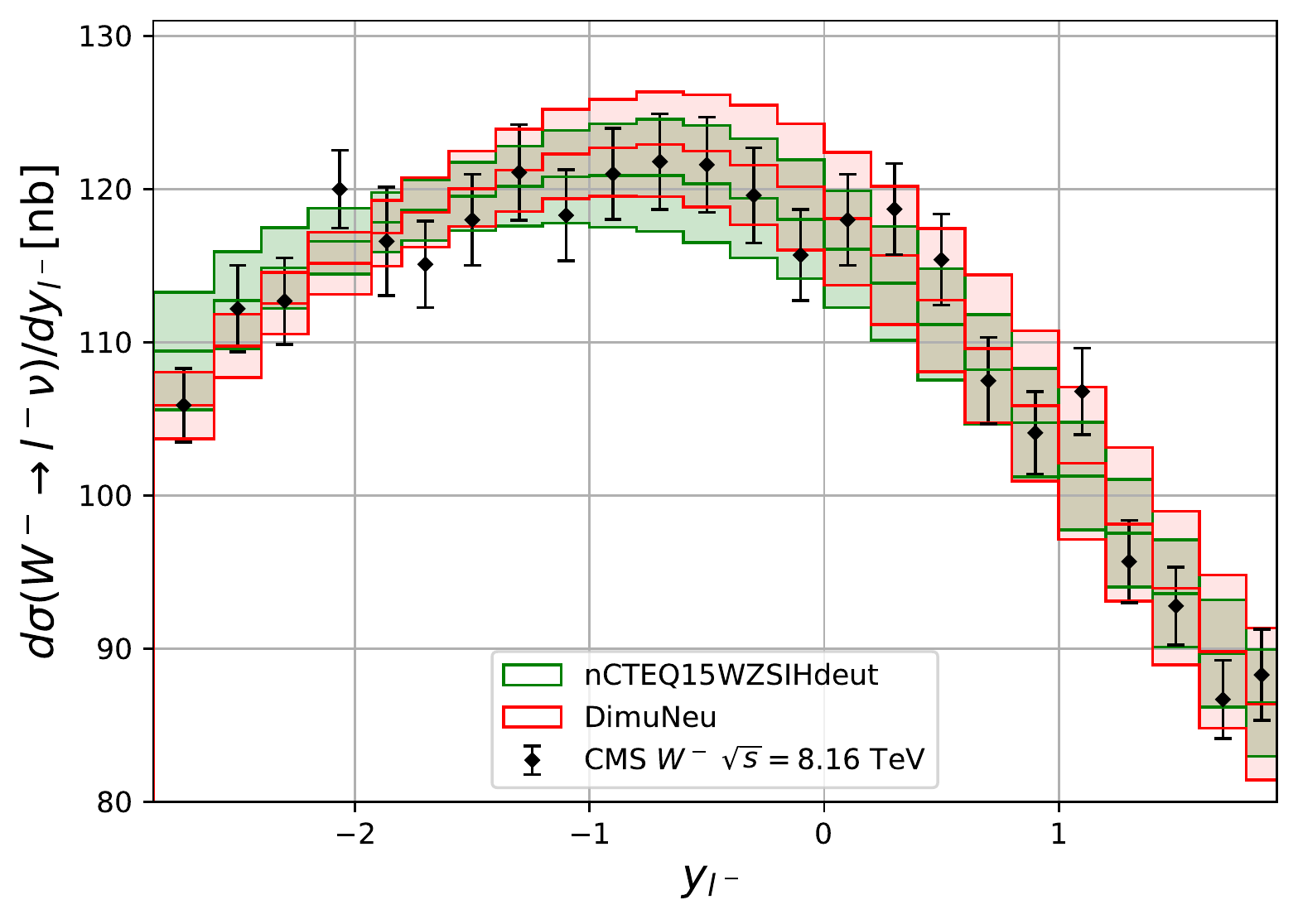}
    \hfil
    \includegraphics[width=0.90\columnwidth]{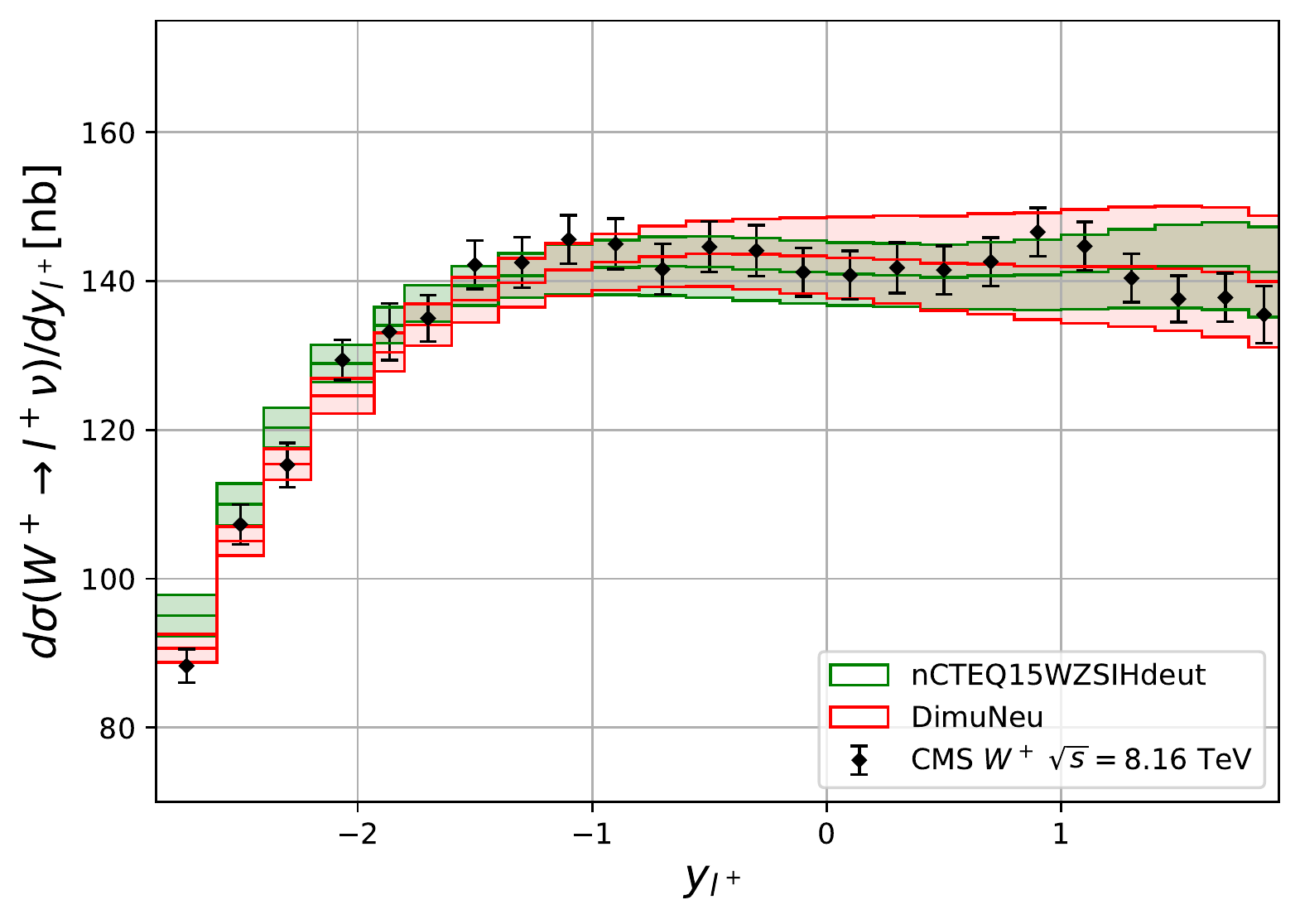}
    \caption{Comparison between CMS $W^\pm$ boson production cross section data with the theory predictions from our fits. The green (red) bands show the theory uncertainties from nCTEQ15WZSIHdeut (DimuNeu) error PDFs. All theory predictions have been shifted by their respective fitted normalization shift.
}
    \label{CMSpred}
\end{figure*}
%
It is instructive to see how the resulting nPDFs from the DimuNeu analysis describe the experimental data. In Fig.~\ref{RF2_CC_dimuneu} we compare the predictions stemming from the DimuNeu analysis for the nuclear correction factor constructed from the $F_2$ structure functions from the neutral or charged current deep inelastic scattering to the corresponding structure function data. There is a subtlety one has to take into account. In the case of the neutral current DIS (see the left panel of Fig.~\ref{RF2_CC_dimuneu}), the data are presented as ratios $F_2^{A}/F_2^D$, where the denominator comes from a measurement on deuterium targets. In the charged current case with neutrino beams (see the right panel of Fig.~\ref{RF2_CC_dimuneu}), deuterium targets are not heavy enough to generate sufficient statistics. Therefore, one uses a nuclear correction factor constructed as
\begin{equation}
    R[F_2^{CC}] = \frac{F_2^{CC}[f_i^A]}{F_2^{CC}[f_{i}^{A,{\rm free}}]},
    \label{rf2cc}
\end{equation}
where the charged current structure function $F_2^{CC}$ is defined as an average $F_2^{CC}= (F_2^{\nu A}+F_2^{\bar{\nu} A})/2$. In the case of the theoretical predictions, the numerator is calculated using the nuclear PDFs, $f_i^{A}$, for the corresponding nucleus $A$, and in the denominator the combination of free proton and neutron PDFs, $f_i^{A,{\rm free}}$, are used instead. In Fig.~\ref{RF2_CC_dimuneu}, the experimental points are obtained by dividing the data on $F_2^{CC}$ by the same "free" PDF denominator as for the theoretical prediction. In Fig.~\ref{CMSpred} we also show predictions from the DimuNeu analysis for the $W^{\pm}$ production at the LHC as a function of the rapidity of the charged lepton $y^{\pm}$.

Based on the total $\chi^2$ in Tab.~\ref{tab:chi2dimuneu}, we see that the DimuNeu result can decently describe all neutrino data. We see however that not all data are described equally well. On one side, both neutrino and anti-neutrino data from CDHSW and CCFR experiments are very well compatible with the DimuNeu prediction. On the other side, all dimuon data and all Chorus data as well as anti-neutrino data from the NuTeV show a mild tension where the $\chi^2{\rm /pt}\sim 1.2$. The neutrino data from the NuTeV collaboration are the most precise and show the largest tension with the DimuNeu analysis. As was stated in previous analyses and verified also in the course of this analysis, NuTeV neutrino data cannot be adequately described in this nPDF framework even if the data are fitted alone.  

In the right panel of Fig.~\ref{RF2_CC_dimuneu}, we see that the predicted nuclear correction factor, coming from the global neutrino DimuNeu analysis, describes the data from NuTeV and CDHSW within their uncertainty. This can be compared to the nuclear correction factor from the nCTEQ15WZSIHdeut analysis where the $x$-shape of the correction factor is completely different and cannot describe the neutrino data at all. We also observe in the left panel of Fig.~\ref{RF2_CC_dimuneu} that the inverse is true for the neutral current data where the nuclear correction factor which describes the neutrino data fails to describe the aforementioned data. This is true almost for any $x$ but the largest deviation can be seen for $x<0.07$. Even for mid-$x$ where the shape of the DimuNeu nuclear correction factor would be consistent with the data, it consistently undershoots all data. Here the situation is reversed and the nuclear correction factor from nCTEQ15WZSIHdeut describes the data well. This apparent inconsistency of the nuclear correction factor determined from neutrino data with the rest of the neutral current data is what prompted the series of studies starting with \cite{Schienbein:2007fs}. In Fig.~\ref{CMSpred} we show that not all observables disagree. In the case of the $W^{\pm}$ production at the LHC we see a nice agreement between the results from the nCTEQ15WZSIHdeut and DimuNeu analyses. This should come as no surprise given that the $W^{\pm}$ production is quite sensitive to the gluon PDF%
    \footnote{Actually, in case of a nPDF fit without jet data the $W/Z$ LHC data provide the most stringent constraints for the gluon.}
which remains fixed and is the same in both analyses.

\begin{figure}[t!]
    \centering
    \includegraphics[width=0.961\columnwidth]{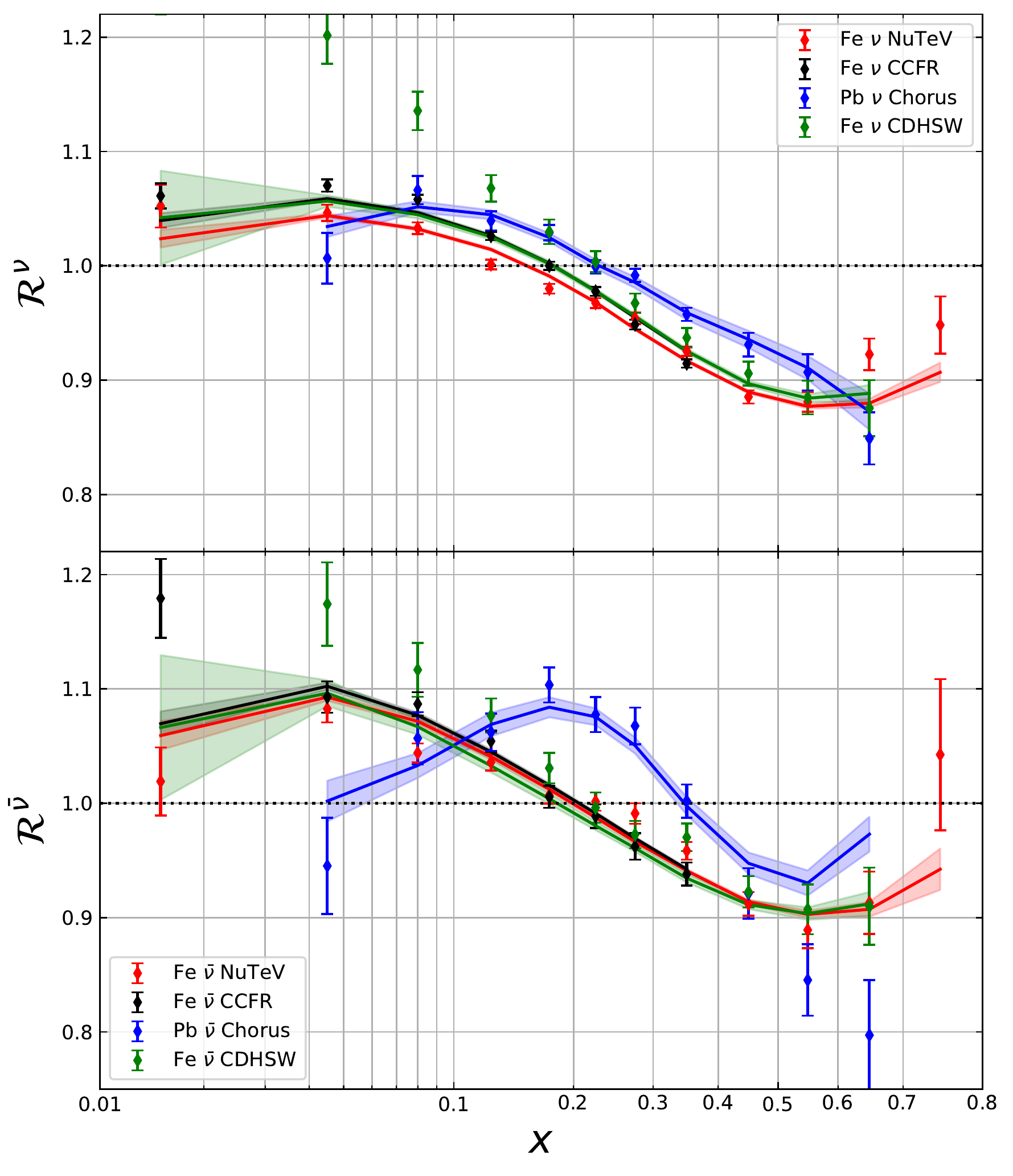}
    \caption{The weighted average of the cross section ratio for individual neutrino and anti-neutrino cross section data from NuTeV, Chorus, CCFR and CDHSW. The solid bands show the prediction from the DimuNeu fit. Note that the plotted points match those presented in Fig.~\ref{fig:Rnunubar}.}
    \label{Rratio}
\end{figure}

Above, we have verified that the prediction from the DimuNeu analysis correctly describes the experimental data on the $F_2^{CC}$ structure function by comparing the nuclear correction factor $R[F_2^{CC}]$. Given that we have not used the structure function data in our analysis, it is also instructive to see how well the cross-section data are being described analogously to the results and discussion of Fig.~\ref{fig:Rnunubar}. For that purpose we return to the weighted average introduced in Sec.~\ref{sec:weightedaverage} and in Fig.~\ref{Rratio} to check how well the DimuNeu analysis fits the data. Even though all data considered in Fig.~\ref{Rratio} correspond to the same observable, the result of the averaging procedure depends on which data set is used in the averaging as different experiments have different ranges in $Q^2$ which are being averaged over. Therefore, separate theoretical predictions for the weighted average for each experiment with the corresponding uncertainties are shown. In constructing the theoretical prediction for the weighted average we have replaced $R_i^\sigma$ and $\Delta R_i^\sigma$ in Eqs.~(\ref{Rsigma}) and (\ref{DRsigma}) by the predicted central value and the theoretical uncertainty stemming from the PDF uncertainty, respectively. We have retained the weights $w_i$ calculated from the corresponding experimental data to ensure the same weighing procedure is used for both data and theory predictions.

We see that in general the theoretical prediction from the DimuNeu analysis fits the cross-section data as well as it did the structure function data. There is a good agreement between the data and the DimuNeu prediction for all experiments in the intermediate Bjorken-$x$ region. In the large-$x$ region, the DimuNeu result is a compromise between the diverging experimental data where the NuTeV measurement starkly differs from the others. For small Bjorken $x$ the fit is also a compromise given that the CDHSW, CCFR and NuTeV show no distinct shadowing in this region whereas the CHORUS data display a shadowing behavior similar to the neutral current DIS data.

Given the noticeable difference between the neutrino data taken on iron and the data taken on lead in Fig.~\ref{Rratio}, one might conclude at first glance that these data are incompatible with each other. However, we see that the DimuNeu analysis can describe both neutrino data on iron and on lead quite successfully within one unified nPDF framework. To investigate the matter a little further, we have performed two separate fits which we label ``DimuNeuIron'' and ``ChorusW''. 
%
%
\begin{figure}[t!]
	\centering
	\includegraphics[width = 0.98\columnwidth]{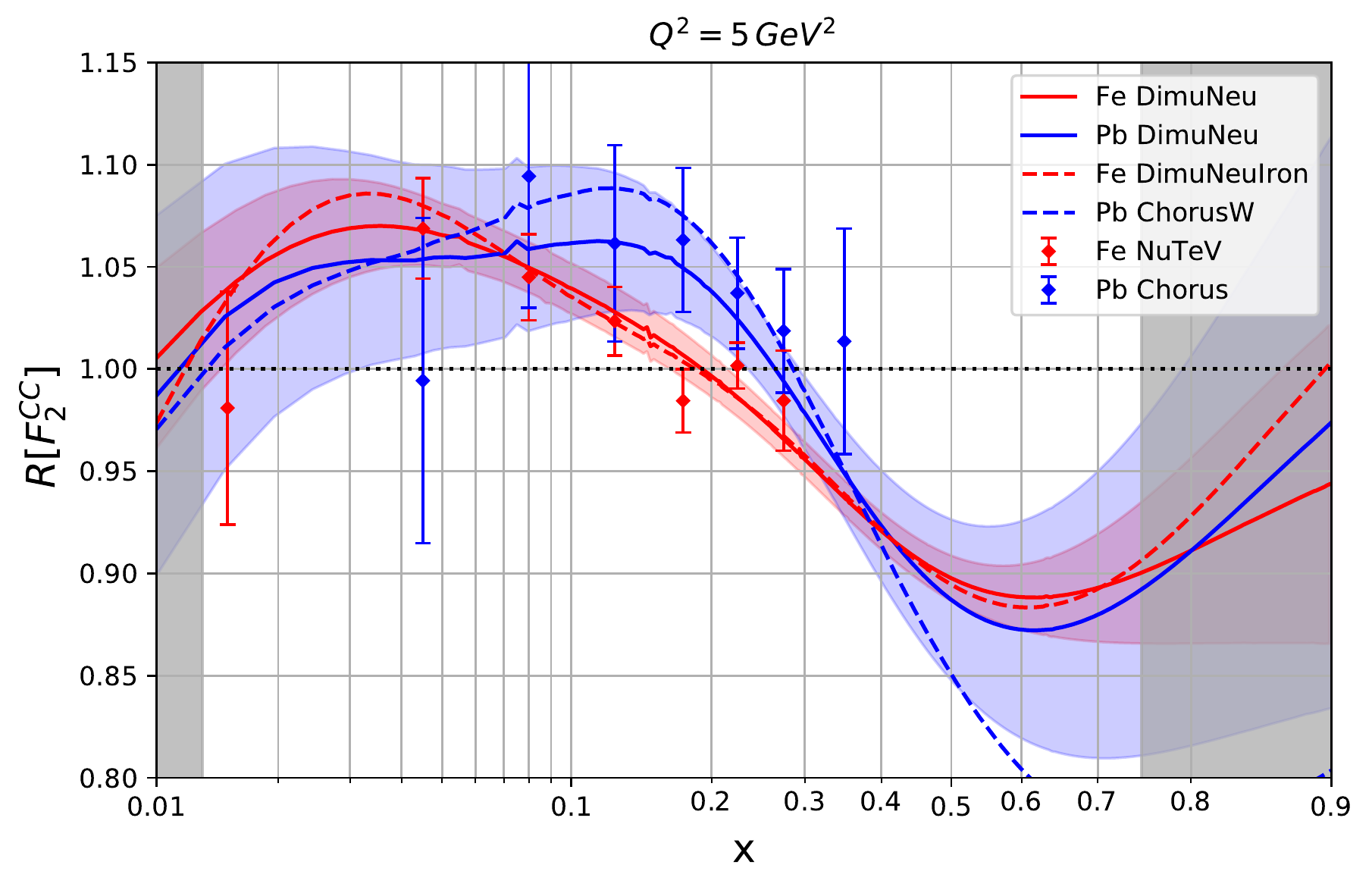}
	\caption{A comparison of predictions from the DimuNeu, DimuNeuIron and ChorusW analyses for the charged-current structure function ratios $R[F_2^{\rm CC}]$ for iron and lead.}
	\label{fig:ironlead_dimuneu}
\end{figure}
%
Both fits use only 14 free parameters and compared to the free parameters of the nCTEQ15WZSIHdeut fit listed in Sec.~\ref{sec:nCTEQ15} all parameters $b_i^x$ corresponding to the $A$-dependence were held fixed. The reason for fixing these parameters is that both fits include data taken only on one nucleus. In the case of the DimuNeuIron analysis, only neutrino data from CDHSW, CCFR and NuTeV taken on iron were included and in the case of the ChorusW analysis only Chorus neutrino data and LHC data on $W$-boson production both taken on lead were used. In Fig.~\ref{fig:ironlead_dimuneu} we compare the predictions for the charged-current structure function ratios for iron (red) and for lead (blue) from these specialized fits (dashed lines) with the predictions from the global DimuNeu neutrino analysis (solid lines). We see that in general the predictions from the specialized fits agree well with the ones from the global DimuNeu analysis with the sole exception of the large-$x$ region where the precise NuTeV data dominate the global analysis.  

The difference in the nuclear correction factor for iron and for lead can come from two sources. The main effect usually comes from the different proton and neutron content of the iron and the lead atoms. The large excess of neutrons in a lead nucleus leads to noticeable differences in predicted observables even though the underlying effective bound proton and bound neutron PDFs are the same as for other elements. The second possible source for the difference is the dependence of the underlying bound nucleon PDFs on the atomic number $A$. The second effect is typically subleading. We can see the impact of the large neutron excess if we compare the predictions for lead in Fig.~\ref{Rratio}, where in accordance with the experimental data it was assumed that $A=208$ and $Z=82$, with the predictions shown in Fig.~\ref{fig:ironlead_dimuneu}, where $A=208$ and $Z=104$ were used given that the structure function data from Chorus are isoscalar corrected. We can therefore conclude that the neutrino data from all experiments irrespective if they are taken on iron or lead show similar behaviour for all but large $x>0.5$.   
%
\section{Neutrino Data Compatibility}
\label{sec:nuglobal}
In this section we will introduce a combined global nuclear PDF analysis including all data from the reference nCTEQ15WZSIHdeut fit (see Sec.~\ref{sec:framework}) and all neutrino data discussed in Sec.~\ref{sec:data}. Extending an existing PDF analysis by including new data is a standard and frequent occurrence. Usually one includes new data in a PDF analysis in order to improve on the precision or on the $x$-$Q^2$ coverage of previously used data or to constrain PDFs of partons which were previously left unconstrained. In order for the new data to provide all that, it has to be possible to consistently describe them in the underlying theoretical framework based on the factorisation theorem, perturbative QCD and on the $x$-parametrization of the PDFs at the input scale. Schematically, if the new data cannot be consistently described in a combined analysis, it can mean one of two things. Either the theoretical framework needs to be extended for example by including small-$x$ resummation effects or the target mass corrections or there was a problem with the data acquisition e.g. the experimental errors were underestimated.  

Based on the preliminary analysis we have performed on the neutrino deep inelastic scattering data in the previous section, we expect possible large tensions between the neutrino data and the rest of the nuclear scattering data. Therefore, we will investigate the compatibility of the neutrino DIS data with the bulk of the nuclear scattering data in detail. We will take a closer look at the compatibility of the results of each neutrino DIS experiment separately. We will also look into the possibility that all neutrino DIS data are showing significant tensions, which, in one interpretation, may indicate incompleteness in the theoretical framework used to describe neutrino scattering in the nPDF analysis.
\subsection{Compatibility Criteria}\label{sec:criteria}
Before we dive into the details of the compatibility discussion, we need to clearly specify the criteria for compatibility which we will be using. In general, we will be discussing the compatibility of two data sets $S$ and $\bar{S}$ in a global fit which includes both of the sets $Z\equiv S\cup \bar{S}$. In our case, the set $S$ will always be the set of data used in the reference fit nCTEQ15WZSIHdeut and the set $\bar{S}$ will be some subset of the newly considered neutrino data. In what follows, we will be using three different criteria.
\begin{table*}[htb]
	\caption{Statistical information such as the total $\chi^2$ and number of data points for all analyses discussed here are presented. Moreover, the $\chi^2$-percentiles with respect to the reference fit nCTEQ15WZSIHdeut (denoted $S$) and to the only neutrino DimuNeu analysis (denoted $\bar{S}$) are also given.}
   \label{tab:statperc}
	\centering
	\begin{tabular}{|c|c|c|c|c|c|}
	\hline
		Analysis name & $\chi^2_S/N$ & $\chi^2_{\bar{S}}/N $ & $\Delta\chi^2_S$ & $\Delta\chi^2_{\bar{S}}$ & $p_S/p_{\bar{S}}$ \\[1mm] \hline \hline
        nCTEQ15WZSIHdeut & 735/940 & - & 0 & - & 0.500 / - \\
        DimuNeu & - & 6383/5689 & - & 0 & - / 0.500\\
        BaseDimuNeu & 866/940 & 6666/5689 & 131 & 283 & 0.99987/0.990\\
        \hline
	\end{tabular}
\end{table*}
%
\newline\newline %
{\bfseries $\Delta\chi^2_S\,$-compatibility} This first criterion for comparison of the compatibility of two data sets $S$ and $\bar{S}$ uses the $\chi^2$ of the global analyses of the data sets $S$ and $Z$. We use the $\chi^2$ to assess whether the nPDFs extracted from the fit to the combined data set $Z$ are within the error bands of the nPDFs from a fit to the baseline data set $S$. It can be shown that in the Hessian error formalism, this happens if and only if the increase of the $\chi^2$ of $S$ before and after including $\bar{S}$ is less then the tolerance $\Delta\chi^2_S$, hence the name of this criterion.

To apply this criterion in our case, we have to define a proper tolerance $\Delta\chi^2_S$ of the global reference fit to the data $S$ which in our case is the analysis nCTEQ15WZSIHdeut discussed in Sec.~\ref{sec:framework}. In the nCTEQ15 analysis, we have used $\Delta\chi^2=35$ with $N=740$ data points. However, the nCTEQ15WZSIHdeut analysis contains significantly more data $N=940$ so an adjustment of $\Delta\chi^2_S$ is required. We will make use of the $\chi^2$-distribution for $N$ degrees of freedom 
\begin{equation}
P(\chi^{2},N)=\frac{(\chi^{2})^{N/2-1}e^{-\chi^{2}/2}}{2^{N/2}\Gamma(N/2)}\,,
\label{eq:chi2dist}
\end{equation}
to define the $\Delta\chi^2_S$. The $\chi^2$-distribution allows us to define the percentiles, $\xi_{p}$, via 
\begin{equation}\label{perc}
\int_{0}^{\xi_{p}}P(\chi^{2},N)\,d\chi^{2}=\frac{p}{100}\quad\;{\rm where}\quad p=\{50,90,99\}\,.
\end{equation}
$\xi_{50}$ serves as an estimate of the mean of the $\chi^2$-distribution and we expect the $\chi^2$ of a good fit to be close to $\xi_{50}$. In the case of nCTEQ15WZSIHdeut analysis where $\chi_0^2=735 < \xi_{50} = 939$, the fit was better than expected. Due to the large discrepancy between $\chi_0^2$ and $\xi_{50}=939$, we have decided to rescale all percentiles by a factor $\gamma_S = \chi_0^2/\xi_{50}$. The new rescaled 90\% percentile then becomes $\chi^2_{90} = \gamma_S\, \xi_{90} = 779$. We can finally define $\Delta\chi^2_S$ as
\begin{equation}
    \Delta\chi^2_S = \chi^2_{90} - \chi^2_{0} = 45\,.
\end{equation}
This is the tolerance we use to define the error PDFs for the nCTEQ15WZSIHdeut analysis.

Assessing compatibility using the $\Delta\chi^2_S\,$-criterion has one obvious drawback. If the reference analysis of data $S$ contains a parameter (or a combination of parameters) which cannot be sufficiently constrained, the uncertainty connected to this parameter is often underestimated. This is due to the fact that in the Hessian approach the unconstrained parameters are connected to very small eigenvalues of the Hessian matrix and the diagonalization of a large matrix where the eigenvalues span multiple orders of magnitude is numerically unstable. If the global analysis of the extended data set $Z\equiv S\cup \bar{S}$ constrains the previously unconstrained combination of parameters, the resulting PDF is often outside of the underestimated error band of the previous analysis. In this case the criterion signals incompatibility even though there is none. Therefore no matter how useful this criterion is, we cannot rely just on this single criterion.
\newline\newline %
{\bfseries $\chi^2_S\,$-compatibility} The second criterion approaches the problem of compatibility slightly differently. Using this criterion, we asses if the data sets $S$ and $\bar{S}$ are described acceptably well in a combined fit to $Z\equiv S\cup \bar{S}$, comparing the quality of the description of the data sets in the combined fit to the fits to the data sets alone. We will consider the data sets $S$ and $\bar{S}$ are $\chi^2_S\,$-compatible if both their $\chi^2$ in a combined fit are within at most 90\% percentile defined in Eq.~(\ref{perc}) from their expected value. To account for the cases where a data set cannot be optimally described even in a fit only to the data set itself, we will define the rescaled percentile $\chi^2_{90} = \gamma_S\, \xi_{90}$ exactly as we did in the case of the $\Delta\chi^2_S\,$-compatibility criterion above.

Similar to the first criterion, using the $\chi^2_S\,$-compatibility criterion also has its issues. In order to properly use this criterion it has to be possible to fit the data set alone. This limits the usefulness of this criterion only to data sets which are sufficiently large to be fit alone.
\newline\newline %
{\bfseries $S_E$-compatibility} The last criterion used in our analysis is yet another alternative to investigate compatibility of data sets in a combined global analysis. Here we will consider only the global analysis of the combined data sets $Z\equiv S\cup \bar{S}$ and investigate the quality of description of each experiment $E$ in this analysis. The comparison of the quality between two different experiments is made difficult by the fact that the $\chi^2$-distribution $P(\chi^2,N)$ (see Eq.~(\ref{eq:chi2dist})) is heavily dependent on the number of data points $N$ of the experiment. Therefore, instead of the $\chi^2$-distribution $P(\chi^2,N)$ we use a variable $S(\chi^2,N)$ 
\begin{equation}\label{SEdef}
    S(\chi^2(N),N) = \sqrt{2\chi^2(N)} - \sqrt{2N-1}
\end{equation}
which is no longer strongly sensitive to the number of data points. Moreover, the variable $S(N)$ is distributed according to the normal distribution with zero mean and unit variance \cite{Kovarik:2019xvh}. We can evaluate $S_E=S(\chi^2_E,N_E)$ for each experiment using the number of data points $N=N_E$ and $\chi^2=\chi^2_E$ and check if the variable for all experiments is distributed according to the normal distribution with the expected mean and variance. This happens if the $\chi^2$ values of all experiments involved in the global analysis are distributed according to the corresponding $\chi^2$-distributions. On top of checking if $S_E$ for the totality of experiments is distributed as expected, we can also identify experiments which are not compatible with this distribution and also quantify to what degree using the standard confidence levels of the normal distribution.
\subsection{Global analysis with neutrino data}

\begin{figure*}[htb]
	\centering
	\includegraphics[width=0.95\textwidth]{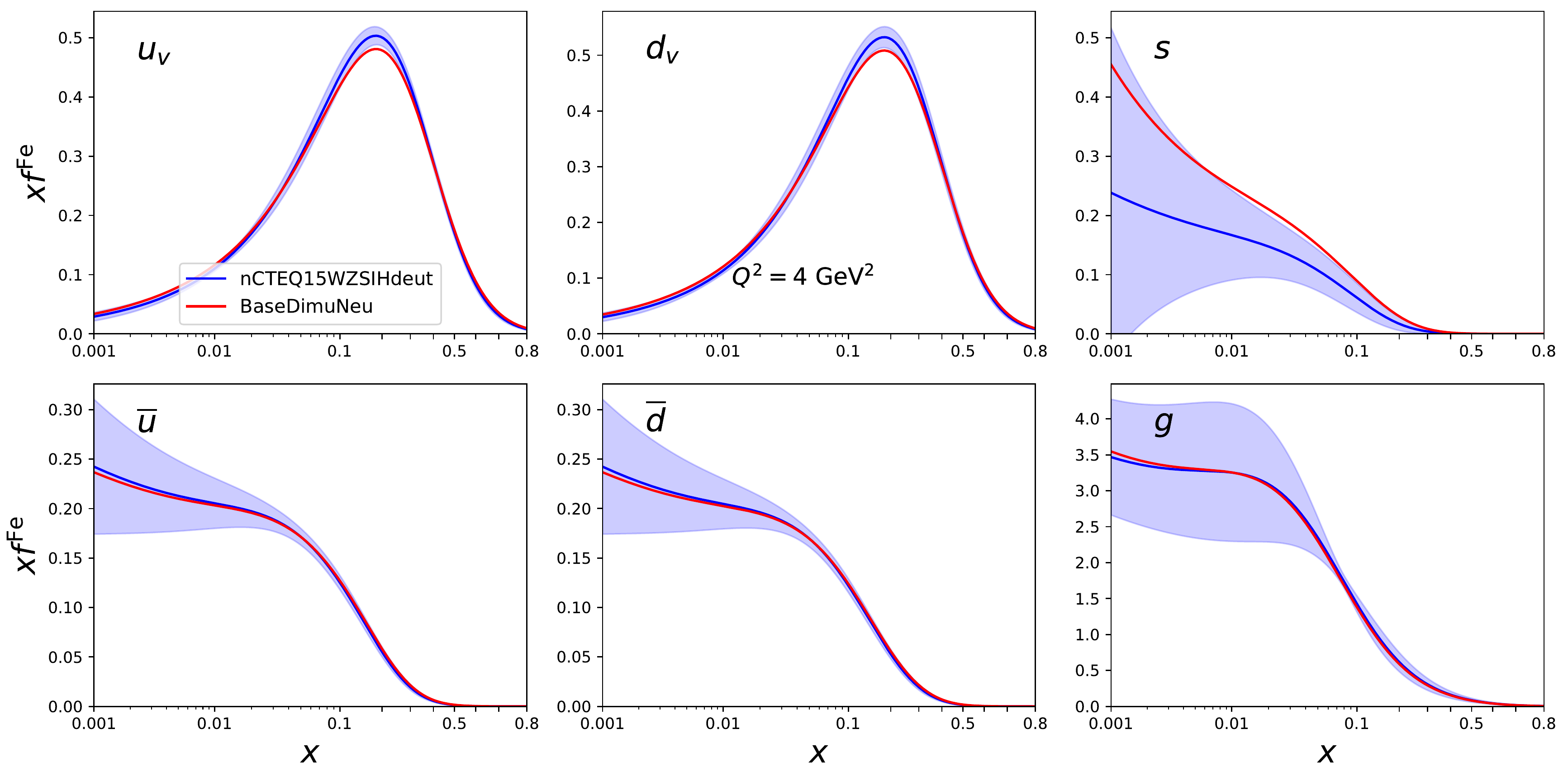}
	\caption{The full iron PDFs at $Q^2=4\ {\rm GeV}^2$. All uncertainty bands are computed using the Hessian method with $\Delta \chi^2=45.$}
	\label{fig.pdf}
\end{figure*}

\begin{figure*}[htb]
	\centering
	\includegraphics[width=0.95\textwidth]{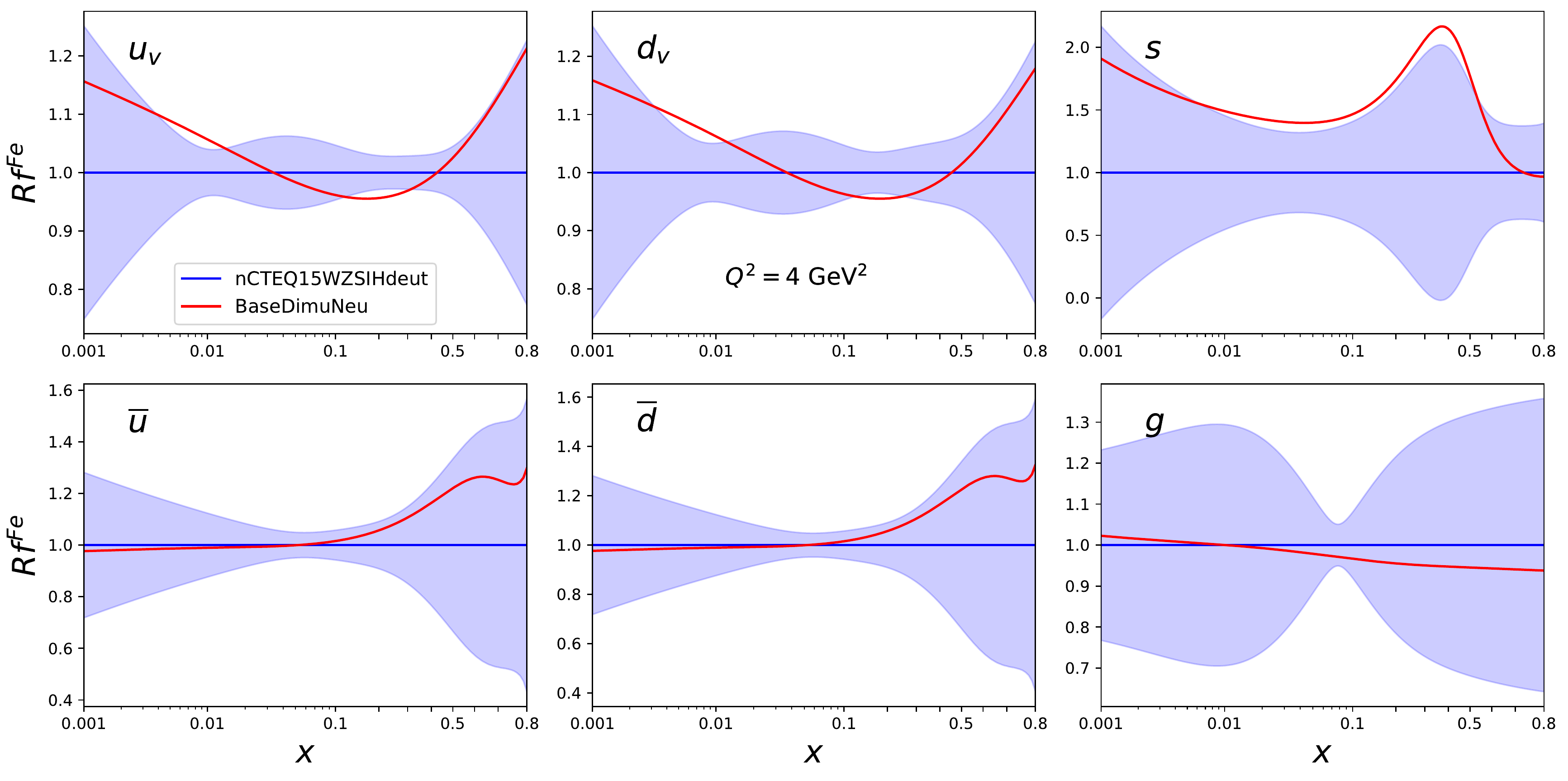}
	\caption{Ratio of the full iron PDFs to the corresponding PDFs from nCTEQ15WZSIHdeut fit at $Q^2=4\ {\rm GeV}^2$. All uncertainty bands are obtained using the Hessian method with $\Delta \chi^2=45$.}
	\label{fig.pdfratio}
\end{figure*}

\begin{figure*}[htb]
	\centering
	\includegraphics[width=0.80\textwidth]{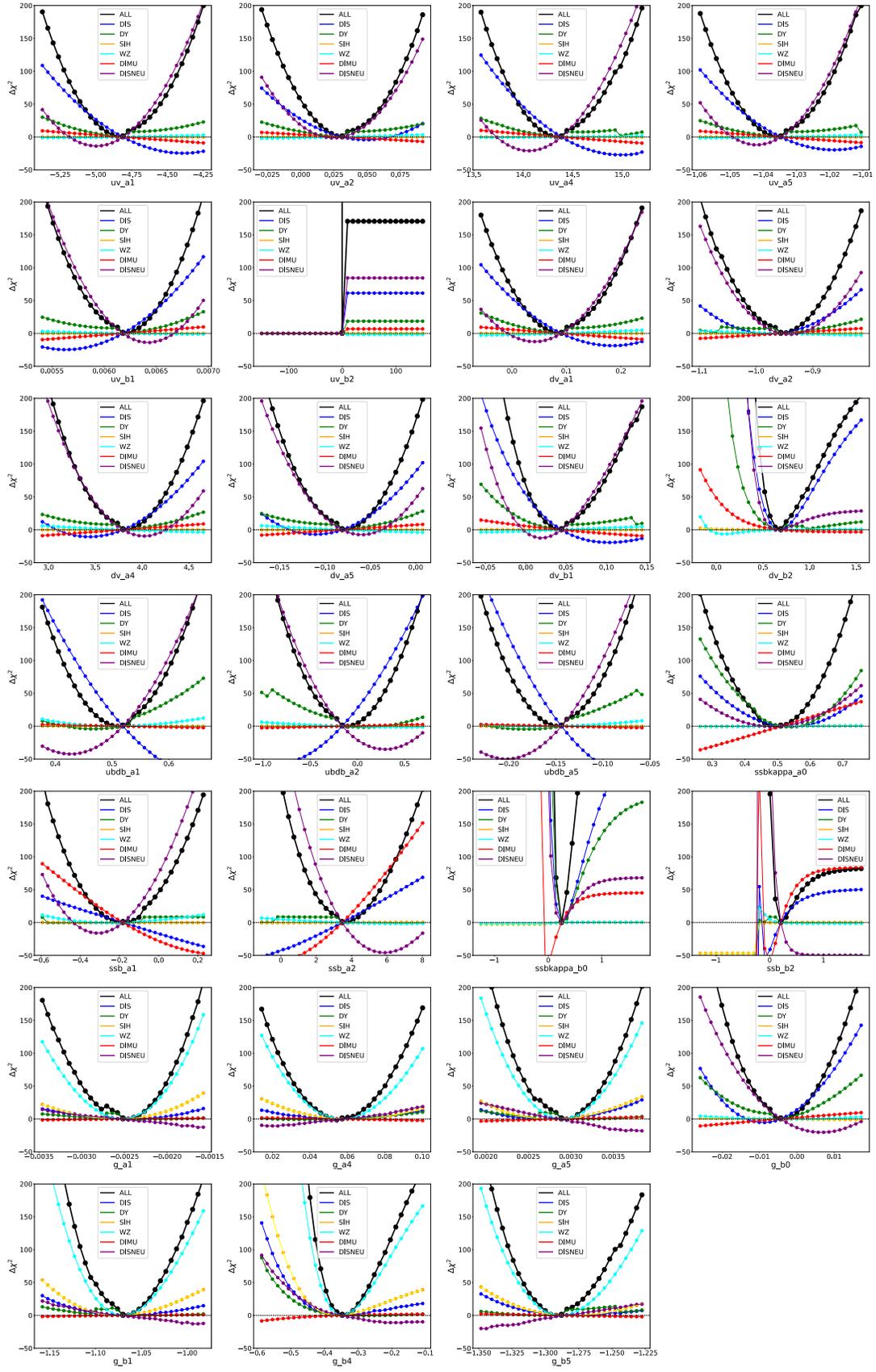}
	\caption{Scans of the $\chi^2$ function along the PDF parameter directions varying always one free parameter at a time while other parameters were left fixed at the global minimum of the BaseDimuNeu analysis. The breakdown into $\chi^2$ for classes of experimental data is also shown.}
	\label{fig.chi2.scan}
\end{figure*}

%
We will start our analysis of the compatibility of the neutrino DIS data with the rest of the nuclear scattering data used so far in the nCTEQ analyses by considering a global analysis which adds all available neutrino data to the rest of the nCTEQ data mentioned in Sec.~\ref{sec:nCTEQ15}. The fit BaseDimuNeu contains all the data from the reference nCTEQ15WZSIHdeut analysis and all inclusive (anti-)neutrino DIS data from the CDHSW, Chorus, CCFR and NuTeV experiments as well as semi-inclusive di-muon data from CCFR and NuTeV. We have to emphasize that there is a disparity between the number of data present in the original nCTEQ15WZSIHdeut analysis ($N=940$) and the number of the new neutrino DIS data added ($N=5689$). Therefore, the neutrino data will dominate the global analysis and we expect that if there is any tension, it can be seen in a different description of the original data of the nCTEQ15WZSIHdeut analysis.

The global analysis BaseDimuNeu uses the same framework discussed in Sec.~\ref{sec:framework} with the same 27 free parameters to determine nuclear PDFs by fitting 6629 data points. We obtain $\chi^2=7532$ or alternatively $\chi^2$/pt = 1.14. Given that all neutrino data could be described with $\chi^2$/pt = 1.12 and we have added nCTEQ15WZSIHdeut data to the analysis which on its own was described with $\chi^2$/pt = 0.78, the result of the global analysis can be considered as the first signal that there may be some tension among the data within the analysis.

\begin{figure*}[htb]
	\centering
	\includegraphics[width=0.48\textwidth]{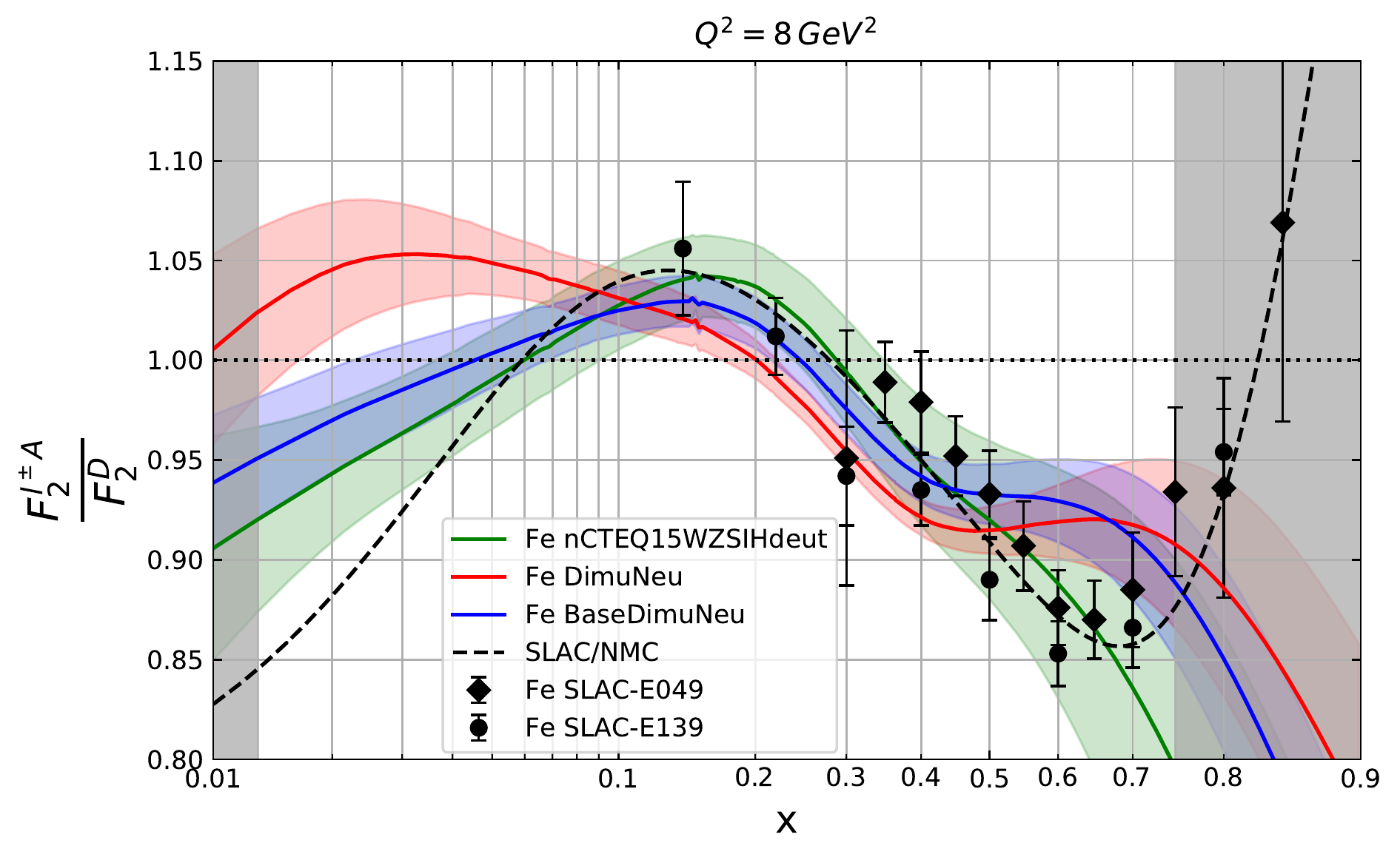}\qquad
	\quad
	\includegraphics[width=0.45\textwidth]{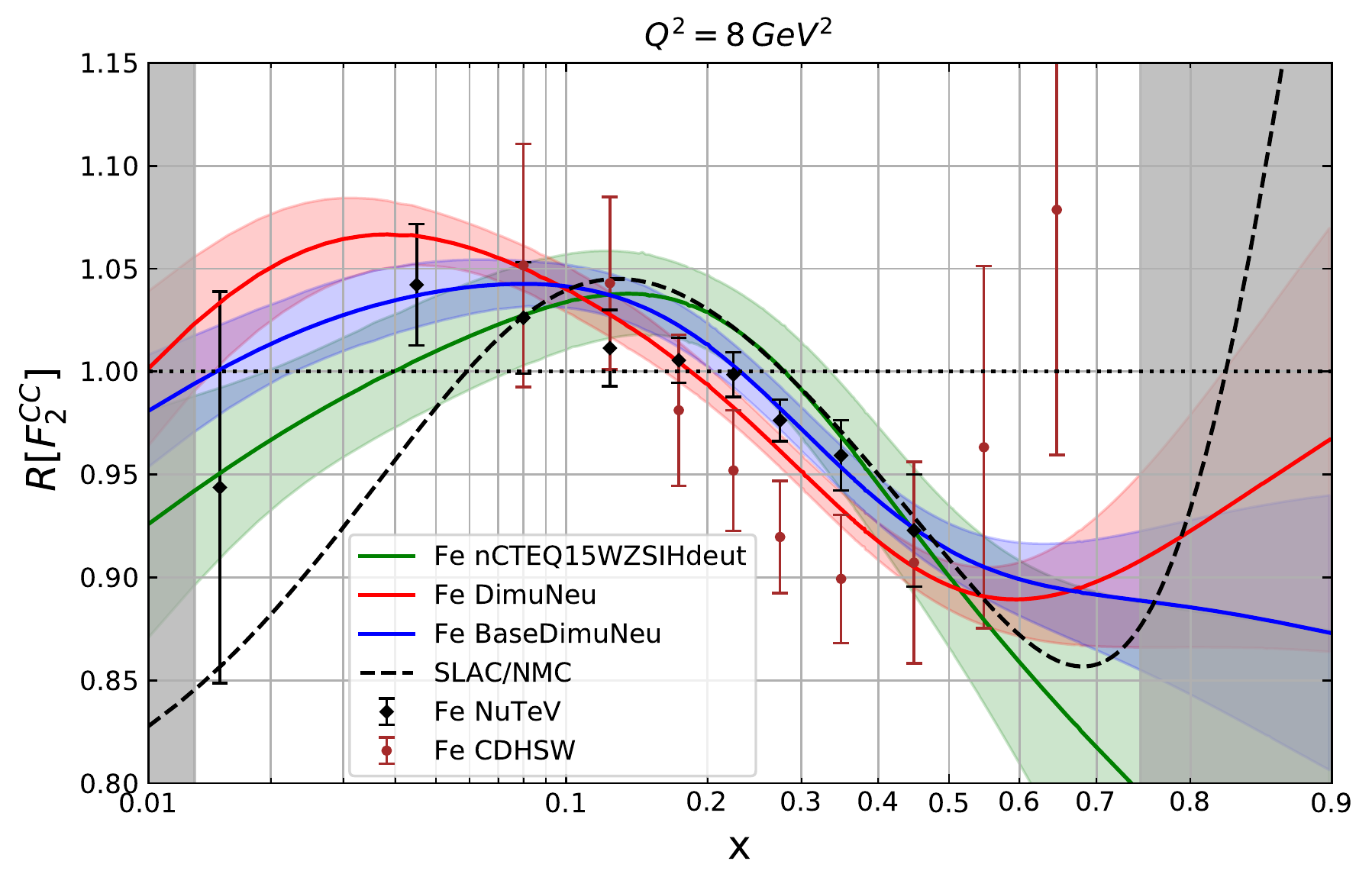}
	\caption{Neutral current nuclear ratio $F_2^{\rm Fe}/F_2^{\rm D}$ (left) and charged current nuclear ratio $R[F_2^{\rm CC}]$ as defined in Eq.~(\ref{rf2cc}) (right) using the fitted nPDFs. Note that we have applied nuclear correction for the neutral current deuterium structure function $F_2^D$ but not for the charged current one.}
	\label{fig.F2.prediction.comp}
\end{figure*}
%
\begin{figure*}[htb]
	\centering
	\includegraphics[width=0.48\textwidth]{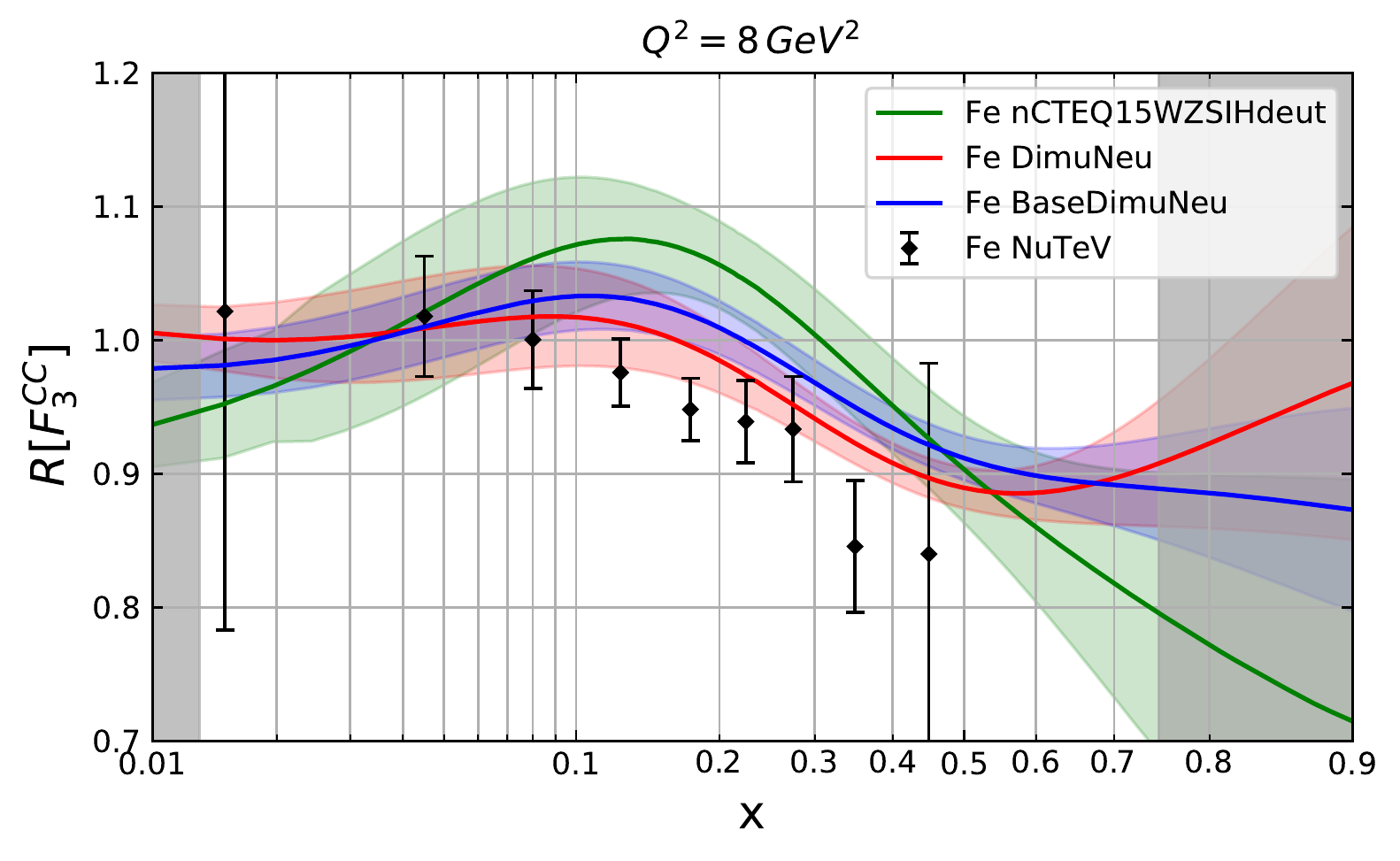}\qquad
	\caption{Charged current nuclear ratio $R[F_3^{\rm CC}]$ defined analogously to $R[F_2^{\rm CC}]$ using the fitted nPDFs.}
	\label{fig.F3.prediction.comp}
\end{figure*}
%
Specifically, when we compare the description of the subset of the data common to both nCTEQ15WZSIHdeut and BaseDimuNeu analyses, we notice a distinct rise from $\chi^2$ = 735 to $\chi^2$ = 866. This is an increase of 131 which is almost three times larger than the $\Delta \chi^2$=45 which was used to generate the error PDFs of the nCTEQ15WZSIHdeut result. This, according to the $\Delta\chi^2_S$ compatibility criterion introduced above, signals that the newly added data are incompatible with the original data of the nCTEQ15WZSIHdeut analysis. All relevant $\chi^2$ values are summarised in Tab.~\ref{tab:statperc}.

As we have stated previously, violating the $\Delta\chi^2_S$ compatibility criterion is also related to large differences in extracted PDFs. In Figs.~\ref{fig.pdf} and \ref{fig.pdfratio} we show the nuclear PDFs for iron resulting from the BaseDimuNeu analysis and compare them to the nPDFs of the nCTEQ15WZSIHdeut fit including the uncertainties. The comparison of both analyses is best seen in Fig.~\ref{fig.pdfratio} where the ratio of BaseDimuNeu and nCTEQ15WZSIHdeut nPDFs is shown. We can clearly see that the up- and down-quark valence PDF distributions as well as the strange-quark nuclear PDF from the global analysis including all neutrino data lie outside or at the edge of the error band of the reference nCTEQ15WZSIHdeut analysis. To exclude the possibility that the newly added neutrino data just constrain previously unconstrained PDF parameters, we investigate also the $\chi^2$ profiles varying the free parameters (see Fig.~\ref{fig.chi2.scan}). In Fig.~\ref{fig.chi2.scan} we see that for many quark parameters the result of the BaseDimuNeu analysis is a compromise between the neutral current DIS data already present in the nCTEQ15WZSIHdeut analysis (labeled DIS in Fig.~\ref{fig.chi2.scan}) and the newly added inclusive neutrino DIS data (labeled DISNEU). The final minima of the $\chi^2$ function lie frequently between the minima preferred by the DIS subsets. 
The DIS and DISNEU subsets show clear sensitivity to the quark valence parameters $a_1^{u_v}$, $a_2^{u_v}$, $a_4^{u_v}$, $a_5^{u_v}$, $a_1^{d_v}$, $a_4^{d_v}$, $a_5^{d_v}$ based on their respective $\chi^2$ growth profiles, but with widely-separated preferred values for those parameters.
This is a clear sign for tensions between these subsets. On the other hand, the situation is slightly different in the case of the strange quark. There, the minima preferred by the same subsets are also distinct but we can also observe that the neutrino DIS data are much more sensitive to the strange quark parameter variations than the neutral current DIS data sets. This leads us to conclude that, in the case of the strange quark, the neutrino DIS is the data set providing the first strong constraint on the strange PDF parameters and hence the discrepancy is not a sign of tension here. However, there is a small caveat. 
The neutrino differential cross-section data prefer a different strange quark PDF compared to the di-muon neutrino data. Moreover, the di-muon data and the neutral current DIS data prefer a similar strange quark. This tension can be later seen in Tab.~\ref{tab:chi2nudata} where the listed $\chi^2$/pt of the di-muon data signify that they are described much worse than in the neutrino only DimuNeu analysis.

The difference between the extracted PDFs from the BaseDimuNeu and nCTEQ15WZSIHdeut analyses translates into different predictions for observables such as the ratio of structure functions $F_2$ and $F_3$ shown in Figs.~\ref{fig.F2.prediction.comp} and \ref{fig.F3.prediction.comp} respectively. Here a similar interpretation is possible where we can clearly see that the results of the BaseDimuNeu analysis are a compromise between the nCTEQ15WZSIHdeut results and the results of the DimuNeu analysis which included only the neutrino data. The compromise predictions of the BaseDimuNeu analysis for the neutral-current nuclear ratio are compatible up to 1-$\sigma$ with the nCTEQ15WZSIHdeut prediction given that the central value lies within the error band of the nCTEQ15WZSIHdeut analysis. In the case of the other observables, the tension is larger. 
In the case of the charged-current nuclear ratio the results of the BaseDimuNeu are incompatible with the nCTEQ15WZSIHdeut result at $x\sim 0.025$ as the difference between the central predictions of the two analyses is larger than the error estimate on either analysis. The same is true if we would compare the predictions from the BaseDimuNeu and from the DimuNeu analyses. The case of the ratio of the structure function $F_3$ is a little different. First of all, there was almost no experimental information directly on the structure function $F_3^{\rm NC}$ from the neutral-current DIS data. Furthermore, the data on the charged-current structure function $F_3^{\rm CC}$ have larger errors compared to the structure function $F_2$. Even with larger errors, the $F_3$ data from NuTeV experiment (see Fig.~\ref{fig.F3.prediction.comp}) are not described particularly well by any of the analyses. Moreover, similar to the case of the structure function $F_2$ the predictions of the nCTEQ15WZSIHdeut and the DimuNeu analyses are incompatible with each other. This time the largest tension is found in the interval $0.1 < x < 0.4$. The central predictions of the global analysis BaseDimuNeu are in turn outside of the error band of the nCTEQ15WZSIHdeut analysis for $0.15 < x < 0.3$. We conclude that the tension which can be observed at the level of extracted PDFs in Figs.~\ref{fig.pdf} and \ref{fig.pdfratio} translates also to the ratios of the charged-current structure functions.

To reach a conclusive picture of the compatibility of neutrino DIS data with the remaining scattering data, we will use the other two criteria introduced in the previous section. The $\chi^2$ of the neutrino and the rest of scattering data subsets in the combined analysis are $\chi^2 = 6666$ and $\chi^2 = 866$, respectively (see Tab.~\ref{tab:statperc}). Using the rescaled percentiles as defined previously, we see that the description of both subsets of data is outside of the 90\% percentile (and even outside of the 99\% percentile in the case of nCTEQ15WZSIHdeut data), making the data sets incompatible according to the $\chi^2_S$-compatibility criterion.

\begin{figure*}[htb]
	\centering
	\includegraphics[width=0.32\textwidth]{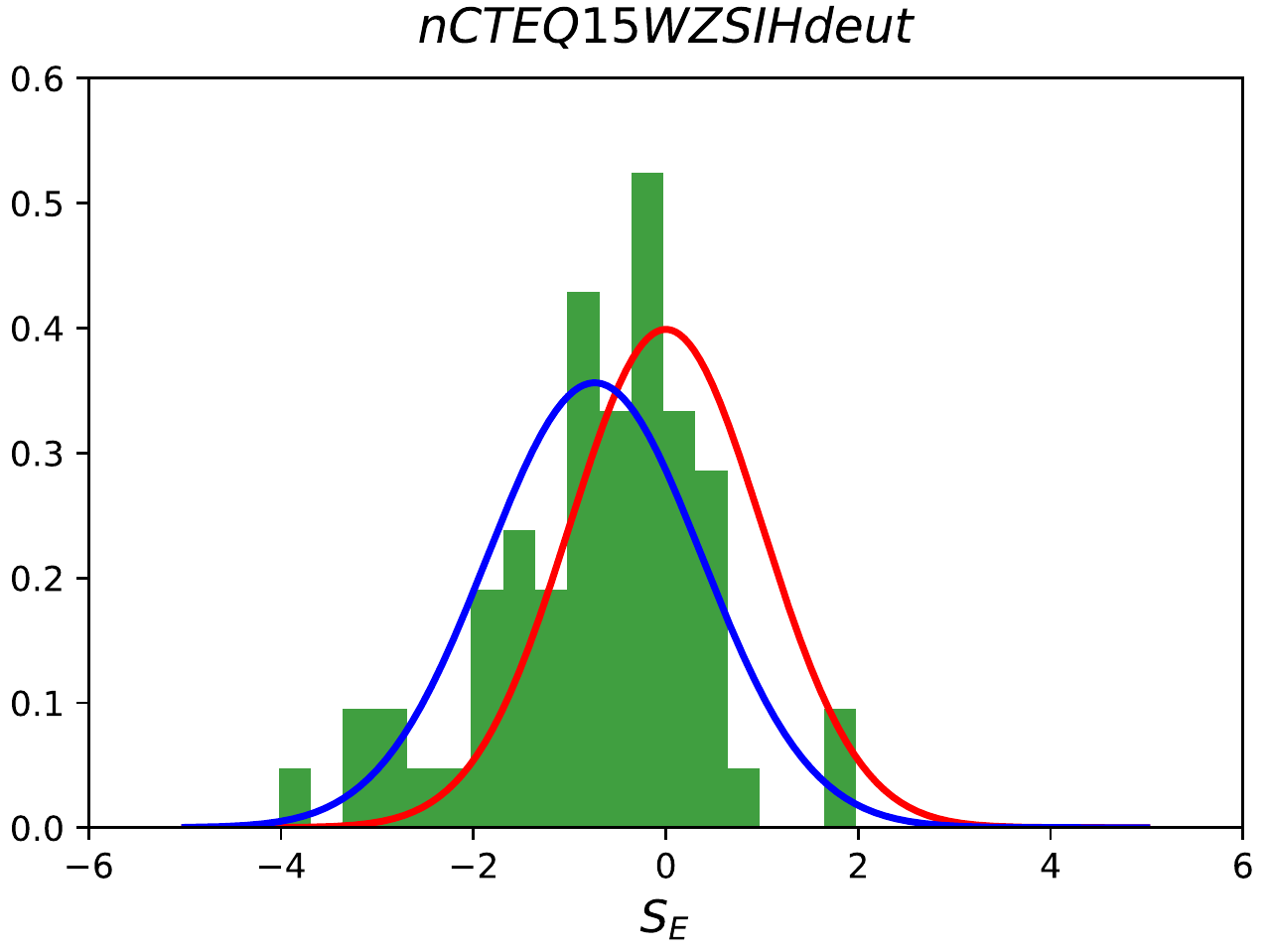}\quad 
	\includegraphics[width=0.32\textwidth]{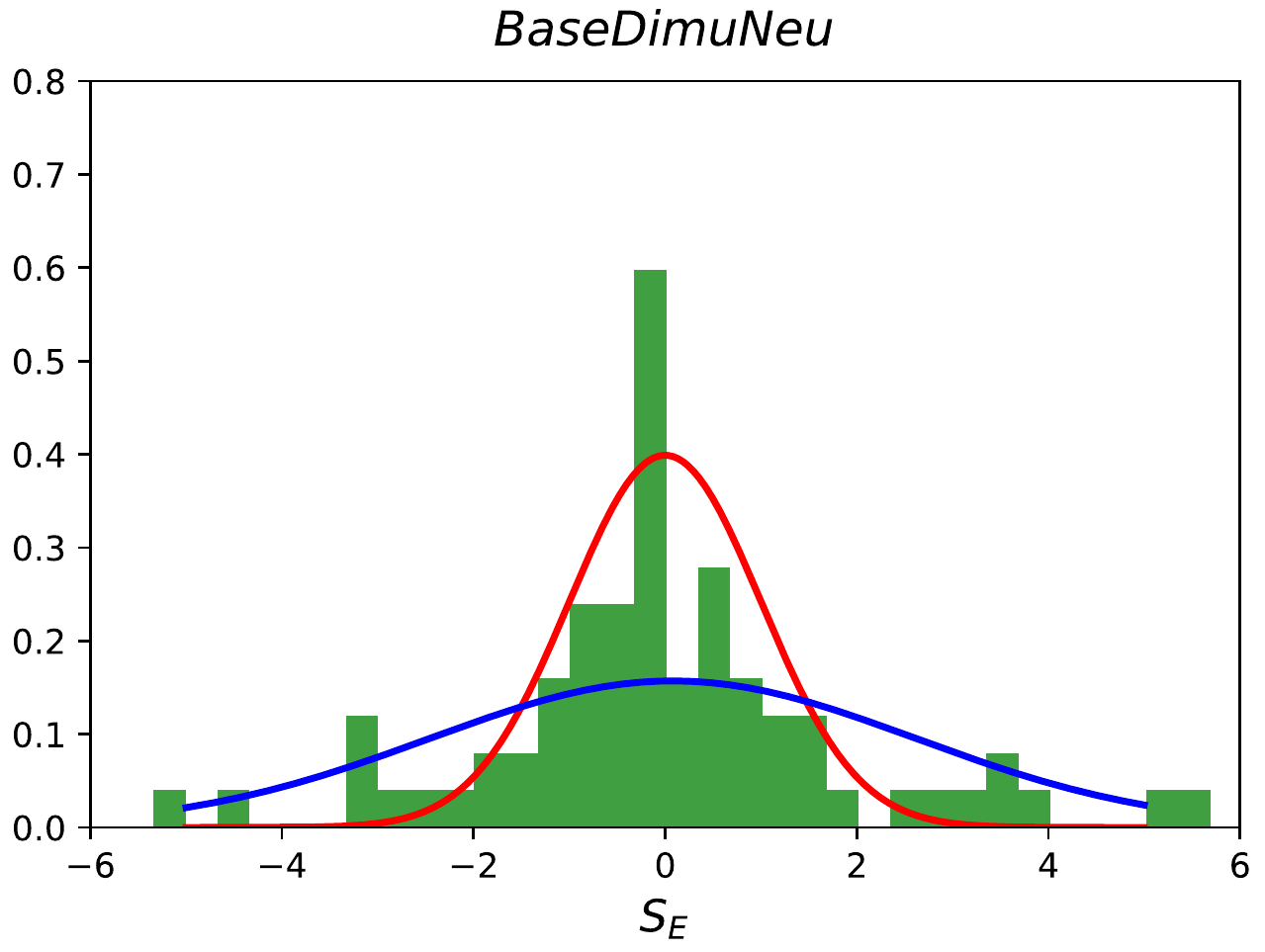}\quad 
	\includegraphics[width=0.32\textwidth]{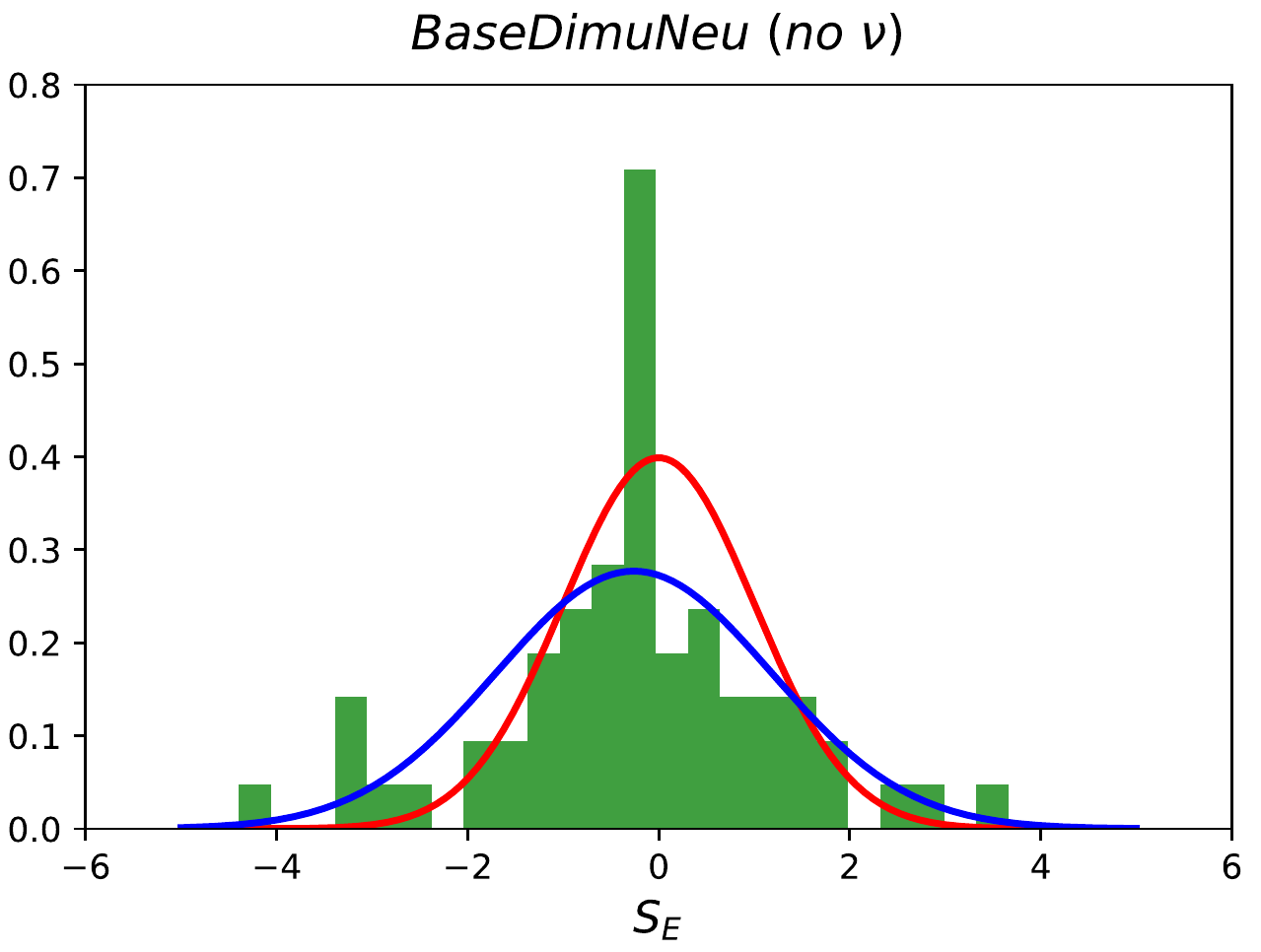}
	\caption{Distribution of the variable $S_E$ for all experiments in the nCTEQ15WZSIHdeut analysis (left) and for all experiments in the BaseDimuNeu analysis (middle). The right panel shows the distribution of the variable $S_E$ from the BaseDimuNeu analysis for experiments in nCTEQ15WZSIHdeut. All panels show the fitted Gaussian distribution to the actual $S_E$ distribution (blue) compared to the ideal Gaussian $S_E$ distribution with $\mu=0$ and $\sigma=1$ (red). Note some of the $S_E$ values lie outside the plot range.}
	\label{fig.SE.comp}
\end{figure*}
Lastly, we will look into the details of how well all experiments are described in the combined global analysis with all neutrino data. In contrast to using the rescaled percentile to account for imperfect description of data, we will use the distribution of the $S(\chi^2,N)$ variable for all the experiments in the combined analysis. Considering the whole distribution allows for the possibility that some experiments in the global analysis are not described well leading to $S_E > 0$ and that some are over fitted ($S_E<0$). Before we investigate the $S_E$ distribution of the combined analysis, we will review the same distribution for the reference nCTEQ15WZSIHdeut analysis which is shown in the left panel of Fig.~\ref{fig.SE.comp}. After analyzing the distribution and determining the mean ($\mu = -0.74$) and the standard deviation ($\sigma = 1.12$), we can see that the nPDF framework with 27 free parameters is describing the data too well on average but the spread is still compatible with the ideal distribution of the  $S(\chi^2,N)$ variable. The distribution of $S_E$ in the case of the BaseDimuNeu analysis is shown in the middle panel of Fig.~\ref{fig.SE.comp} and from the characteristics of the distribution, it is clear that on average experiments are still described well ($\mu = 0.08$). However, this time the standard deviation $\sigma = 2.54$ signifies that there are more outlier experiments. Our interest is twofold. First, we would like to compare the description of the experiments contained in nCTEQ15WZSIHdeut and the subset of the same experiments in the combined analysis BaseDimuNeu. We show the distribution of the nCTEQ15WZSIHdeut experiments in the BaseDimuNeu analysis in the right panel of Fig.~\ref{fig.SE.comp}. Comparing how these two analyses describe the same set of experiments, clearly points to the BaseDimuNeu analysis being a compromise given that the description of this subset of experimental data is worse than in the reference analysis ($\mu = -0.26$ and $\sigma = 1.44$). As expected the worse description can be traced back to the neutral current DIS experimental data which are very sensitive to the up- and down-quark PDF which is one of the PDFs mostly shifted in the combined analysis. The reason why the previous two compatibility criteria signal a problem is hidden in the description of neutrino data. The large standard deviation is mostly caused by the NuTeV neutrino and anti-neutrino cross-section data having extremely large $|S_E|$-values, $S_E=13.05$ for neutrino (not shown on plot) and $S_E=5.5$ for anti-neutrino data. The other contribution to the large standard deviation comes from the di-muon data from both CCFR and NuTeV experiments and from the overfitted CDHSW neutrino cross-section data.
%
\begin{table}[tb]
	\caption{Statistical information on the description of the neutrino data sets used in different analyses.}\label{tab:chi2nudata}
\begin{center}
	\begin{tabular}{|l|c|c|c|}
	\hline
		\multirow{2}{*}{Data set} & \multirow{2}{*}{\#pts} & $\chi^2$/pt ($S_E$) & $\chi^2$/pt ($S_E$)\\
		 &   & {\rm DimuNeu} & BaseDimuNeu \\
		\hline\hline
		CDHSW $\nu$ & 465 & 0.68 (-5.29) & 0.59 (-7.01) \\
		CDHSW $\bar{\nu}$ & 464 & 0.73 (-4.47) & 0.69 (-5.22)\\ \hline
		CCFR $\nu$ & \ 824\ & 0.99 (-0.09) & 1.03 (0.56)\\
		CCFR $\bar{\nu}$ & 826 & 1.00 (0.07) & 1.02 (0.45)\\ \hline
		NuTeV $\nu$ & 1170 & 1.51 (11.12) & 1.61 (13.05)\\
		NuTeV $\bar{\nu}$ & 966 & 1.25 (5.16) & 1.27 (5.50)\\ \hline
		Chorus $\nu$ & 412 & 1.21 (2.85) & 1.25 (3.40)\\
		Chorus $\bar{\nu}$ & 412 & 1.09 (1.26) & 1.25 (3.35)\\ \hline
		CCFR dimuon $\nu$ & 40 & 1.70 (2.79) & 2.52 (5.32)\\
		CCFR dimuon $\bar{\nu}$ & 38 & 0.79 (-0.89) & 0.64 (-1.68)\\ \hline
		NuTeV dimuon $\nu$ & 38 & 0.98 (-0.06) & 2.11 (4.01)\\
		NuTeV dimuon $\bar{\nu}$ & 34 & 0.73 (-1.16) & 1.16 (0.70)\\
		\hline
	\end{tabular}
\end{center}
\end{table}
\begin{table}[tb]
	\caption{Statistical information on the description of the selected neutral current DIS data sets used in the reference nCTEQ15WZSIHdeut and BaseDimuNeu analyses.}\label{tab:chi2data}
\begin{center}
	\begin{tabular}{|l|c|c|c|c|c|}
	\hline
		\multirow{2}{*}{Experiment} & \multirow{2}{*}{Target} & \multirow{2}{*}{ID} & \multirow{2}{*}{\#pts} & $\chi^2$/pt ($S_E$) & $\chi^2$/pt ($S_E$)\\
		 & & & & {\rm Reference} & BaseDimuNeu \\
		\hline\hline
		NMC-95 & C/D & 5113 & 12 & 0.88 (-0.20) & 1.70 (1.59) \\
		NMC-95,re & C/D & 5114 & 12 & 1.18 (0.53) & 2.16 (2.40)\\ \hline
		NMC-95 & Ca/D & 5121 & \ 12\ & 1.15 (0.46) & 2.98 (3.66)\\ \hline
		BCDMS & Fe/D & 5101 & 10 & 0.63 (-0.81) & 2.00 (1.97)\\
		BCDMS & Fe/D & 5102 & 6 & 0.48 (-0.93) & 1.62 (1.09)\\
		\hline
	\end{tabular}
\end{center}
\end{table}

Comparing the statistical results for the nCTEQ15WZSIHdeut and DimuNeu analyses with the combined analysis BaseDimuNeu (see Fig.~\ref{fig.SE.comp} and Tab.~\ref{tab:chi2nudata}), we can identify the origin of the inconsistencies signaled by the first two compatibility criteria. For the $\chi^2$/pt and $S_E$ data for all neutrino experiments shown in Tab.~\ref{tab:chi2nudata}, we can see that the description of the NuTeV cross-section data, Chorus cross-section data and above all the di-muon data in the compromise fit of the BaseDimuNeu analysis is much worse than in the reference only neutrino DimuNeu analysis. Moreover, if one examines the shifts in the description of the experiments in the reference nCTEQ15WZSIHdeut analysis seen in Fig.~\ref{fig.SE.comp} more closely, we can discover large shifts in $\chi^2$/pt or alternatively in the $S_E$ variable especially in precise DIS experiments (for details see Tab.~\ref{tab:chi2data}). These facts all together lead us to conclude that the inconsistency signalled by the other criteria is justified and there is indeed a large tension between the neutrino data and the rest of the scattering data.

The crucial question which we will address in the final part of this paper is if there is a way to include the neutrino DIS data in a combined analysis while at the same time avoiding large tensions and incompatibilities.
%
\section{Consistent global nPDF analysis with neutrino data}
\label{sec:nufinal}
In the previous section, we have shown that incorporating neutrino data into the nCTEQ framework can produce significant tensions among key data sets. Moreover, we have observed in Sec.~\ref{sec:data} that these include tensions among different neutrino scattering measurements, most notably among the ones taken on iron from the CDHSW, CCFR and NuTeV collaborations and those taken on lead from the Chorus collaboration.\footnote{The inconsistency between the CCFR and NuTeV data at large Bjorken $x$ was resolved by accepting the reasoning in Ref.~\cite{Tzanov:2005kr} and not including any CCFR data in the region of $x>0.4$.} To complicate matters even more, the neutrino inclusive DIS data and the neutrino di-muon data each prefer a different strange quark PDF, leading to substantial tensions as well. The goal of this section is to explore ways to include neutrino data in a global analysis so that these large tensions can be avoided or mitigated. 

Before we consider a global analysis, we will introduce a series of fits where on top of all data from the reference nCTEQ15WZSIHdeut analysis, we include neutrino and anti-neutrino data from one single experiment. This way we can explore tensions of neutrino data from every single neutrino experiment with the reference analysis without considering any tensions among the neutrino data themselves. We show the statistical results of four analyses (BaseChorus, BaseCDHSW, BaseCCFR and BaseNuTeV) in Tab.~\ref{tab:statperc2}. The results show that apart from the data from the Chorus experiment, adding the other neutrino experimental data causes tension with the neutral current scattering data. This should come as no surprise in light of the nuclear correction factors extracted from the neutrino and anti-neutrino data shown in Fig.~\ref{fig:Rnunubar}, where only the nuclear correction factor from the Chorus neutrino and anti-neutrino data has a shape similar to the one preferred by the neutral current scattering data. Given the results shown in Fig.~\ref{fig:Rnunubar}, we clearly expect the tensions for the other experiments to come from neutrino data in the low-$x$ and/or in the high-$x$ kinematic region. We will use this information in the following.

Aiming for a global analysis without large tensions among data sets, there are several possible approaches one can take: 
\begin{enumerate}
    \item If the tensions can be attributed to a specific kinematic region, they can be removed by imposing a kinematic cut on the neutrino data.
    \item Large tensions can often be caused by very precise experimental data, and a compromise can be reached if it is believed that the estimate of the experimental errors is underestimated. In such a case, the errors might be artificially enlarged.
    \item The last option is to identify experiments which are still consistent with the bulk of the original data and include only those in our analysis.
\end{enumerate}
We will investigate all of these approaches in the following.

\begin{table*}[htb]
	\caption{Statistical information such as the total $\chi^2$ and the number of data points for all analyses discussed here are presented. Moreover, the $\chi^2$-percentiles with respect to the default data sets of the reference fit nCTEQ15WZSIHdeut (denoted $S$) and to the DimuChorus analysis (denoted $\bar{S}$) are also given if applicable.}
   \label{tab:statperc2}
	\centering
	\begin{tabular}{|c|c|c|c|c|c|c|c|}
	\hline
		Analysis name & $\chi^2_S/N$ & $\chi^2_S/pt$ & $\chi^2_{\bar{S}}/N $ & $\chi^2_{\bar{S}}/pt $ & $\Delta\chi^2_S$ & $\Delta\chi^2_{\bar{S}}$ & $p_S/p_{\bar{S}}$ \\[1mm] \hline \hline
        nCTEQ15WZSIHdeut & 735/940 & 0.78 & - & - & 0 & - & 0.500 / - \\
        DimuChorus & - & - & 1059/974 & 1.09 & - & 0 & - / 0.500 \\
        \hline
        BaseChorus & 737/940 & 0.78 & 969/824 & 1.18 & 2 & - & 0.530 / -\\
        BaseCDHSW & 778/940 & 0.83 & 584/929 & 0.63 & 43 & - & 0.895 / -\\
        BaseCCFR & 815/940 & 0.87 & 2119/2207 & 0.96 & 80 & - & 0.989 / -\\
        BaseNuTeV & 807/940 & 0.86 & 3049/2136 & 1.43 & 72 & - & 0.981 / -\\
        BaseNuTeVU & 787/940 & 0.84 & 1984/2136 & 0.93 & 52 & - & 0.933 / -\\
        \hline
        BaseDimuNeuU & 861/940 & 0.92 & 5569/5689 & 0.98 & 126 & - & 0.99978 / - \\
        BaseDimuNeuX & 781/940 & 0.83 & 5032/4644 & 1.08 & 46 & - & 0.908 / - \\
        BaseDimuChorus & 740/940 & 0.79 & 1117/974 & 1.15 & 5 & 58 & 0.559 / 0.885  \\
        \hline
	\end{tabular}
\end{table*}
\begin{figure*}[htb]
	\centering
	\includegraphics[width=0.32\textwidth]{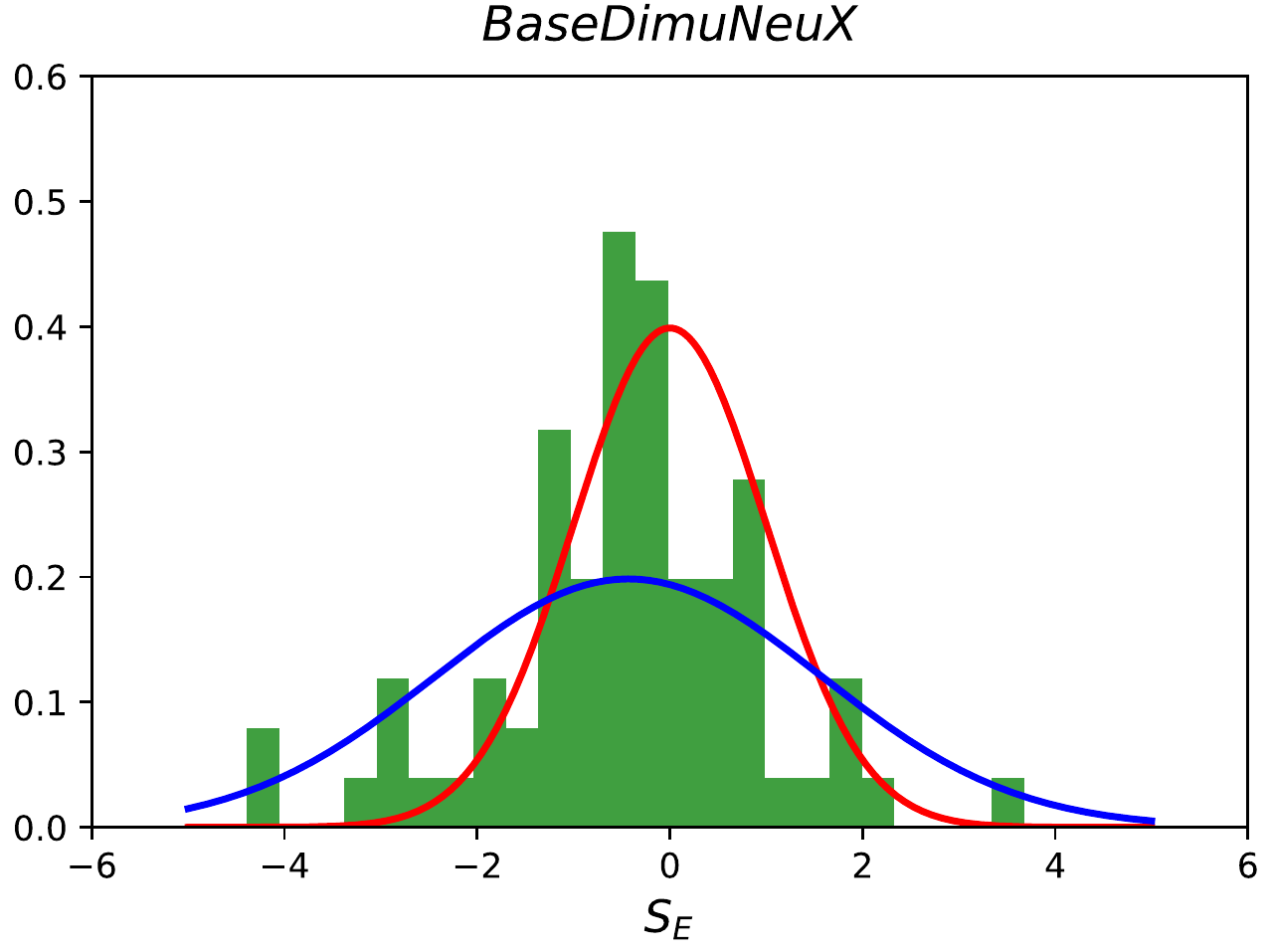}\quad 
	\includegraphics[width=0.32\textwidth]{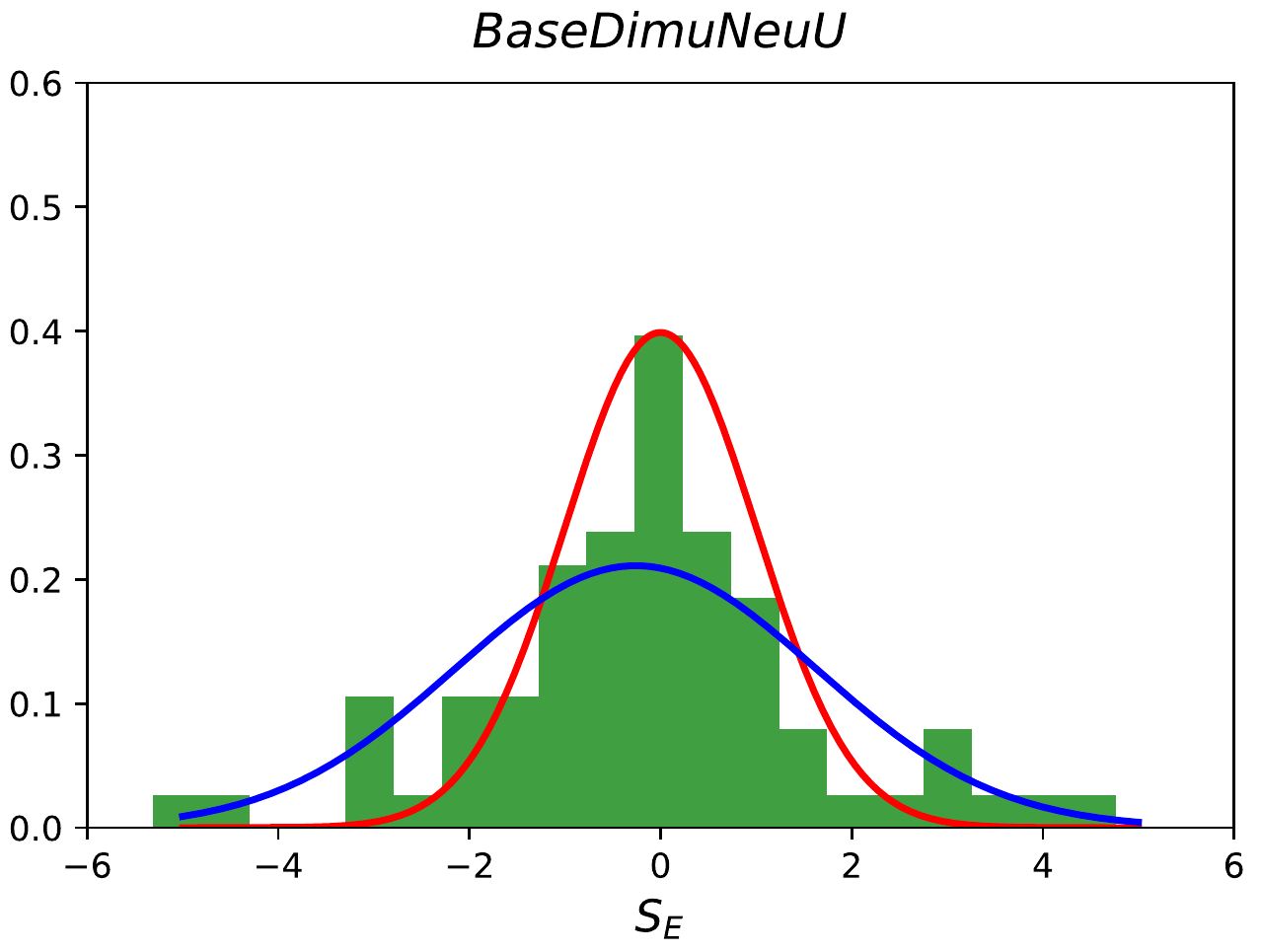}\quad 
	\includegraphics[width=0.32\textwidth]{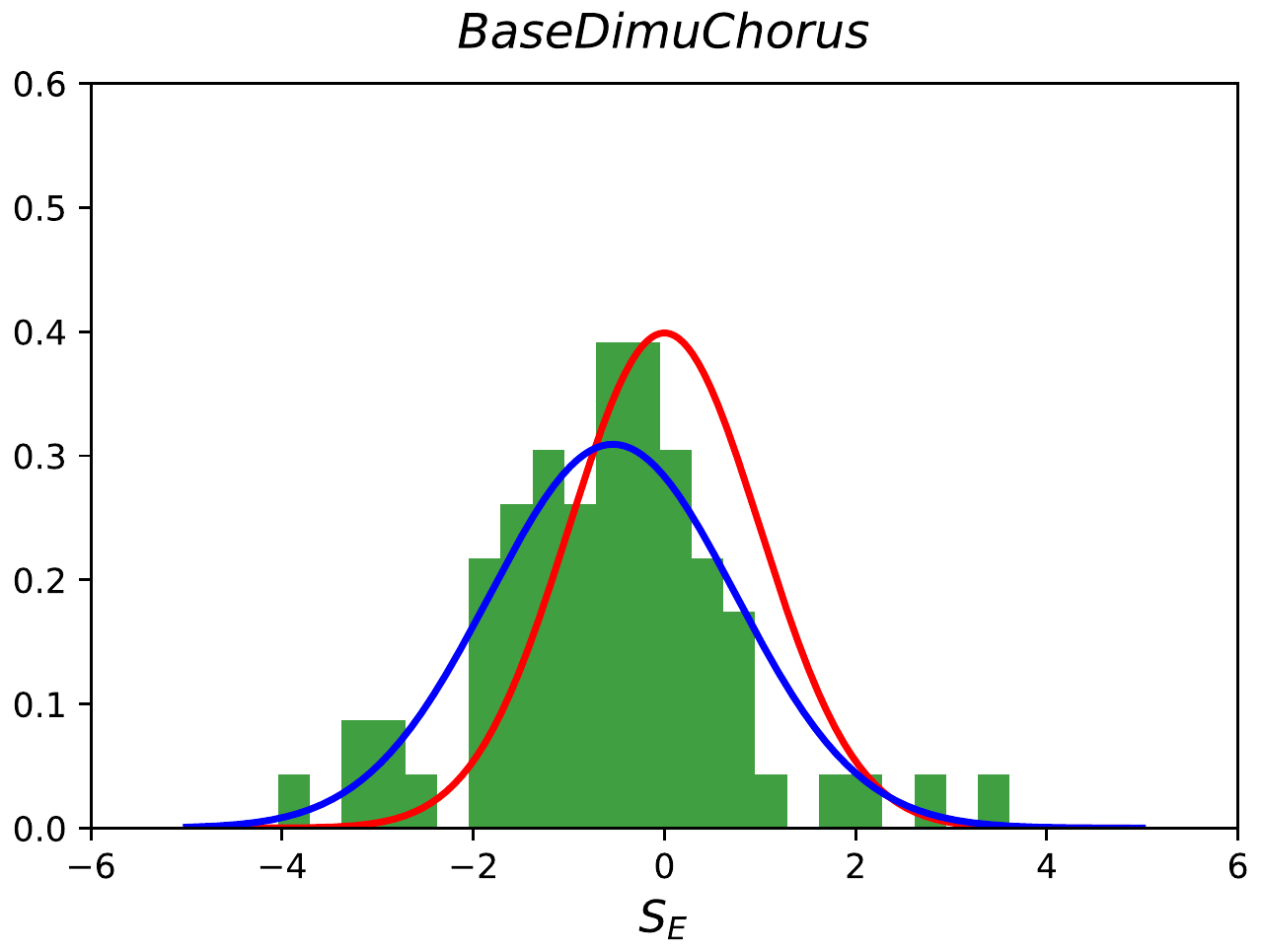}
	\caption{Distribution of the variable $S_E$ for all experiments in the BaseDimuNeuX analysis (left) and for all experiments in the BaseDimuNeuU analysis (middle). The right panel shows the distribution of the variable $S_E$ from the BaseDimuChorus analysis. All panels show the fitted Gaussian distribution to the actual $S_E$ distribution (blue) compared to the ideal Gaussian $S_E$ distribution with $\mu=0$ and $\sigma=1$ (red). Note that in the case of the BaseDimuNeuX analysis we do not show a bin with $S_E$=9.72 which corresponds to the NuTeV neutrino data.}
	\label{fig.SE.comp2}
\end{figure*}
\subsection{Neutrino DIS data with $x>0.1$}
\label{sec:xcut}
%
\begin{figure*}[htb]
	\centering
	\includegraphics[width=0.95\textwidth]{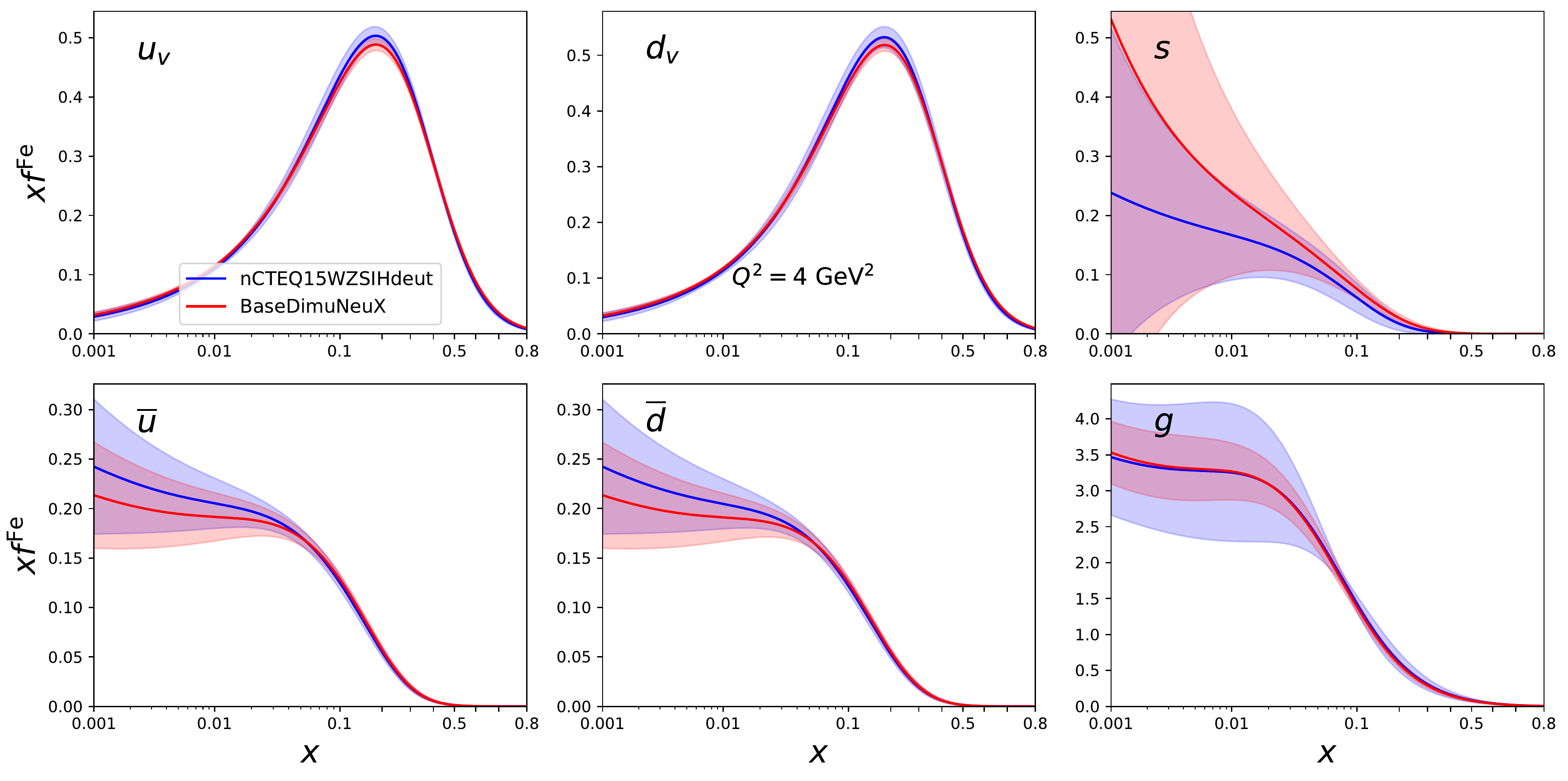}
	\caption{The full iron PDFs at $Q^2=4\ {\rm GeV}^2$. All uncertainty bands are computed using the Hessian method with $\Delta \chi^2=45.$}
	\label{fig.pdfbasedimuneux}
\end{figure*}
\begin{figure*}[htb]
	\centering
	\includegraphics[width=0.95\textwidth]{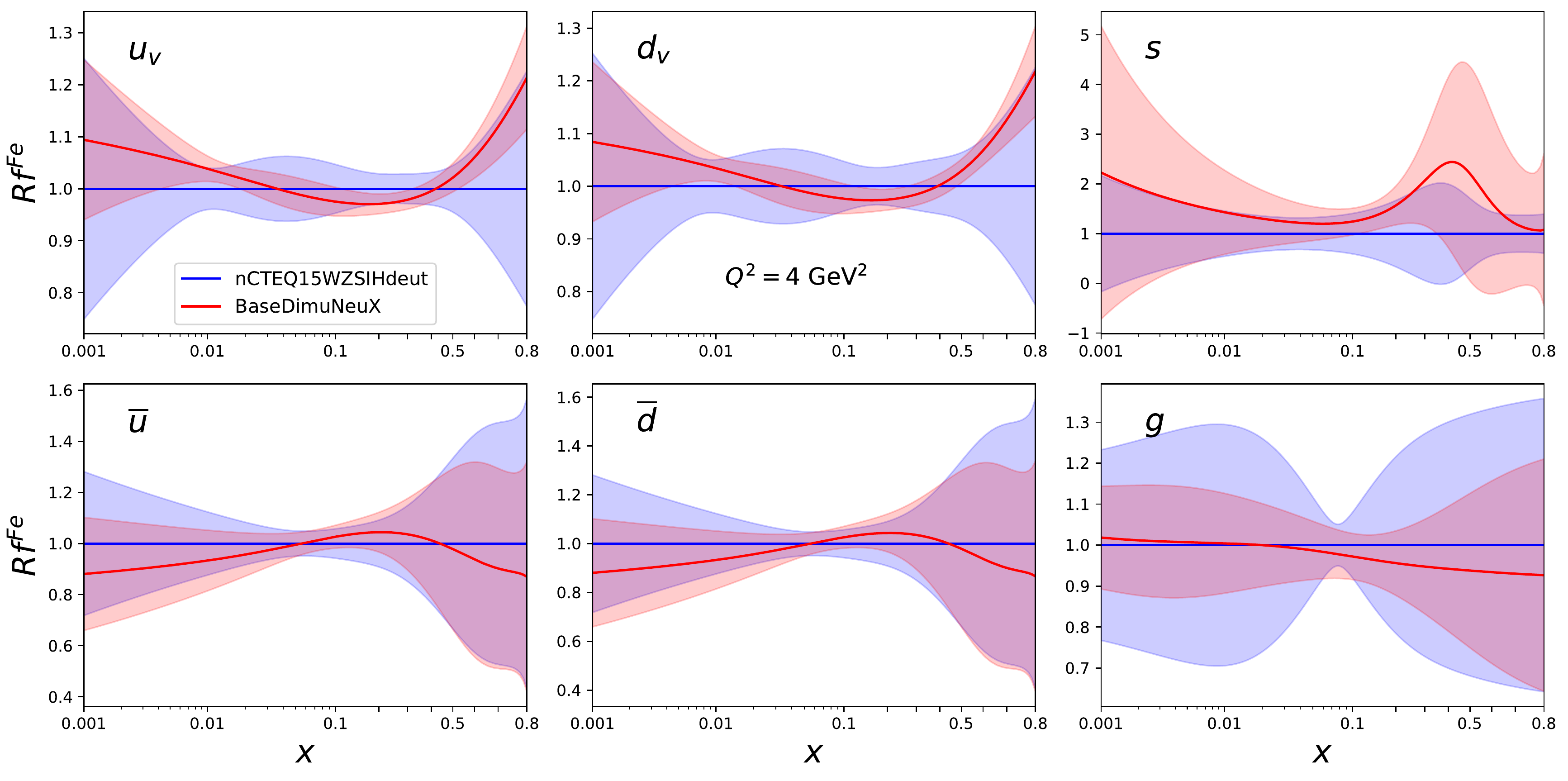}
	\caption{The fitted iron PDF ratio to nCTEQ15WZSIHdeut. All uncertainty bands are obtained using the Hessian method with $\Delta \chi^2=45$.}
	\label{fig.rpdfbasedimuneux}
\end{figure*}
The large tensions and incompatibilities observed in the previous section were not completely surprising considering the ratio we have extracted from the cross-section data in Sec.~\ref{sec:weightedaverage} and which as shown in Fig.~\ref{fig:Rnunubar}, shows a markedly different shape of the nuclear correction factor, especially in the small-$x$ and in the very large-$x$ regions. Given that our conservative kinematic cuts on $Q^2$ and $W^2$ are already effectively restricting the large $x$ region, the only way how we can resolve the tension using a kinematic cut is to exclude the low-$x$ neutrino data.

Using arbitrary cuts to remove the data which cause the largest tensions in each experiment is not in line with the philosophy of a global analysis, because it introduces a bias which such an analysis tries to avoid. One possible motivation for using a cut to remove data could be signs that the theoretical description of the data in a specific region is inadequate. In this section, we will assume that the large tensions in the low-$x$ region may be due to e.g. a different mechanism for nuclear shadowing in charged current DIS \cite{Kopeliovich:2012kw} which is not properly included in our theoretical framework. A different reason one could have to justify a kinematic cut is an internal tension between all neutrino data in this region. We can see in Fig.~\ref{fig:Rnunubar} that there is indeed such a tension in the low-$x$ region, especially between the NuTeV and CCFR data on one side and Chorus data on the other. 

Citing any of these reasons, we will employ an arbitrary constraint, $x>0.1$, which the charged current DIS data have to fulfill. This applies to all inclusive DIS and di-muon data. 

To show the impact of such a cut, we have performed an analysis similar to the global BaseDimuNeu analysis requiring that all neutrino data satisfy the constraint, $x>0.1$. This analysis, called in the following BaseDimuNeuX, uses the same number of free parameters and, similarly to the previous analysis, also fits the normalisation of all neutrino experiments. 

The kinematic cut removes 1045 data points from the low-$x$ region of neutrino scattering data. The result of this analysis has $\chi^2$/pt = 1.04. Further details and the breakdown of the $\chi^2$ for the usual data subsets are listed in Tab.~\ref{tab:statperc2}. 

Analyzing the statistical properties, we see that with $\Delta\chi^2_S = 46$, which is approximately within the 91-percentile, the analysis BaseDimuNeuX is barely consistent with the original data in the nCTEQ15WZSIHdeut analysis. A closer look at Fig.~\ref{fig.SE.comp2} reveals that most experiments are fitted well except only a few outliers. The tensions are experienced by the NuTeV neutrino cross-section data ($S_E$ = 9.72 largest not shown) and by the NuTeV anti-neutrino data ($S_E$ = 3.37). Without these data the $S_E$ distribution would be very similar to the one of the reference analysis shown in Fig.~\ref{fig.SE.comp}. 

In Figs.~\ref{fig.pdfbasedimuneux} and \ref{fig.rpdfbasedimuneux} we compare the extracted nuclear PDFs to the ones of nCTEQ15WZSIHdeut. If we first focus on the central values of the nuclear PDFs extracted in the BaseDimuNeuX analysis, we observe that except for the strange quark PDF, the central values are within the error bands of the reference analysis. This nicely highlights the usefulness of the $\Delta\chi^2$-criterion. Comparing the results with those of the BaseDimuNeu analysis shown in Fig.~\ref{fig.pdfratio}, we see that the shapes of the central values are very similar. This indicates that the tensions from the original global analysis are not completely removed but just reduced in size. The unexpected part of the result which can be seen in Fig.~\ref{fig.pdfbasedimuneux} is the uncertainty band of the strange quark and gluon PDFs. The large uncertainty of the strange quark is a result of two competing preferences for the strange quark PDF from the di-muon and the neutrino inclusive cross-section data. As a result the uncertainty is enlarged to account for this tension. The reduced gluon PDF uncertainty is due to adding a large number of precise DIS data, constraining the gluon via the NLO sensitivity of the DIS process to the gluon PDF.

The predictions for the nuclear correction factors from the neutral and charged current DIS are shown in Fig.~\ref{fig.F2.prediction.comp-2} and compared with those of the reference analysis. There we can see that, as expected, the much different behavior of the theoretical prediction in the low-$x$ region, which was present in Fig.~\ref{fig.F2.prediction.comp} for the BaseDimuNeu analysis, is largely gone and the prediction has a larger uncertainty band for the charged current nuclear correction factor. Moreover, excluding neutrino data with $x<0.1$ from the analysis significantly effects the prediction of the nuclear correction factor also in other regions in $x$. In Fig.~\ref{fig.F2.prediction.comp-2} we see that the structure function data from NuTeV and CDHSW are not correctly described even in the intermediate $x$ region and only the large $x$ behavior is driven by the NuTeV data and remains very different from the predictions of the reference analysis. 

Overall, we see that employing the cut $x>0.1$ to all neutrino data reduces the tensions just enough for this fit to be considered consistent. However, some problems still remain. The tension in the previously well determined valence quark PDFs is still present and the NuTeV cross-section data is still badly described. Moreover, all this has been achieved after removing the small-$x$ and large-$x$ data where the tensions are the largest. It needs to be stressed once more that this analysis can be considered the final result only if a plausible explanation for the additional kinematic cut is put forward.
%
\subsection{NuTeV with uncorrelated systematic errors}
\label{sec:uncorr}
The second possible approach to lessen the tensions we consider is to enlarge the errors of the experimental data causing the tension. An equivalent to enlarging the errors of all data of a data set is to introduce a weight for this data set in the calculation of the $\chi^2$-function. We have investigated this option in our previous analysis \cite{Kovarik:2010uv} and found no acceptable way to include the neutrino DIS data in a global analysis. 

In a similar spirit, previous analyses \cite{Kovarik:2010uv, Paukkunen:2010hb} enlarged the errors of the NuTeV cross-section data by not considering the correlated systematic errors. Let us therefore explore the effect of neglecting these correlations on the combined analysis. First, we have performed a fit with the data from the nCTEQ15WZSIHdeut analysis and only from the NuTeV experiment using uncorrelated systematic errors. The analysis BaseNuTeVU clearly shows that with uncorrelated systematic errors the framework we use to fit the experimental data can, for the first time, describe the NuTeV data well with $\chi^2$/pt=0.93. Moreover, comparing to the BaseNuTeV analysis which used correlated systematic errors, we see that the tension with the neutral current data is reduced but still present (for details see Tab.~\ref{tab:statperc2}). This shows that the inconsistencies cannot be attributed solely to the use of correlated systematic errors. For completeness, we have also performed a global analysis much like BaseDimuNeu but without correlations in the case of the NuTeV data (called BaseDimuNeuU). Here a similar picture emerges. The neutrino data are described much better ($\chi^2$/pt=0.98), but the tension with the neutral current data is unchanged. Some details of the tensions are again visible in the $S_E$-distribution shown in Fig.~\ref{fig.SE.comp2}, where the standard deviation of the distribution is much larger than unity ($\sigma$ = 1.89). Large $S_E$ contributions can be traced back to the neutrino di-muon data from both CCFR ($S_E$=4.77) and NuTeV ($S_E$=3.19) which as we have seen before prefer a different strange quark PDF compared to the inclusive neutrino data. The tensions with the neutral current DIS data have also not improved but rather got worse compared to the BaseDimuNeu analysis (see Tab.~\ref{tab:chi2data}). The largest $S_E$ contributions still come from the Ca/D and C/D data from the NMC collaboration ($S_E$=3.91 and $S_E$=2.45 respectively). Therefore, we conclude that the use of correlated systematic errors for the NuTeV data has no effect on the compatibility of the neutrino data with the rest of the scattering data and neglecting the correlations does not reduce the tensions, even though the neutrino data seem to be described well overall.

\begin{figure*}[htb]
	\centering
	\includegraphics[width=0.95\textwidth]{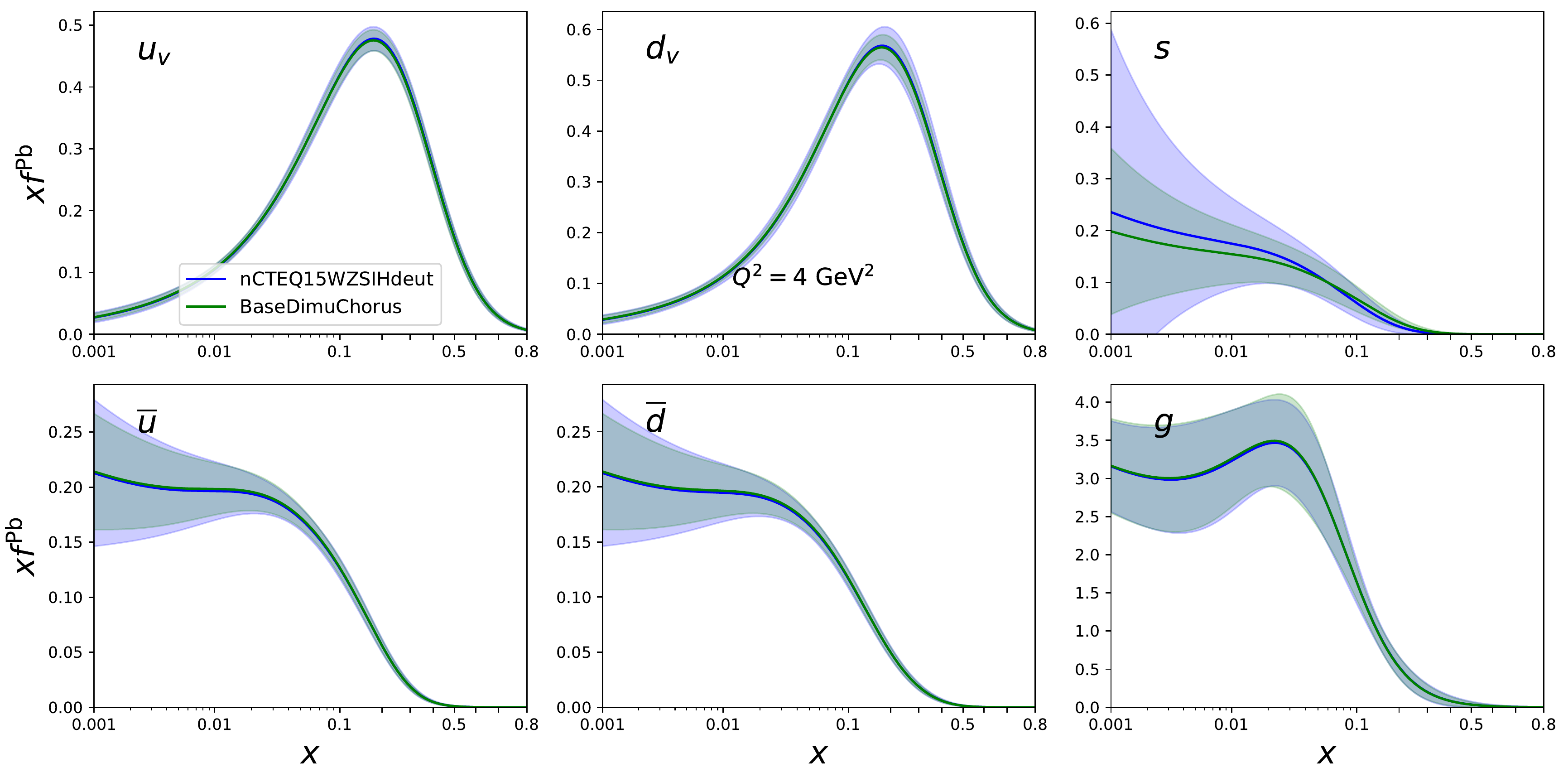}
	\caption{The full lead PDFs at $Q^2=4\ {\rm GeV}^2$. All uncertainty bands are computed using the Hessian method with $\Delta \chi^2=45.$}
	\label{fig.pdfchorus}
\end{figure*}
\begin{figure*}[htb]
	\centering
	\includegraphics[width=0.95\textwidth]{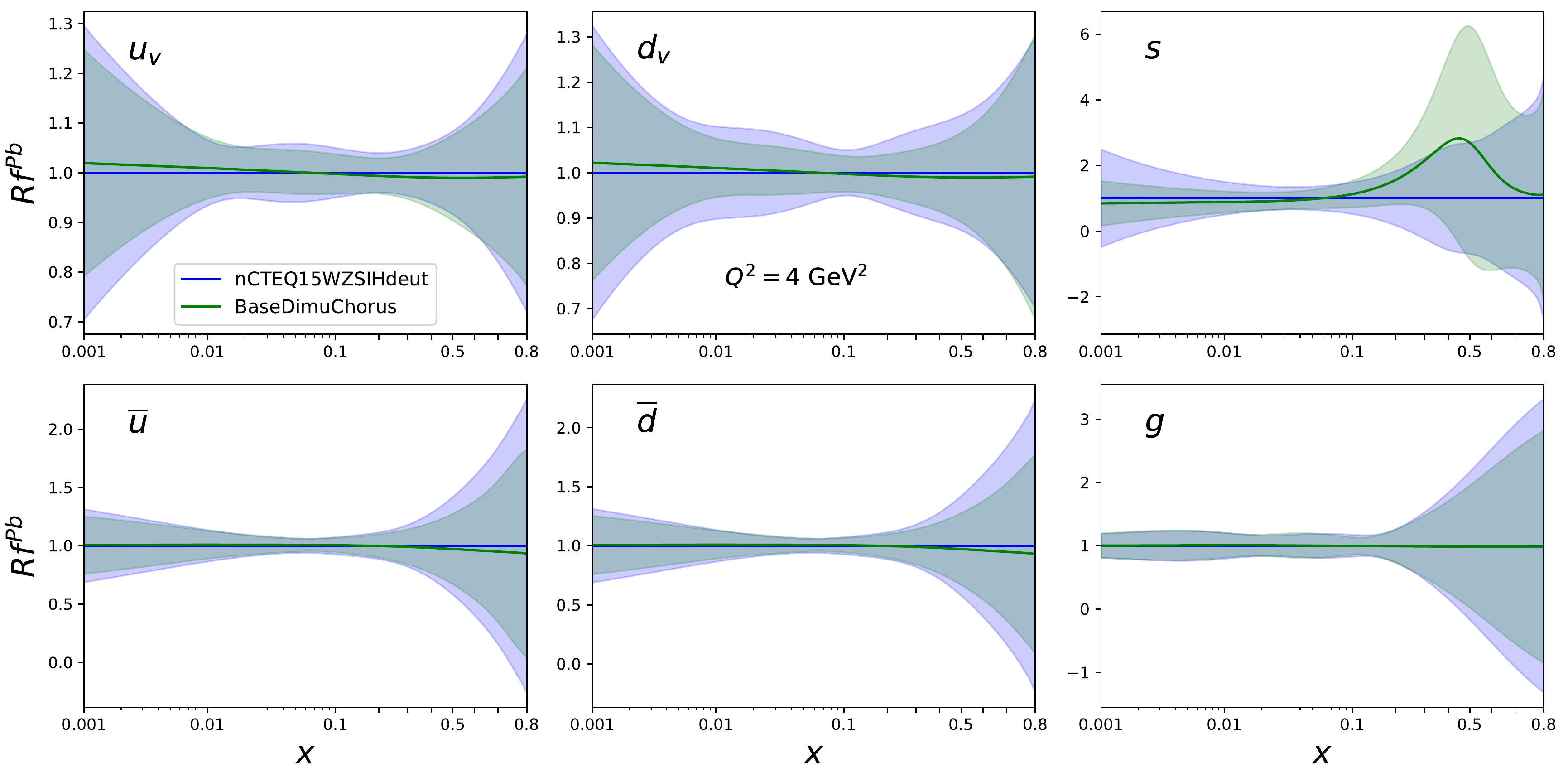}
	\caption{The fitted lead PDF ratio to nCTEQ15WZSIHdeut. All uncertainty bands are obtained using the Hessian method with $\Delta \chi^2=45$.}
	\label{fig.pdfratiochorus}
\end{figure*}
\begin{figure*}[htb]
	\centering
	\includegraphics[width=0.80\textwidth]{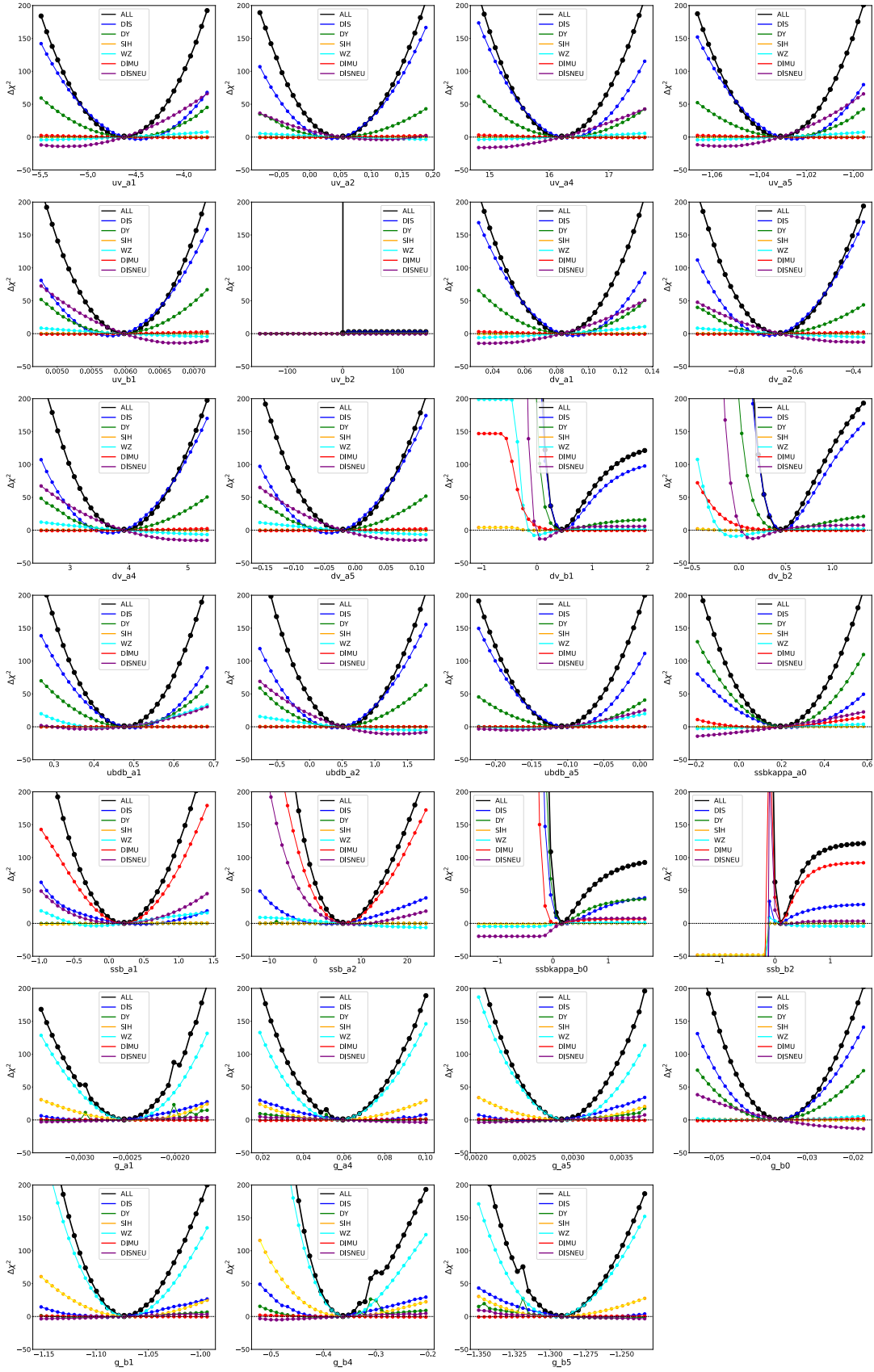}
	\caption{Scans of the $\chi^2$ function along the PDF parameter directions varying always one free parameter at a time while other parameters were left fixed at the global minimum of the BaseDimuChorus analysis. The breakdown into $\chi^2$ for classes of experimental data is also shown. We note that in this case "DISNEU" refers to the Chorus data which is the only inclusive neutrino data used in this fit.}
	\label{fig.chi2.scan.chorus}
\end{figure*}
%
\section{Global analysis with Chorus and di-muon data}
\label{sec:ncteqnu}
%
\begin{figure*}[htb]
	\centering
	\includegraphics[width=0.48\textwidth]{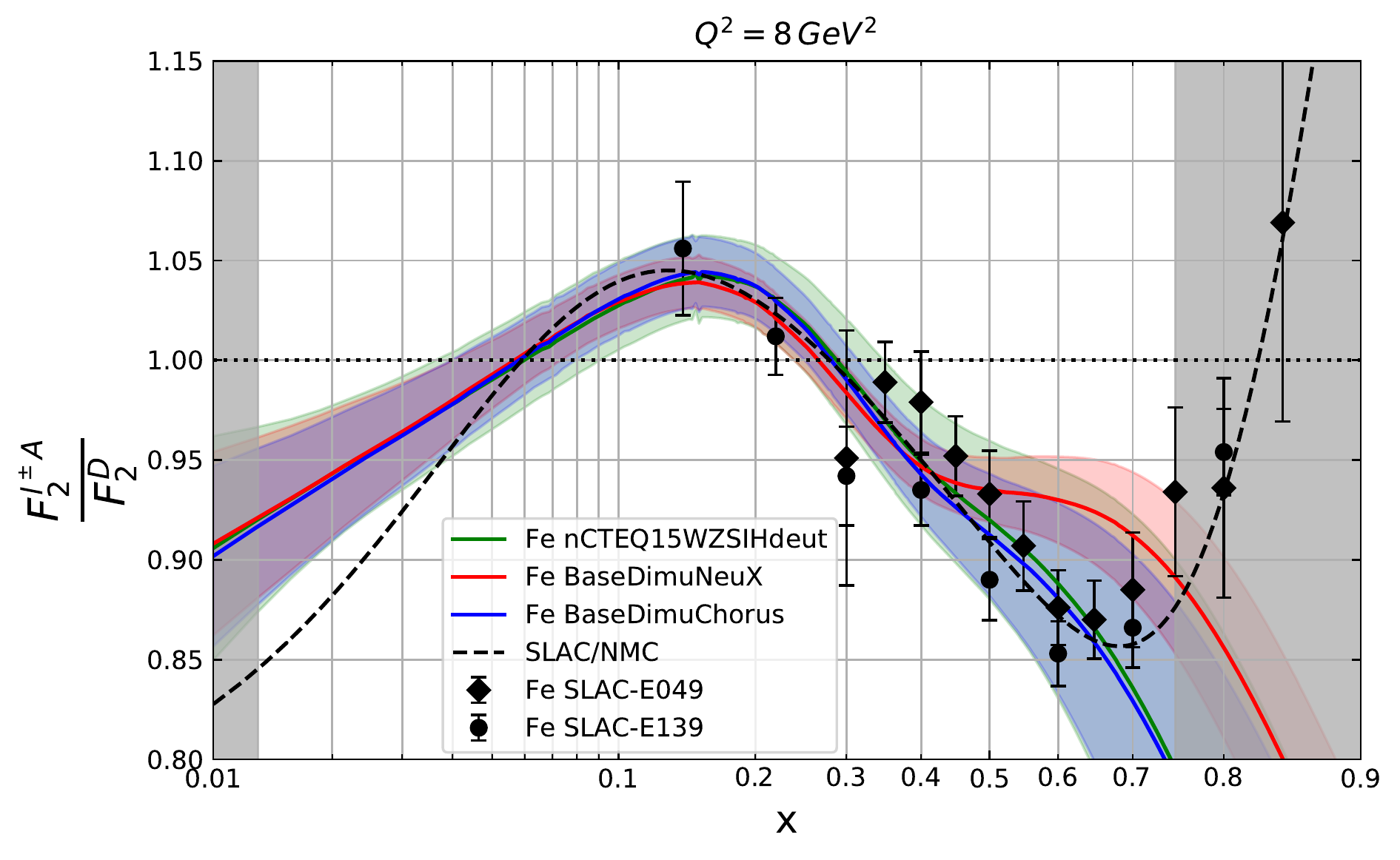}\qquad
	\quad
	\includegraphics[width=0.45\textwidth]{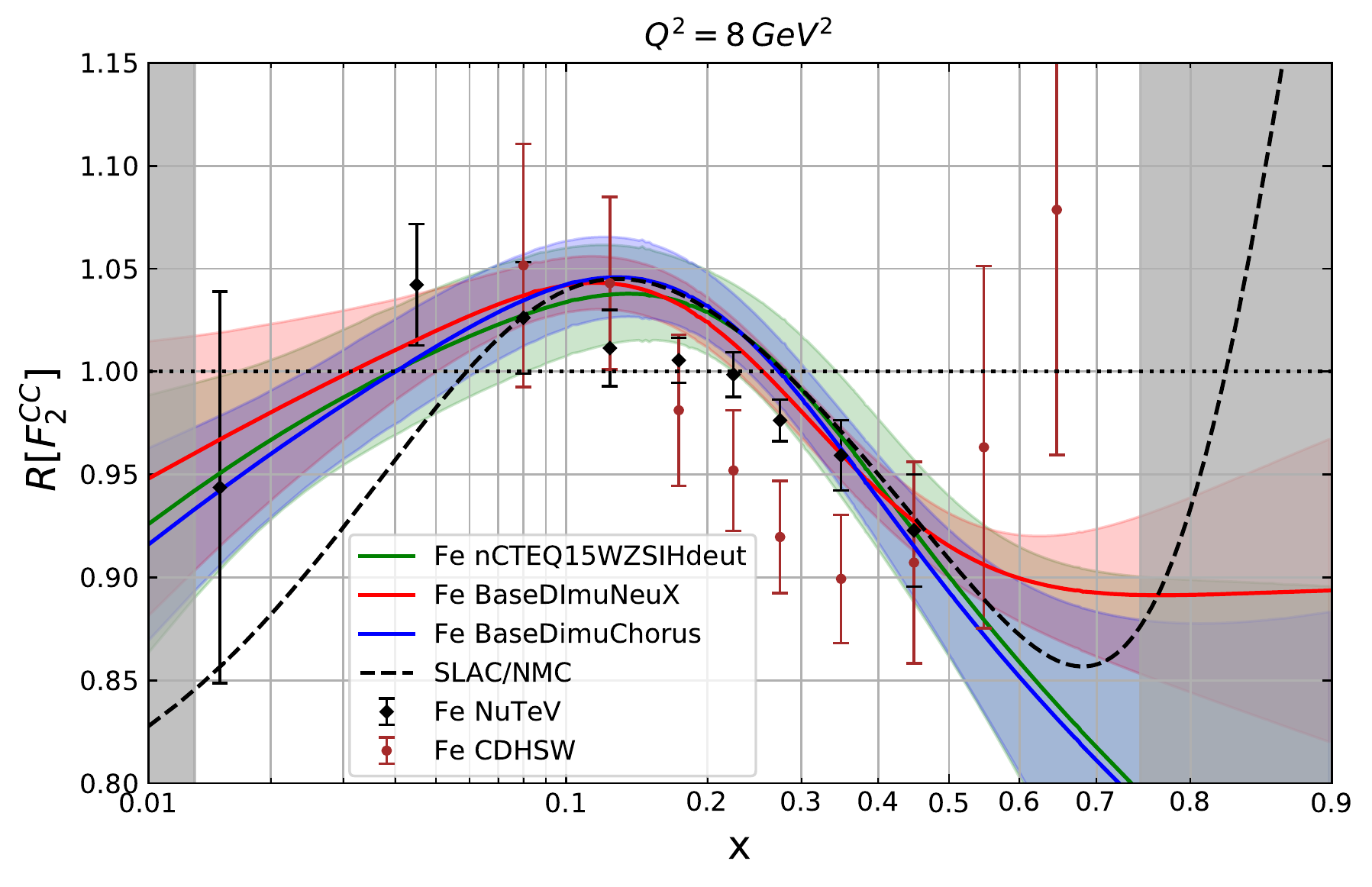}
	\caption{Neutral current nuclear ratio $F_2^{\rm Fe}/F_2^{\rm D}$ (left) and charged current nuclear ratio $R[F_2^{\rm CC}]$ as defined in Eq.~(\ref{rf2cc}) (right) using the fitted nPDFs. Note that we have applied nuclear corrections for the neutral current deuterium structure function $F_2^D$, but not for the charged current one.}
	\label{fig.F2.prediction.comp-2}
\end{figure*}
%
\begin{figure*}[htb]
	\centering
	\includegraphics[width=0.5\textwidth]{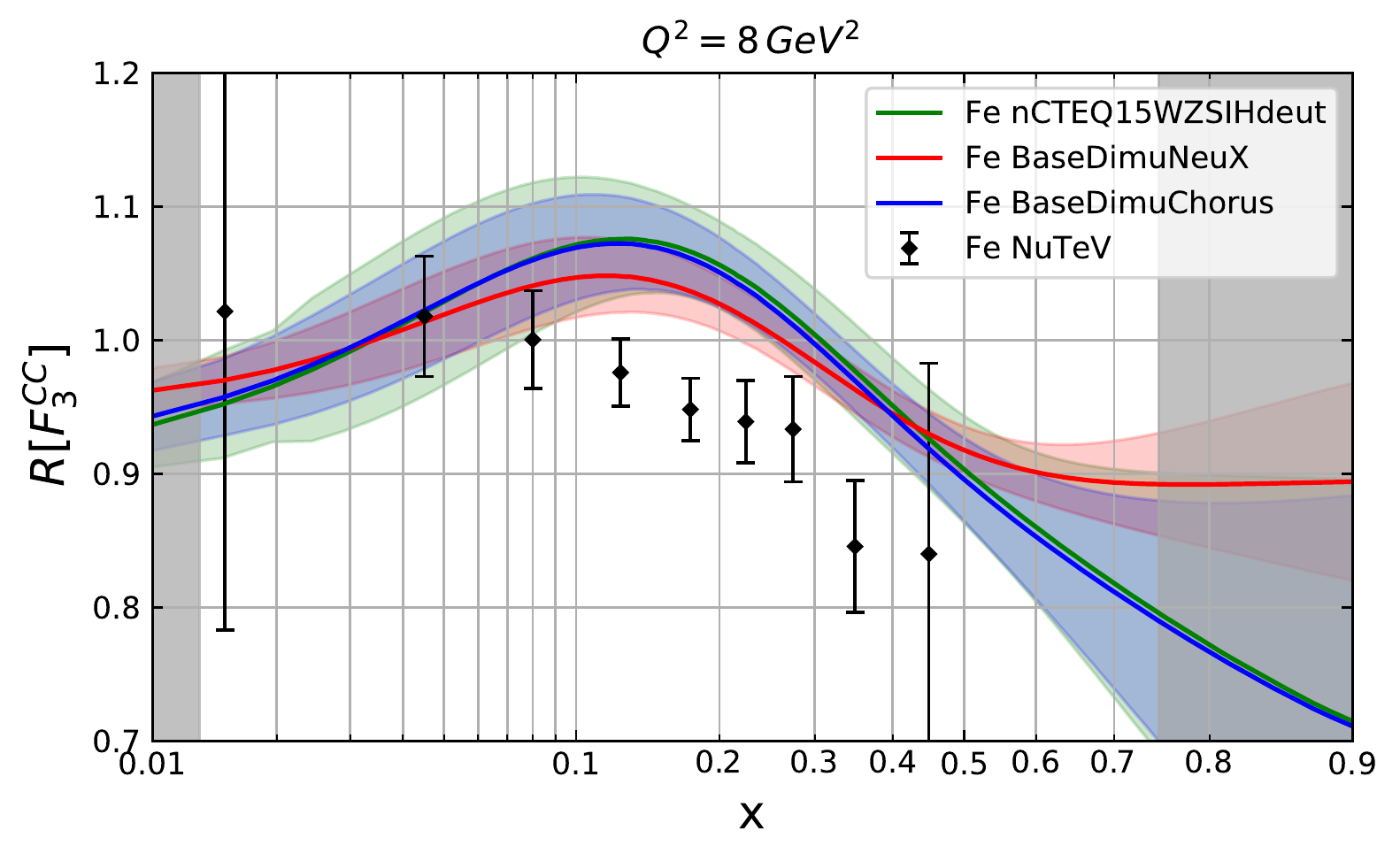}\qquad
	\caption{Charged current nuclear ratio $R[F_3^{\rm CC}]$ defined analogously to $R[F_2^{\rm CC}]$ using the fitted nPDFs.}
	\label{fig.F3.prediction.comp-2}
\end{figure*}
%
\begin{figure}[htb]
	\centering
	\includegraphics[width=0.45\textwidth]{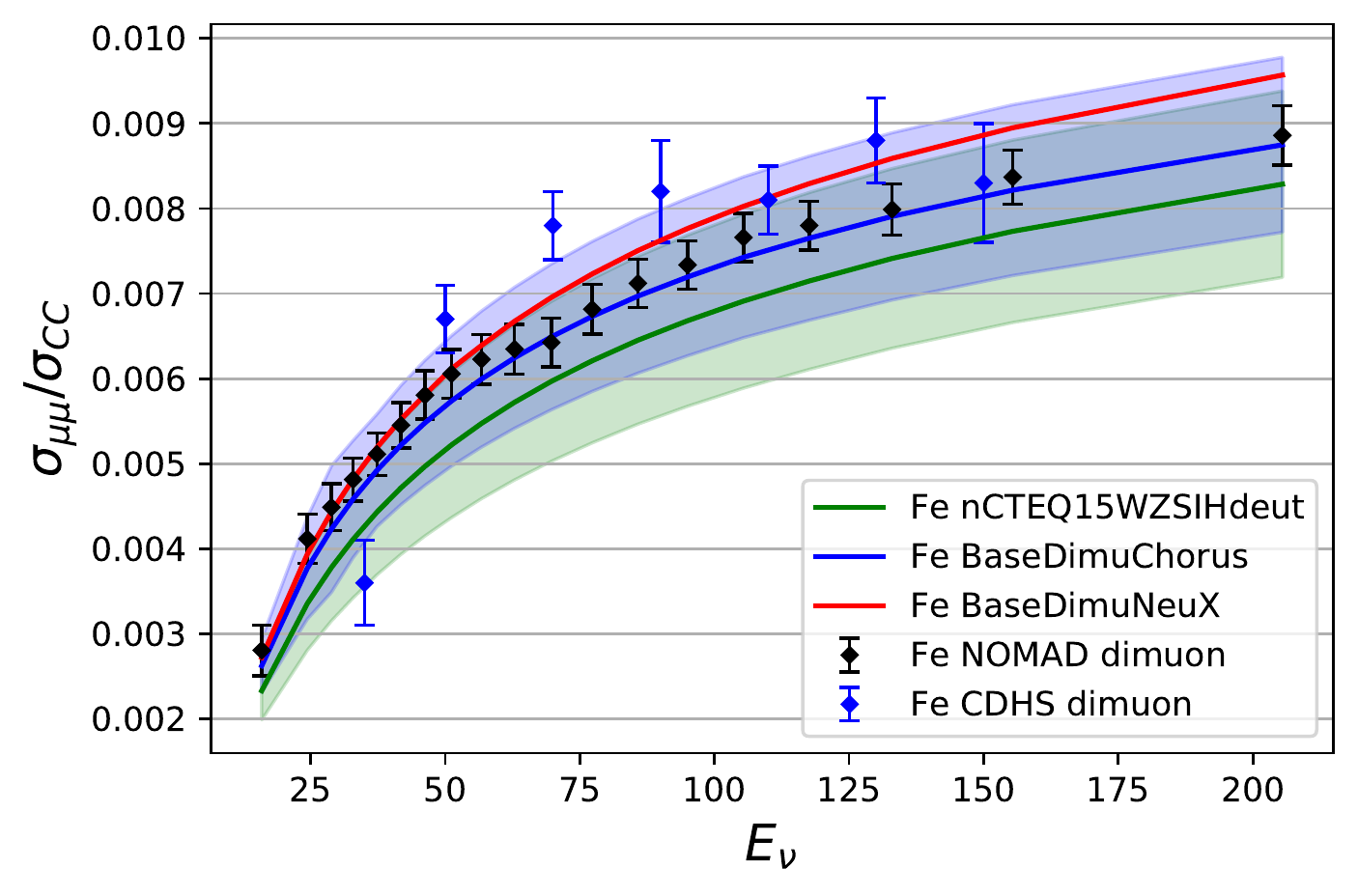}\qquad
	\quad
	\caption{Comparison between the data from the NOMAD experiment \cite{NOMAD:2013hbk} and our theory predictions using our fitted PDFs for the ratio of the di-muon production and the total charged current DIS cross-section.}
	\label{fig.dimuon.prediction.comp}
\end{figure}
As we have shown in Sec.~\ref{sec:nuglobal}, the global analysis of all available data where also all neutrino data are included leads to large tensions. Furthermore, we have shown that these cannot be sufficiently removed by introducing a kinematic cut or by neglecting the correlations of the systematic errors of the neutrino experiment where the tensions are the largest. One option which we have not yet explored is to try to identify a subset of the neutrino data which shows no or little tension. Based on what we have observed in previous analyses, we will add all di-muon and both Chorus neutrino and anti-neutrino scattering data to our global analysis and disregard all other (anti-)neutrino data. We will refer to this global analysis as BaseDimuChorus. The statistical results of this analysis are also given in Tab.~\ref{tab:statperc2} and the total $\chi^2$/pt = 0.97. As can be seen from the details in Tab.~\ref{tab:statperc2}, in this combined analysis, all data from the reference nCTEQ15WZSIHdeut analysis as well as all neutrino data are described well. We have performed a dedicated analysis of only di-muon and Chorus data (DimuChorus analysis) so that we can assess how well these data are described in the combined analysis. Using the rescaled percentiles defined above, we see that the descriptions of both the nCTEQ15WZSIHdeut data and di-muon and Chorus data are both within the 90\% percentile of the $\chi^2$-distribution. In Fig.~\ref{fig.SE.comp2} we also show the $S_E$ distribution and clearly see that on average the data are over fitted ($\mu$ = -0.54) and that the standard deviation of the distribution is larger than the one for nCTEQ15WZSIHdeut ($\sigma$ = 1.28). This is due to the new neutrino data given that the neutrino cross-section data from Chorus are fitted to $\chi^2$/pt = 1.27 ($S_E$ = 3.61) and also the di-muon data from CCFR to $\chi^2$/pt = 1.68 ($S_E$ = 2.70). However, we can see that these data were not described much better in any other analysis and given that all other criteria do not signal inconsistencies, we can look at these results as statistical fluctuations.

In Figs.~\ref{fig.pdfchorus} and \ref{fig.pdfratiochorus} we show the extracted nuclear PDFs from this analysis and compare them to those extracted from the reference nCTEQ15WZSIHdeut analysis. We can see that the central values are almost identical for all but the strange quark PDFs, where the addition of new neutrino data leads to a shift in the central value of the strange quark PDF. Moreover, the neutrino data are also more sensitive to the strange quark, which is reflected in the noticeably reduced uncertainty. The effect of better flavor separation of the quark PDFs thanks to the addition of the (anti-)neutrino cross-section data from Chorus can be also observed in the reduced uncertainties of the valence quark PDFs. From the scans of the $\chi^2$ function along the free parameters and the breakdown into separate contributions to the global $\chi^2$ stemming from different experiment classes shown in Fig.~\ref{fig.chi2.scan.chorus} we can read off the details, which subset of experiments is responsible for constraining specific parameters. We can infer from Fig.~\ref{fig.chi2.scan.chorus} that the valence quark and the anti-quark parameters are mainly constrained by the neutral current DIS experiments while the gluon parameters are constrained by the vector boson production processes at the LHC and from the single inclusive hadron production processes. Most importantly for this analysis, we see that the strange quark parameters are constrained by the di-muon data and also from the Chorus inclusive data alike. 

The predictions for the nuclear correction factors for the neutral and charged current DIS are shown in Figs.~\ref{fig.F2.prediction.comp-2} and \ref{fig.F3.prediction.comp-2}. The predictions from the BaseDimuChorus and the reference nCTEQ15WZSIHdeut analyses are almost identical and we can observe a reduction in the uncertainties after adding the Chorus and di-muon data. In the case of the charged current nuclear correction factor for the structure function $F_2$, we see that the theoretical prediction from the BaseDimuChorus analysis does not describe the structure function data from NuTeV or CDHSW well. This is to be expected as we have omitted the corresponding NuTeV, CCFR and CDHSW cross-section data from the fit as they were the source of inconsistencies. In the case of the structure function $F_3$, neither the predictions from the BaseDimuChorus or from the BaseDimuNeuX analysis can describe the $F_3$ data from NuTeV well. We should note that even though the normalization of the cross-section data from NuTeV (and also from the other collaborations) was allowed to vary as a part of the fitting procedure, no shift was applied to the structure function data shown in Figs.~\ref{fig.F2.prediction.comp-2} and \ref{fig.F3.prediction.comp-2}. Shifting the NuTeV data by the normalization of 3.6\% determined in the BaseDimuNeuX analysis would improve the tensions between the data and the theoretical prediction for both structure functions from this analysis.

Finally, in Fig.~\ref{fig.dimuon.prediction.comp} we also compare the theoretical predictions for the ratio of di-muon and charged current total cross-sections measured by the NOMAD collaboration as a function of the incoming neutrino energy. We see that the prediction from the BaseDimuChorus analysis where the strange quark PDF is largely determined by the CCFR and NuTeV di-muon data, describes the NOMAD di-muon data very well for all incoming neutrino energies. We also observe that the uncertainty on the prediction is much larger than the experimental errors indicating that including this data in our future analysis can lead to a substantially more precise extraction of the strange quark PDF. Given the large uncertainties on all theoretical predictions shown in Fig.~\ref{fig.dimuon.prediction.comp}, we can consider the NOMAD data to be described well enough even by the nCTEQ15WZSIHdeut and BaseDimuNeuX analyses. This is an indication of a realistic estimation of the uncertainty of the strange quark PDF in these analyses. 

Out of all possible approaches listed at the beginning of Sec.~\ref{sec:nufinal}, only the last one presented here led to a combined analysis compatible with the reference analysis nCTEQ15WZSIHdeut. Moreover, the neutrino data included in this analysis provided a much improved description of the strange quark PDF.
\section{Conclusions and outlook}
\label{sec:conclusion}
The aim of this analysis was to take a second look at the (anti-)neutrino deep inelastic scattering data and see if, after all the developments of recent years, a conclusion different to the one presented in our analysis \cite{Kovarik:2010uv} can be reached. As our previous study of the neutrino data predates the nCTEQ15 analysis and any updates thereafter, one could have imagined a shift in the outcome. Moreover, compared to our previous analysis, we were now in a position to use different tools to analyse the compatibility of the neutrino DIS data. We have also added other neutrino data sets to make the current analysis much more comprehensive. 

The analysis presented in this paper starts by collecting all relevant updates to the nCTEQ15 analysis to form the reference fit to use in comparing the compatibility of neutrino data. This is then followed by reviewing the neutrino data and presenting the extraction of effective nuclear correction factors from the cross-section data. On top of that, a fit to all neutrino data is performed and the results are compared with the reference analysis.

In the main part of this analysis in Sec.~\ref{sec:nuglobal} we have performed a global fit (BaseDimuNeu) where we have added all neutrino data to the extended nCTEQ15 analysis. We have observed large tensions in the previously well determined valence quark PDFs and, even in the strange quark PDF determination, tension among the neutrino data is visible. Therefore, the first important conclusion of this analysis is that, due to the large tensions, the bulk of neutrino data is considered incompatible with the data of the baseline analysis or even among each other. 

In an effort to recover at least a subset of neutrino data to be used in a global analysis, we have proposed three strategies to alleviate the tensions between the neutrino DIS data and all the data in the reference analysis. We have analyzed the possibility of neglecting the correlations in the systematic errors of the NuTeV experiment, which are responsible for a substantial part of the tensions in the neutrino data itself. This yielded a much better description of the neutrino data, but the tensions with the original data of the nCTEQ15WZSIHdeut analysis remained. 

Since the neutrino data introducing tension at high Bjorken x had already been removed by initial global kinematic cuts, the next possibility we investigated was the introduction of an arbitrary kinematic cut to remove the remaining problematic neutrino data in the region of low Bjorken x (BaseDimuNeuX).   As expected after the removal of this data, which causes most of the remaining tension, the description of all the data improved and this can in principle be considered a way to go. However, for this possibility to be viable, a reason for introducing such a cut has to be provided. It was hypothesised (e.g., in \cite{Kopeliovich:2012kw}) that shadowing in neutrino scattering on nuclei works differently than in the neutral-current DIS, which is the cornerstone of the reference PDF analysis. If this were indeed the case, one would have to modify the theoretical predictions for neutrino scattering. Alternatively, before doing so a cut might be introduced to remove data which do not have a proper theoretical description. In such a case, the results of the BaseDimuNeuX analysis might be considered the final result of this study. 

Given, however, that an alternative mechanism for shadowing in neutrino-nuclei interactions is not yet completely established and is merely hypothesised, we have put forward a different final result of our compatibility analysis. We have identified a subset of neutrino data which has no tension with the data in the reference analysis and so it can be safely included in a combined global analysis. Unfortunately, the majority of neutrino DIS cross-section data have been left unused in the process. Including just the scattering data from Chorus and the di-muon data from NuTeV and CCFR, we have performed the analysis called BaseDimuChorus. The result of including the new data is a much improved description of the strange quark PDF.

Even though we have found a way to include some neutrino data in our analysis in order to improve the determination of the strange quark PDF, the fact that the bulk of the DIS neutrino cross-section data is incompatible with the neutral current DIS data is established. Without new experimental data on neutrino-nucleus interactions in the DIS regime, there is no way to decide if this inconsistency is due to a different mechanism for the neutrino-nucleus interaction or simply a sign of problems in the acquisition of the current neutrino experimental data. The resolution could have come from the high-statistics NOMAD experiment but even after more than 20 years only the results of the di-muon analysis were publicly released so far. Unfortunately, after plans for a new neutrino scattering experiment were not followed-up on \cite{NuSOnG:2009rcm}, no new high-energy neutrino scattering experiment is currently in planning. Nevertheless, there is potential to obtain new crucial data from novel ideas or experiments such as the proposed Forward Physics Facility \cite{Anchordoqui:2021ghd} at the LHC or from precise measurements of charged current DIS processes at the future Electron-Ion-Collider \cite{Accardi:2012qut,AbdulKhalek:2021gbh}.

\section*{Acknowlegments}
We are pleased to thank Un-ki Yang for providing us the CCFR differential cross section data. We are also grateful to Alberto Accardi, Chlo\'e L\'eger, and Peter Risse for useful discussions.
The work of P.D., T.J., M.K. and K.K. was funded by the Deutsche Forschungsgemeinschaft (DFG, German Research Foundation) – project-id 273811115 – SFB 1225. P.D., T.J., K.F.M., M.K. and K.K. also acknowledge support of the DFG through the Research Training Group GRK 2149.
\\
This manuscript has been authored by Fermi Research Alliance, LLC under Contract No.~DEAC02-07CH11359 with the U.S.~Department of Energy, Office of Science, Office of High Energy Physics.
\\
F.O. was supported by the U.S. Department of Energy Grant No. DE-SC0010129.
\\
A.K. and R.R. acknowledge the support of Narodowe Centrum Nauki under Sonata Bis Grant No. 2019/34/E/ST2/00186.
\\
R.R. acknowledges the support of the Polska Akademia Nauk (grant agreement 
PAN.BFD.S.BDN. 613. 022. 2021 - PASIFIC 1, POPSICLE). This work has 
received funding from the European Union's Horizon 2020 research and 
innovation program under the Sk{\l}odowska-Curie grant agreement No. 
847639 and from the Polish Ministry of Education and Science. 
\\
The work of I. S. was supported in part by the French National Centre for
Scientific Research CNRS through IN2P3 Project GLUE@NLO.

\clearpage

\appendix
\section{Results of all fits}\label{sec:fitresults}
In this appendix we collect the values of the PDF parameters obtained in all the fits presented in the paper. The parameters are collected in Tabs.~\ref{tab:uv}-\ref{tab:ubardbar}. For ease of comparison each table contains values for different flavour or flavour combination. The values indicated in bold were allowed to change in the fitting procedure the normal font indicates that the values were fixed.
The parameters for the remaining flavour combinations: $\bar{d}/\bar{u}$ and $s-\bar{s}$ were not changed compared to our previous analyses and are correspondingly given in Tab.~V of ref.~\cite{Kovarik:2015cma} or in case of strange asymmetry they are all zero as we used symmetric strange.
\begin{table*}[h!]
	\caption{Values of all parameters of the up-quark valence distribution $u_v$ in all fits quoted here. Values in bold belong to free parameters which were allowed to vary in the corresponding analysis.}	\centering
	\begin{tabular}{|l|l|l|l|l|l|l|l|l|l|l|}
	\hline
		Analysis  & $a_1^{u_v}$ & $a_2^{u_v}$ & $a_3^{u_v}$ & $a_4^{u_v}$ & $a_5^{u_v}$  & $b_1^{u_v}$ & $b_2^{u_v}$ & $b_3^{u_v}$ & $b_4^{u_v}$ & $b_5^{u_v}$\\ \hline 
		nCTEQ15WZSIHdeut  & \textbf{-4.568} &\textbf{0.059}& 0.018 & \textbf{16.265} &\textbf{-1.028} & \textbf{0.006} & \textbf{0.524} & 0.073 & 0.038& 0.615  \\ 
		DimuNeu & \textbf{-4.373} &\textbf{2.039}& 0.018 & \textbf{13.802} &\textbf{ -1.044} & \textbf{0.0052} & \textbf{-0.025} & 0.073 & 0.038& 0.615   \\ 
		BaseDimuNeu & \textbf{-4.811} &\textbf{ 0.031}& 0.018 & \textbf{ 14.386} &\textbf{-1.035} & \textbf{0.006} & \textbf{-0.187} & 0.073 & 0.038& 0.615  \\ 
		BaseDimuNeuUncorr & \textbf{-4.809} &\textbf{0.072}& 0.018 & \textbf{14.381} &\textbf{-1.035} & \textbf{0.006} & \textbf{-0.221} & 0.073 & 0.038& 0.615  \\ 
		BaseDimuNeuX & \textbf{-4.501} &\textbf{ 0.088}& 0.018 & \textbf{ 15.290} &\textbf{-1.031} & \textbf{0.006} & \textbf{-0.205} & 0.073 & 0.038& 0.615  \\ 
		BaseDimuChorus & \textbf{-4.622} &\textbf{0.054}& 0.018 & \textbf{16.209} &\textbf{-1.031} & \textbf{0.006} & \textbf{0.524} & 0.073 & 0.038& 0.615  \\ 
		\hline
	\end{tabular}
	\label{tab:uv}
\end{table*}

\begin{table*}[h!]
	\caption{Values of all parameters of the down-quark valence distribution $d_v$ in all fits quoted here. Values in bold belong to free parameters which were allowed to vary in the corresponding analysis.}	\centering
\begin{tabular}{|l|l|l|l|l|l|l|l|l|l|l|}
	\hline
		Analysis  & $a_1^{d_v}$ & $a_2^{d_v}$ & $a_3^{d_v}$ & $a_4^{d_v}$ & $a_5^{d_v}$  & $b_1^{d_v}$ & $b_2^{d_v}$ & $b_3^{d_v}$ & $b_4^{d_v}$ & $b_5^{d_v}$\\ \hline 
		nCTEQ15WZSIHdeut  & \textbf{0.086} & \textbf{-0.064 }& 0.085 & \textbf{3.874} &\textbf{-0.023} & \textbf{0.466 }& \textbf{0.44} & 0.107 & -0.018 & -0.236 \\ 
		DimuNeu & \textbf{-0.116} & \textbf{-1.012}& 0.085 & \textbf{ 4.164} &\textbf{ 0.224} & \textbf{0.1}& \textbf{1.109} & 0.107 & -0.018 & -0.236 \\ 
		BaseDimuNeu & \textbf{0.091} & \textbf{-0.957 }& 0.085 & \textbf{ 3.794} &\textbf{-0.081} & \textbf{0.044 }& \textbf{ 0.690} & 0.107 & -0.018 & -0.236 \\  
		BaseDimuNeuUncorr & \textbf{ 0.087} & \textbf{-0.947 }& 0.085 & \textbf{ 3.801} &\textbf{0.071} & \textbf{ 0.081 }& \textbf{0.648} & 0.107 & -0.018 & -0.236 \\  
		BaseDimuNeuX & \textbf{0.137} & \textbf{-0.958 }& 0.085 & \textbf{4.854} &\textbf{0.072} & \textbf{0.067 }& \textbf{0.528} & 0.107 & -0.018 & -0.236 \\ 
		BaseDimuChorus & \textbf{0.083} & \textbf{-0.065 }& 0.085 & \textbf{3.917} &\textbf{-0.020} & \textbf{0.466 }& \textbf{0.44} & 0.107 & -0.018 & -0.236 \\ 
		\hline
	\end{tabular}
	\label{tab:dv}
\end{table*}

\begin{table*}[h!]
	\caption{Values of all parameters of the gluon distribution $g$ in all fits quoted here. Values in bold belong to free parameters which were allowed to vary in the corresponding analysis.}	\centering
	\begin{tabular}{|l|l|l|l|l|l|l|l|l|l|l|l|l|}
	\hline
		Analysis & $a_0^{g}$ & $a_1^{g}$ & $a_2^{g}$ & $a_3^{g}$ & $a_4^{g}$ & $a_5^{g}$ & $b_0^{g}$ & $b_1^{g}$ & $b_2^{g}$ & $b_3^{g}$ & $b_4^{g}$ & $b_5^{g}$\\ \hline
		nCTEQ15WZSIHdeut & -0.256 & \textbf{-0.0025} & 0.0 & 0.383 & \textbf{0.059} & \textbf{0.0029} & \textbf{-0.036} & \textbf{-1.073} & 0.0 & 0.52 &\textbf{-0.364}&  \textbf{-1.293}\\
		DimuNeu  & -0.256 & \textbf{-0.00256} & 0.0 & 0.383 & \textbf{0.059} & \textbf{0.0029} & \textbf{-0.036} & \textbf{-1.0738} & 0.0 & 0.52 &\textbf{-0.364}&  \textbf{-1.293}\\
		BaseDimuNeu  & -0.256 & \textbf{-0.0025} & 0.0 & 0.383 & \textbf{0.057} & \textbf{0.0029} & \textbf{-0.004} & \textbf{-1.070} & 0.0 & 0.52 &\textbf{-0.348}&  \textbf{-1.288}\\
		BaseDimuNeuUncorr  & -0.256 & \textbf{-0.0026} & 0.0 & 0.383 & \textbf{0.042} & \textbf{0.0028} & \textbf{ -0.008} & \textbf{-1.058} & 0.0 & 0.52 &\textbf{-0.340}&  \textbf{-1.266}\\
	    BaseDimuNeuX  & -0.256 & \textbf{-0.0026} & 0.0 & 0.383 & \textbf{0.050} & \textbf{0.003} & \textbf{-0.009} & \textbf{-1.075} & 0.0 & 0.52 &\textbf{-0.359}&  \textbf{-1.290}\\
	    BaseDimuChorus  & -0.256 & \textbf{-0.0025} & 0.0 & 0.383 & \textbf{0.059} & \textbf{0.0029} & \textbf{-0.036} & \textbf{-1.073} & 0.0 & 0.52 &\textbf{-0.363}&  \textbf{-1.292}\\
		\hline
	\end{tabular}
	\label{tab:g}
\end{table*}

\begin{table*}[h!]
	\caption{Values of all parameters of the $s+\bar{s}$ distribution in all fits quoted here. Values in bold belong to free parameters which were allowed to vary in the corresponding analysis.}	\centering
	\begin{tabular}{|l|l|l|l|l|l|l|l|l|l|l|l|l|}
	\hline
		Analysis & $a_0^{s+\bar{s}}$ & $a_1^{s+\bar{s}}$ & $a_2^{s+\bar{s}}$ & $a_3^{s+\bar{s}}$ & $a_4^{s+\bar{s}}$ & $a_5^{s+\bar{s}}$ & $b_0^{s+\bar{s}}$ & $b_1^{s+\bar{s}}$ & $b_2^{s+\bar{s}}$ & $b_3^{s+\bar{s}}$ & $b_4^{s+\bar{s}}$ & $b_5^{s+\bar{s}}$\\ \hline
		nCTEQ15WZSIHdeut & \textbf{0.152} &\textbf{0.1639} & \textbf{6.82} & 0.0 & 0.0 & 0.0 & \textbf{0.104} & \textbf{0.109} & 0.290 & 0.0 & 0.0 & 0.0\\
		DimuNeu & \textbf{2.289} &\textbf{0.555} & \textbf{ 4.710} & 0.0 & 0.0 & 0.0 & \textbf{-1.876} & \textbf{0.493} & 0.290 & 0.0 & 0.0 & 0.0\\
		BaseDimuNeu & \textbf{0.510} &\textbf{-0.183} & \textbf{3.466} & 0.0 & 0.0 & 0.0 & \textbf{0.246} & \textbf{0.206} & 0.290 & 0.0 & 0.0 & 0.0\\
		BaseDimuNeuUncorr & \textbf{ 0.480} &\textbf{-0.173} & \textbf{3.395} & 0.0 & 0.0 & 0.0 & \textbf{ 0.221} & \textbf{ 0.188} & 0.290 & 0.0 & 0.0 & 0.0\\
		BaseDimuNeuX & \textbf{  0.405} &\textbf{ -0.413} & \textbf{1.482} & 0.0 & 0.0 & 0.0 & \textbf{ 0.274} & \textbf{0.159} & 0.290 & 0.0 & 0.0 & 0.0\\
		BaseDimuChorus & \textbf{0.194} &\textbf{0.217} & \textbf{6.012} & 0.0 & 0.0 & 0.0 & \textbf{0.158} & \textbf{0.099} & 0.290 & 0.0 & 0.0 & 0.0\\
		\hline
	\end{tabular}
	\label{tab:ssbar}
\end{table*}

\begin{table*}[h!]
	\caption{Values of all parameters of the $\bar{u}+\bar{d}$ distribution in all fits quoted here. Values in bold belong to free parameters which were allowed to vary in the corresponding analysis.}	\centering
	\begin{tabular}{|l|l|l|l|l|l|l|l|l|l|l|}
	\hline
		Analysis & $a_1^{\bar{u}+\bar{d}}$ & $a_2^{\bar{u}+\bar{d}}$ & $a_3^{\bar{u}+\bar{d}}$ & $a_4^{\bar{u}+\bar{d}}$ & $a_5^{\bar{u}+\bar{d}}$ & $b_1^{\bar{u}+\bar{d}}$ & $b_2^{\bar{u}+\bar{d}}$ & $b_3^{\bar{u}+\bar{d}}$ & $b_4^{\bar{u}+\bar{d}}$ & $b_5^{\bar{u}+\bar{d}}$\\ \hline
		nCTEQ15WZSIHdeut & \textbf{0.471} & \textbf{0.435} &  -0.759 & -0.203 & \textbf{-0.105 }& 0.172 & 0.290 & 0.298 &0.888  & 1.35312 \\
		DimuNeu& \textbf{0.961} & \textbf{-1.485} &  -0.759 & -0.203 & \textbf{-0.629 }& 0.172 & 0.290 & 0.298 &0.888  & 1.35312 \\
		BaseDimuNeu & \textbf{0.519} & \textbf{ -0.163} &  -0.759 & -0.203 & \textbf{-0.144 }& 0.172 & 0.290 & 0.298 &0.888  & 1.35312 \\
		BaseDimuNeuUncorr & \textbf{0.491} & \textbf{-0.173} &  -0.759 & -0.203 & \textbf{-0.138}& 0.172 & 0.290 & 0.298 &0.888  & 1.35312 \\
		BaseDimuNeuX & \textbf{  0.580} & \textbf{ 0.832} &  -0.759 & -0.203 & \textbf{-0.083}& 0.172 & 0.290 & 0.298 &0.888  & 1.35312 \\
		BaseDimuChorus & \textbf{0.475} & \textbf{0.509} &  -0.759 & -0.203 & \textbf{-0.108 }& 0.172 & 0.290 & 0.298 &0.888  & 1.35312 \\
		\hline
	\end{tabular}
	\label{tab:ubardbar}
\end{table*}
\section{Treatment of Normalization Uncertainties}\label{sec:app_norm_unc}
The normalization uncertainty is a scale uncertainty that affects both the central data and its uncertainties. The conventional way which is still often used to include the normalization uncertainty in a $\chi^2$ fitting procedure is by constructing a covariance matrix in the following way:
\begin{align}\label{cij}
  C_{D, ij} &= C_{ij} + \sigma_{norm}^2 D_iD_j \\
  C_{ij} &= \sigma_i^2 \delta_{ij}+ \sum_\alpha \bar{\sigma}_{i\alpha}\bar{\sigma}_{j\alpha}
\end{align}
where $D_i$ is the $i$-th data point, $\sigma_i$, $\bar{\sigma}_{i\alpha}$ and $\sigma_{norm}$ are the statistical uncertainty, systematic uncertainty from $\alpha$-th source, and the normalization uncertainty. Using Eq.~\eqref{cij} during $\chi^2$ fitting can lead to d' Agostini bias~\cite{DAgostini:1993arp} which causes the fitted theory to be much lower than expected. Furthermore, the bias becomes worse as the number of data points increases~\cite{DAgostini:1993arp,Stump:2001gu}. 

Using Sherman-Morrison formula~\cite{10.1214/aoms/1177729893} to write the inverse of $C_{D}$ : 
\begin{equation}
C_D^{-1} = C^{-1}- \frac{\sigma_{norm}^2 C^{-1} D D^T C^{-1}}{1+ \sigma_{norm}^2 D^T C^{-1} D},
\end{equation}
it is straightforward to prove that using the covariance matrix (\ref{cij}) is equivalent to using the following $\chi^2$ function : 
\begin{equation}\label{chi2d}
\chi^2_D (a, r) = (rD-T(a))^T C^{-1} (rD-T(a)) + \frac{(1-r)^2}{\sigma_{norm}^2}.
\end{equation}
Here, $T(a)$ is the theory prediction and both the theory parameters $a$ and the normalization one $r$ are to be fitted to the data. The equivalence means that 
\begin{equation}
    \min_r \chi^2_D (a, r)  = (D-T)^T C_D^{-1} (D-T)
\end{equation}
Hence, using (\ref{chi2d}) will also lead to same d' Agostini bias.

\begin{figure}[htb]   
	\centering
	\includegraphics[width=0.45\textwidth]{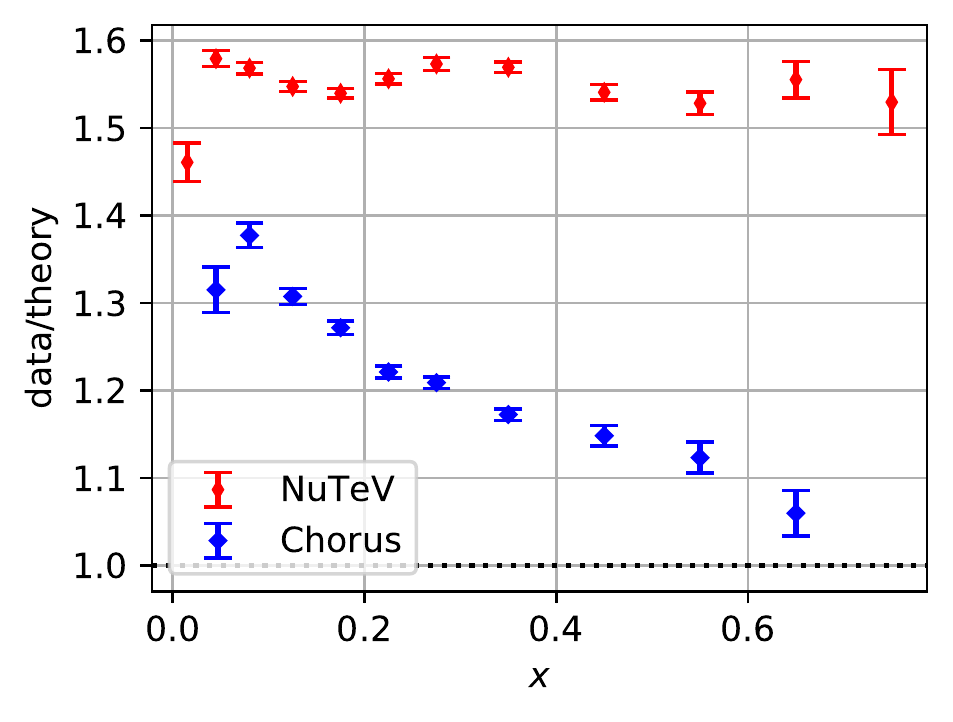}
	\includegraphics[width=0.45\textwidth]{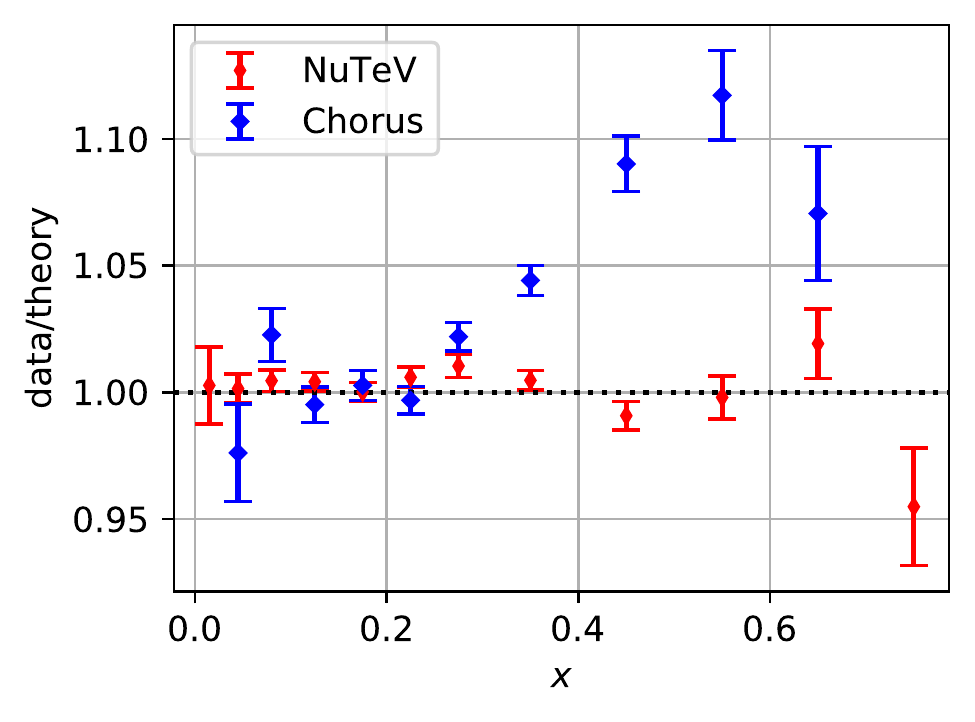}
	\caption{The weighted average of the data/theory from fits with  NuTeV and Chorus data where the normalization uncertainties are treating using (\ref{chi2d}) (top panel) and using the method adopted in this work (\ref{chi2a}) (bottom panel). }
	\label{Rdagshift}
\end{figure}

To illustrate how the bias could really affect fits with high statistic neutrino data such as NuTeV and Chorus, we have performed fits using Eq.\eqref{chi2d} with the individual NuTeV and Chorus data. During the fit, we open 12 parameters and fixing the gluon parameters to the same values as in nCTEQ15 analysis\cite{Kovarik:2015cma}. We obain $\chi^2/N = 0.86$  and $\chi^2/N = 0.95$ for the NuTeV and Chorus fit respectively. 
We plot the weighted average of data/theory in top panel of Fig.~\ref{Rdagshift}. The figure shows that the theory is severely below the data. Even though the normalisation uncertainties in both experiments are the same ($2.1\%)$, the bias in the NuTeV fit is more severe than in the Chorus fit. The reason for this is the much larger number of data points in NuTeV (2136 points) than in Chorus (824 points). 

To avoid d' Agostini bias, several prescriptions exist in the literature. The first method is to use the following $\chi^2$ function~\cite{DAgostini:1993arp} : 
\begin{equation}
\label{dag}
\chi^2_{1/r}( a, r) = \sum_{i, j} \left(D_i-\frac{T_i}{r}\right)C_{ij}^{-1}\left(D_j-\frac{T_j}{r}\right)+\left(\frac{1-r}{\sigma_{norm}}\right)^2 \; .
\end{equation}
This method requires to fit the normalization fluctuation, $r$, directly to the data. The main drawback of this approach is that the number of normalization parameters can become large, and in case there are many data sets in the global fit, even comparable to the number of PDF parameters. This causes the fit to be prone to numerical problems, such as saddle point or local minimum trap. The larger number of parameters also means the computing cost will increase. 

It is worth mentioning that as the normalization fluctuation parameters are basically nuisance parameters fitted to the data, thus their uncertainties must be taken into account when estimating the uncertainties of the fitted PDFs. An easy way for an error estimation with nuisance parameters by freezing them to the minimum point of the $\chi^2$ will result in an underestimation of the true uncertainty. A consistent way to include the uncertainty of the nuisance parameters is given by profile likelihood method. For a $\chi^2(a_\mu, r_i)$ function, where $a_\mu, \, \mu =1,..., N$ denotes the parameters of interest (PDF parameters) and $r_i, \, i=1,..., M$ are the nuisance parameters, one defines the 'profile` $\chi^2$ function as
\begin{equation}
    \chi^2_p (a):= \min_{r} \chi^2( a, r)
\end{equation}
which is a function of PDF parameters only. The Hessian-based error PDF determination can be done using this profile $\chi^2_p$. However, the computation of $\chi^2_p$ is expensive as there is no closed-form solution for $\chi^2_p$, hence this method is impractical. An alternative, but equivalent, method is to use the full $\chi^2(a, r)$, but in the Hessian error estimation, the inverse of the $N\times N$ effective Hessian matrix is given by $N\times N$-submatrix of the inverse of the full $(N+M)\times (N+M)$ Hessian matrix~\cite{cox_2006}. To prove this, let $H^p_{\mu\nu}$ be the second derivative of $\chi^2_p(a)$ with respect to the theory parameters $a_\mu$ and $a_\nu$, where $\mu, \nu = 1,..., N$. Let $H_{\mu i }, H_{\mu\nu}$, and $H_{ij}$ be the second deriative of $\chi^2$ with respect to $a_\mu$ and $r_i$, $a_\mu$ and $r_i$ and $r_j$. Here, $i=1, ..., M$. By implicit differentiation, the Hessian $H^p_{\mu\nu}$ can be written as 
\begin{equation}\label{hessp}
   H^p_{\mu\nu} = H_{\mu\nu}+ H_{\mu i} \frac{\partial \hat{r}}{\partial a_\nu}
\end{equation}
where $\hat{r}(a) = \arg \min_r \chi^2(a, r)$. The derivative $\partial \hat{r}/\partial a_\nu$ evaluated at any $a$ is hard to be calculated as the explicit function $\hat{r}(a)$ is unknown. However, for $a=\hat{a} = \arg \min_a \chi^2_p(a)$, we can express the derivative as 
\begin{eqnarray}
 \frac{\partial\hat{r_i}(\hat{a})}{\partial a_\nu } = -{{H_r}^{-1}}_{ij} H_{j\nu}
\end{eqnarray}
where $H_r$ is an $M\times M$ matrix whose components are the same as $H_{ij}$. Note that all the Hessian matrices on the RHS are evaluated at the minimum $\hat{a}$. Inserting this to (\ref{hessp}), we obtain 
\begin{equation}\label{hessp2}
   H^p_{\mu\nu} = H_{\mu\nu}- H_{\mu i} {{H_r}^{-1}}_{ij} H_{i\nu} 
\end{equation}
For any block matrix : 
\begin{equation}
    P=\begin{pmatrix} A & B\\ C& D\end{pmatrix}
\end{equation}
with $A, B, C, D$ are $N\times N$, $N\times M$, $M\times N$, and $M\times M$ matrices, the first (upper left) $N\times N$ component of $P^{-1}$ is given by $(A-BD^{-1}C)^{-1}$. Therefore, one immediately see that 
\begin{equation}
    {H^p}^{-1} = \left . H^{-1}\right|_{N\times N},
\end{equation}
as stated before.


An alternative method to include normalization uncertainties in a global fit is to use $t_0$-method as explained in detail in \cite{Ball:2009qv}. This method basically set the covariance matrix: 
\begin{equation}\label{cijt0}
  C_{t_0, ij} = C_{ij}+ \sigma_{norm}^2 {T_0}_i{T_0}_j 
\end{equation}
where ${T_0}_i$ is the theory prediction from previous iteration of the fit and $C_{ij}$ is the original covariance matrix without normalization uncertainties.  This method eliminates the nuisance parameters from the $\chi^2$ function and hence their uncertainties are automatically included. As the normalization is eliminated, it is not clear how one can obtain the estimated normalization parameters. Knowing the estimated normalization is important for data-theory plotting purpose and for sanity check if its value is close to unity.

In this work, in order to treat normalization uncertainties, we adopt the following prescription 
\begin{equation}
    \chi^2_r(a, r) = \sum_{i, j} (D_i-rT_i) C^{-1}_{ij}(D_j-rT_j)+ \frac{(1-r)^2}{\sigma_{norm}^2} \; .
    \label{chi2a}
\end{equation}
We will see that this method is equivalent to the $t$-method discussed in~\cite{Ball:2009qv}. (\ref{chi2a}) can be rewritten as 
\begin{equation}
    \chi^2_r(a, r)  =\frac{1}{\sigma_{norm}^2}  \left[A \left(r-\frac{B}{A}\right)^2 + \left(E-\frac{B^2}{A}\right)\right]
\end{equation}
where 
\begin{align}
    A = 1+\sigma_{norm}^2 T^TC^{-1} T\\
    B = 1+\sigma_{norm}^2 D^TC^{-1} T\\
    E = 1+ \sigma_{norm}^2 D^TC^{-1} D
\end{align}
It is clear now that the fitted normalization is given by : 
\begin{equation}\label{rhat}
    \hat{r}(a) = \arg \min_{r} \chi^2_r(a, r) = \frac{B}{A}
\end{equation}
Furthermore, the $\chi^2$ at $\hat{r}$ is given by : 
\begin{align}
    \chi^2_T(a) &\equiv \min_{r} \chi^2_r(a, r) = \frac{1}{\sigma_{norm}^2} \left(E-\frac{B^2}{A}\right)\nonumber \\
    & = (D-T)^T C_T^{-1} (D-T)
\end{align}
where
\begin{equation}
  C_{T, ij}(a) = C_{ij}+  \sigma_{norm}^2 T_i(a) T_j(a)
\end{equation}
and we have used the following formula for the inverse of $C_T$ :
\begin{equation}
    C_T^{-1} = C^{-1 }- \frac{\sigma_{norm}^2 C^{-1} T T^T C^{-1}}{1+\sigma_{norm}^2 T^TC^{-1} T}
\end{equation}
This equation follows from Sherman-Morrison formula~\cite{10.1214/aoms/1177729893}. Thus, the fitting normalization uncertainty in this way is equivalent to using an effective covariance matrix $C_T$. The advantage of using this approach is that the nuisance parameters are now completely eliminated and the Hessian errors automatically take into account the uncertainty of the nuisance parameters into the estimation of error PDFs. As the difference between formula~\eqref{chi2a} and~\eqref{dag} essentially comes from the penalty term, then this method is equivalent to~\eqref{dag} if the optimal normalization parameter $r$ is not far from unity, which is usually the case.

It is trivial to generalize this method to a case where there are more than one data sets that share the same normalization. In such case, the fitted normalization formula (\ref{rhat}) still hold, but $A, B$ and $C$ are modified as 
\begin{align}
     A = 1+\sum_s \sigma_{norm}^2 {T^{s}}^TC^{-1}_s T^s\\
    B = 1+\sum_s \sigma_{norm}^2 {D^{s}}^TC^{-1}_s T^s\\
    E = 1+ \sum_s \sigma_{norm}^2 {D^s}^TC^{-1}_s D^s
\end{align}
where $s$ denotes the data set $s$ and the sum is done over all data sets that share the same normalization. 

In order to contrast the fit results obtained with formula~\eqref{chi2d} leading to the d'Agostini bias, we performed analogical fits using formula~\eqref{chi2a}. In the bottom panel of Fig.~\ref{Rdagshift}, we show the weighted average of the data/theory for fits with the individual NuTeV and Chorus data, where now the (\ref{chi2a}) is used. We obtain $\chi^2/N = 1.36$ and $\chi^2/N=1.07$ for the NuTeV and Chorus fits respectively.
We can see that for both NuTeV and Chorus fits the ratio becomes much closer to unity, as one could expected having in mind that the normalization uncertainty for these data is $\sim2\%$. The relatively high data/theory values for the Chorus fit at $x>0.4$ is related to large systematic uncertainties (hence large systematic theory shifts).
Comparing the upper and lower panels of Fig.~\ref{Rdagshift} the difference in the results of the NuTeV fit is especially striking. It also confirms that d'Agostini bias is getting larger with the number of data points.

\bibliographystyle{utphys}
\bibliography{refs,extra}

\providecommand{\href}[2]{#2}\begingroup\raggedright\begin{thebibliography}{10}

\bibitem{Hou:2019efy}
T.-J. Hou {\em et~al.}, ``{New CTEQ global analysis of quantum chromodynamics
  with high-precision data from the LHC},''
\href{http://arxiv.org/abs/1912.10053}{{\ttfamily arXiv:1912.10053 [hep-ph]}}.

\bibitem{Accardi:2016qay}
A.~Accardi, L.~T. Brady, W.~Melnitchouk, J.~F. Owens, and N.~Sato,
  ``{Constraints on large-$x$ parton distributions from new weak boson
  production and deep-inelastic scattering data},''
  \href{http://dx.doi.org/10.1103/PhysRevD.93.114017}{{\em Phys. Rev. D}
  {\bfseries 93} no.~11, (2016) 114017},
  \href{http://arxiv.org/abs/1602.03154}{{\ttfamily arXiv:1602.03154
  [hep-ph]}}.

\bibitem{Bailey:2020ooq}
S.~Bailey, T.~Cridge, L.~A. Harland-Lang, A.~D. Martin, and R.~S. Thorne,
  ``{Parton distributions from LHC, HERA, Tevatron and fixed target data:
  MSHT20 PDFs},'' \href{http://dx.doi.org/10.1140/epjc/s10052-021-09057-0}{{\em
  Eur. Phys. J. C} {\bfseries 81} no.~4, (2021) 341},
  \href{http://arxiv.org/abs/2012.04684}{{\ttfamily arXiv:2012.04684
  [hep-ph]}}.

\bibitem{Abramowicz:2015mha}
{\bfseries H1, ZEUS} Collaboration, H.~Abramowicz {\em et~al.}, ``{Combination
  of measurements of inclusive deep inelastic ${e^{\pm }p}$ scattering cross
  sections and QCD analysis of HERA data},''
  \href{http://dx.doi.org/10.1140/epjc/s10052-015-3710-4}{{\em Eur. Phys. J. C}
  {\bfseries 75} no.~12, (2015) 580},
  \href{http://arxiv.org/abs/1506.06042}{{\ttfamily arXiv:1506.06042
  [hep-ex]}}.

\bibitem{Ball:2017nwa}
{\bfseries NNPDF} Collaboration, R.~D. Ball {\em et~al.}, ``{Parton
  distributions from high-precision collider data},''
  \href{http://dx.doi.org/10.1140/epjc/s10052-017-5199-5}{{\em Eur. Phys. J.}
  {\bfseries C77} no.~10, (2017) 663},
\href{http://arxiv.org/abs/1706.00428}{{\ttfamily arXiv:1706.00428 [hep-ph]}}.

\bibitem{Alekhin:2017kpj}
S.~Alekhin, J.~Bl\"umlein, S.~Moch, and R.~Placakyte, ``{Parton distribution
  functions, $\alpha_s$, and heavy-quark masses for LHC Run II},''
  \href{http://dx.doi.org/10.1103/PhysRevD.96.014011}{{\em Phys. Rev. D}
  {\bfseries 96} no.~1, (2017) 014011},
  \href{http://arxiv.org/abs/1701.05838}{{\ttfamily arXiv:1701.05838
  [hep-ph]}}.

\bibitem{deFlorian:2011fp}
D.~de~Florian, R.~Sassot, P.~Zurita, and M.~Stratmann, ``{Global Analysis of
  Nuclear Parton Distributions},''
  \href{http://dx.doi.org/10.1103/PhysRevD.85.074028}{{\em Phys. Rev.}
  {\bfseries D85} (2012) 074028},
\href{http://arxiv.org/abs/1112.6324}{{\ttfamily arXiv:1112.6324 [hep-ph]}}.

\bibitem{Kovarik:2015cma}
K.~Kova\v{r}\'{i}k {\em et~al.}, ``{nCTEQ15 - Global analysis of nuclear parton
  distributions with uncertainties in the CTEQ framework},''
  \href{http://dx.doi.org/10.1103/PhysRevD.93.085037}{{\em Phys. Rev.}
  {\bfseries D93} no.~8, (2016) 085037},
\href{http://arxiv.org/abs/1509.00792}{{\ttfamily arXiv:1509.00792 [hep-ph]}}.

\bibitem{Kusina:2016fxy}
A.~Kusina, F.~Lyonnet, D.~B. Clark, E.~Godat, T.~Jezo, K.~Kova\v{r}\'{i}k,
  F.~I. Olness, I.~Schienbein, and J.~Y. Yu, ``{Vector boson production in pPb
  and PbPb collisions at the LHC and its impact on nCTEQ15 PDFs},''
  \href{http://dx.doi.org/10.1140/epjc/s10052-017-5036-x}{{\em Eur. Phys. J.}
  {\bfseries C77} no.~7, (2017) 488},
\href{http://arxiv.org/abs/1610.02925}{{\ttfamily arXiv:1610.02925 [nucl-th]}}.

\bibitem{Kusina:2020lyz}
A.~Kusina {\em et~al.}, ``{Impact of LHC vector boson production in heavy ion
  collisions on strange PDFs},''
  \href{http://dx.doi.org/10.1140/epjc/s10052-020-08532-4}{{\em Eur. Phys. J.
  C} {\bfseries 80} no.~10, (2020) 968},
  \href{http://arxiv.org/abs/2007.09100}{{\ttfamily arXiv:2007.09100
  [hep-ph]}}.

\bibitem{Duwentaster:2021ioo}
P.~Duwent\"aster, L.~A. Husov\'a, T.~Je\v{z}o, M.~Klasen, K.~Kova\v{r}\'\i{}k,
  A.~Kusina, K.~F. Muzakka, F.~I. Olness, I.~Schienbein, and J.~Y. Yu,
  ``{Impact of inclusive hadron production data on nuclear gluon PDFs},''
  \href{http://arxiv.org/abs/2105.09873}{{\ttfamily arXiv:2105.09873
  [hep-ph]}}.

\bibitem{Eskola:2016oht}
K.~J. Eskola, P.~Paakkinen, H.~Paukkunen, and C.~A. Salgado, ``{EPPS16: Nuclear
  parton distributions with LHC data},''
  \href{http://dx.doi.org/10.1140/epjc/s10052-017-4725-9}{{\em Eur. Phys. J.}
  {\bfseries C77} no.~3, (2017) 163},
\href{http://arxiv.org/abs/1612.05741}{{\ttfamily arXiv:1612.05741 [hep-ph]}}.

\bibitem{AbdulKhalek:2020yuc}
R.~Abdul~Khalek, J.~J. Ethier, J.~Rojo, and G.~van Weelden, ``{nNNPDF2.0: Quark
  Flavor Separation in Nuclei from LHC Data},''
  \href{http://arxiv.org/abs/2006.14629}{{\ttfamily arXiv:2006.14629
  [hep-ph]}}.

\bibitem{Walt:2019slu}
M.~Walt, I.~Helenius, and W.~Vogelsang, ``{Open-source QCD analysis of nuclear
  parton distribution functions at NLO and NNLO},''
  \href{http://dx.doi.org/10.1103/PhysRevD.100.096015}{{\em Phys. Rev.}
  {\bfseries D100} no.~9, (2019) 096015},
\href{http://arxiv.org/abs/1908.03355}{{\ttfamily arXiv:1908.03355 [hep-ph]}}.

\bibitem{Khanpour:2020zyu}
H.~Khanpour, M.~Soleymaninia, S.~Atashbar~Tehrani, H.~Spiesberger, and
  V.~Guzey, ``{Nuclear parton distribution functions with uncertainties in a
  general mass variable flavor number scheme},''
  \href{http://dx.doi.org/10.1103/PhysRevD.104.034010}{{\em Phys. Rev. D}
  {\bfseries 104} no.~3, (2021) 034010},
  \href{http://arxiv.org/abs/2010.00555}{{\ttfamily arXiv:2010.00555
  [hep-ph]}}.

\bibitem{Kovarik:2019xvh}
K.~Kova\v{r}\'{i}k, P.~M. Nadolsky, and D.~E. Soper, ``{Hadron structure in
  high-energy collisions},'' \href{http://arxiv.org/abs/1905.06957}{{\ttfamily
  arXiv:1905.06957 [hep-ph]}}.

\bibitem{Ethier:2020way}
J.~J. Ethier and E.~R. Nocera, ``{Parton Distributions in Nucleons and
  Nuclei},'' \href{http://dx.doi.org/10.1146/annurev-nucl-011720-042725}{{\em
  Ann. Rev. Nucl. Part. Sci.} no.~70, (2020) 1--34},
\href{http://arxiv.org/abs/2001.07722}{{\ttfamily arXiv:2001.07722 [hep-ph]}}.

\bibitem{Zyla:2020zbs}
{\bfseries Particle Data Group} Collaboration, P.~A. Zyla {\em et~al.},
  ``{Review of Particle Physics},''
  \href{http://dx.doi.org/10.1093/ptep/ptaa104}{{\em PTEP} {\bfseries 2020}
  no.~8, (2020) 083C01}.

\bibitem{Goncharov:2001qe}
{\bfseries NuTeV} Collaboration, M.~Goncharov {\em et~al.}, ``{Precise
  Measurement of Dimuon Production Cross-Sections in $\nu_{\mu}$ Fe and
  $\bar{\nu}_{\mu}$ Fe Deep Inelastic Scattering at the Tevatron.},''
  \href{http://dx.doi.org/10.1103/PhysRevD.64.112006}{{\em Phys. Rev.}
  {\bfseries D64} (2001) 112006},
\href{http://arxiv.org/abs/hep-ex/0102049}{{\ttfamily arXiv:hep-ex/0102049
  [hep-ex]}}.

\bibitem{Kusina:2012vh}
A.~Kusina, T.~Stavreva, S.~Berge, F.~I. Olness, I.~Schienbein,
  K.~Kova\v{r}\'{i}k, T.~Jezo, J.~Y. Yu, and K.~Park, ``{Strange Quark PDFs and
  Implications for Drell-Yan Boson Production at the LHC},''
  \href{http://dx.doi.org/10.1103/PhysRevD.85.094028}{{\em Phys. Rev.}
  {\bfseries D85} (2012) 094028},
\href{http://arxiv.org/abs/1203.1290}{{\ttfamily arXiv:1203.1290 [hep-ph]}}.

\bibitem{Faura:2020oom}
F.~Faura, S.~Iranipour, E.~R. Nocera, J.~Rojo, and M.~Ubiali, ``{The Strangest
  Proton?},'' \href{http://dx.doi.org/10.1140/epjc/s10052-020-08749-3}{{\em
  Eur. Phys. J. C} {\bfseries 80} no.~12, (2020) 1168},
  \href{http://arxiv.org/abs/2009.00014}{{\ttfamily arXiv:2009.00014
  [hep-ph]}}.

\bibitem{Schienbein:2007fs}
I.~Schienbein, J.~Yu, C.~Keppel, J.~Morfin, F.~Olness, and J.~Owens, ``{Nuclear
  parton distribution functions from neutrino deep inelastic scattering},''
  \href{http://dx.doi.org/10.1103/PhysRevD.77.054013}{{\em Phys. Rev. D}
  {\bfseries 77} (2008) 054013},
  \href{http://arxiv.org/abs/0710.4897}{{\ttfamily arXiv:0710.4897 [hep-ph]}}.

\bibitem{Tzanov:2005kr}
{\bfseries NuTeV} Collaboration, M.~Tzanov {\em et~al.}, ``{Precise measurement
  of neutrino and anti-neutrino differential cross sections},''
  \href{http://dx.doi.org/10.1103/PhysRevD.74.012008}{{\em Phys. Rev.}
  {\bfseries D74} (2006) 012008},
\href{http://arxiv.org/abs/hep-ex/0509010}{{\ttfamily arXiv:hep-ex/0509010
  [hep-ex]}}.

\bibitem{Kulagin:2004ie}
S.~A. Kulagin and R.~Petti, ``{Global study of nuclear structure functions},''
  \href{http://dx.doi.org/10.1016/j.nuclphysa.2005.10.011}{{\em Nucl. Phys. A}
  {\bfseries 765} (2006) 126--187},
  \href{http://arxiv.org/abs/hep-ph/0412425}{{\ttfamily arXiv:hep-ph/0412425}}.

\bibitem{Abramowicz:1991xz}
H.~Abramowicz, E.~M. Levin, A.~Levy, and U.~Maor, ``{A Parametrization of
  sigma-T (gamma* p) above the resonance region Q**2 \ensuremath{>}= 0},''
  \href{http://dx.doi.org/10.1016/0370-2693(91)90202-2}{{\em Phys. Lett. B}
  {\bfseries 269} (1991) 465--476}.

\bibitem{Kovarik:2010uv}
K.~Kova\v{r}\'{i}k, I.~Schienbein, F.~I. Olness, J.~Y. Yu, C.~Keppel, J.~G.
  Morfin, J.~F. Owens, and T.~Stavreva, ``{Nuclear Corrections in
  Neutrino-Nucleus DIS and Their Compatibility with Global NPDF Analyses},''
  \href{http://dx.doi.org/10.1103/PhysRevLett.106.122301}{{\em Phys. Rev.
  Lett.} {\bfseries 106} (2011) 122301},
\href{http://arxiv.org/abs/1012.0286}{{\ttfamily arXiv:1012.0286 [hep-ph]}}.

\bibitem{Onengut:2005kv}
{\bfseries CHORUS} Collaboration, G.~Onengut {\em et~al.}, ``{Measurement of
  nucleon structure functions in neutrino scattering},''
  \href{http://dx.doi.org/10.1016/j.physletb.2005.10.062}{{\em Phys. Lett. B}
  {\bfseries 632} (2006) 65--75}.

\bibitem{Paukkunen:2010hb}
H.~Paukkunen and C.~A. Salgado, ``{Compatibility of neutrino DIS data and
  global analyses of parton distribution functions},''
  \href{http://dx.doi.org/10.1007/JHEP07(2010)032}{{\em JHEP} {\bfseries 07}
  (2010) 032},
\href{http://arxiv.org/abs/1004.3140}{{\ttfamily arXiv:1004.3140 [hep-ph]}}.

\bibitem{Paukkunen:2013grz}
H.~Paukkunen and C.~A. Salgado, ``{Agreement of Neutrino Deep Inelastic
  Scattering Data with Global Fits of Parton Distributions},''
  \href{http://dx.doi.org/10.1103/PhysRevLett.110.212301}{{\em Phys. Rev.
  Lett.} {\bfseries 110} no.~21, (2013) 212301},
\href{http://arxiv.org/abs/1302.2001}{{\ttfamily arXiv:1302.2001 [hep-ph]}}.

\bibitem{Kalantarians:2017mkj}
N.~Kalantarians, C.~Keppel, and M.~E. Christy, ``{Comparison of the Structure
  Function F2 as Measured by Charged Lepton and Neutrino Scattering from Iron
  Targets},'' \href{http://dx.doi.org/10.1103/PhysRevC.96.032201}{{\em Phys.
  Rev.} {\bfseries C96} no.~3, (2017) 032201},
\href{http://arxiv.org/abs/1706.02002}{{\ttfamily arXiv:1706.02002 [hep-ph]}}.

\bibitem{Eskola:2021nhw}
K.~J. Eskola, P.~Paakkinen, H.~Paukkunen, and C.~A. Salgado, ``{EPPS21: A
  global QCD analysis of nuclear PDFs},''
  \href{http://arxiv.org/abs/2112.12462}{{\ttfamily arXiv:2112.12462
  [hep-ph]}}.

\bibitem{Khalek:2022zqe}
R.~A. Khalek, R.~Gauld, T.~Giani, E.~R. Nocera, T.~R. Rabemananjara, and
  J.~Rojo, ``{nNNPDF3.0: Evidence for a modified partonic structure in heavy
  nuclei},'' \href{http://arxiv.org/abs/2201.12363}{{\ttfamily arXiv:2201.12363
  [hep-ph]}}.

\bibitem{Nakamura:2016cnn}
S.~X. Nakamura {\em et~al.}, ``{Towards a Unified Model of Neutrino-Nucleus
  Reactions for Neutrino Oscillation Experiments},''
  \href{http://dx.doi.org/10.1088/1361-6633/aa5e6c}{{\em Rept. Prog. Phys.}
  {\bfseries 80} no.~5, (2017) 056301},
  \href{http://arxiv.org/abs/1610.01464}{{\ttfamily arXiv:1610.01464
  [nucl-th]}}.

\bibitem{AtlasWpPb}
{\bfseries ATLAS} Collaboration, ``{Measurement of $W\rightarrow\mu\nu$
  production in $p$+Pb collision at $\sqrt{s_{_\text{NN}}}=5.02$ TeV with ATLAS
  detector at the LHC},''
{\em {ATLAS-CONF-2015-056}} .

\bibitem{Aad:2015gta}
{\bfseries ATLAS} Collaboration, G.~Aad {\em et~al.}, ``{$Z$ boson production
  in $p+$Pb collisions at $\sqrt{s_{NN}}=5.02$ TeV measured with the ATLAS
  detector},'' \href{http://dx.doi.org/10.1103/PhysRevC.92.044915}{{\em Phys.
  Rev.} {\bfseries C92} no.~4, (2015) 044915},
\href{http://arxiv.org/abs/1507.06232}{{\ttfamily arXiv:1507.06232 [hep-ex]}}.

\bibitem{Khachatryan:2015hha}
{\bfseries CMS} Collaboration, V.~Khachatryan {\em et~al.}, ``{Study of W boson
  production in pPb collisions at $\sqrt{s_{\mathrm{NN}}} =$ 5.02 TeV},''
  \href{http://dx.doi.org/10.1016/j.physletb.2015.09.057}{{\em Phys. Lett.}
  {\bfseries B750} (2015) 565--586},
\href{http://arxiv.org/abs/1503.05825}{{\ttfamily arXiv:1503.05825 [nucl-ex]}}.

\bibitem{Khachatryan:2015pzs}
{\bfseries CMS} Collaboration, V.~Khachatryan {\em et~al.}, ``{Study of Z boson
  production in pPb collisions at $\sqrt {s_{NN}} = 5.02$ TeV},''
  \href{http://dx.doi.org/10.1016/j.physletb.2016.05.044}{{\em Phys. Lett.}
  {\bfseries B759} (2016) 36--57},
\href{http://arxiv.org/abs/1512.06461}{{\ttfamily arXiv:1512.06461 [hep-ex]}}.

\bibitem{Sirunyan:2019dox}
{\bfseries CMS} Collaboration, A.~M. Sirunyan {\em et~al.}, ``{Observation of
  nuclear modifications in W$^\pm$ boson production in pPb collisions at
  $\sqrt{s_\mathrm{NN}} =$ 8.16 TeV},''
  \href{http://dx.doi.org/10.1016/j.physletb.2019.135048}{{\em Phys. Lett.}
  {\bfseries B800} (2020) 135048},
\href{http://arxiv.org/abs/1905.01486}{{\ttfamily arXiv:1905.01486 [hep-ex]}}.

\bibitem{ALICE:2016rzo}
{\bfseries ALICE} Collaboration, J.~Adam {\em et~al.}, ``{W and Z boson
  production in p-Pb collisions at $\sqrt{s_{\rm NN}}$ = 5.02 TeV},''
  \href{http://dx.doi.org/10.1007/JHEP02(2017)077}{{\em JHEP} {\bfseries 02}
  (2017) 077}, \href{http://arxiv.org/abs/1611.03002}{{\ttfamily
  arXiv:1611.03002 [nucl-ex]}}.

\bibitem{Aaij:2014pvu}
{\bfseries LHCb} Collaboration, R.~Aaij {\em et~al.}, ``{Observation of $Z$
  production in proton-lead collisions at LHCb},''
  \href{http://dx.doi.org/10.1007/JHEP09(2014)030}{{\em JHEP} {\bfseries 09}
  (2014) 030},
\href{http://arxiv.org/abs/1406.2885}{{\ttfamily arXiv:1406.2885 [hep-ex]}}.

\bibitem{Adler:2006wg}
{\bfseries PHENIX} Collaboration, S.~Adler {\em et~al.}, ``{Centrality
  dependence of pi0 and eta production at large transverse momentum in
  s(NN)**(1/2) = 200-GeV d+Au collisions},''
  \href{http://dx.doi.org/10.1103/PhysRevLett.98.172302}{{\em Phys. Rev. Lett.}
  {\bfseries 98} (2007) 172302},
  \href{http://arxiv.org/abs/nucl-ex/0610036}{{\ttfamily
  arXiv:nucl-ex/0610036}}.

\bibitem{PHENIX:2013kod}
{\bfseries PHENIX} Collaboration, A.~Adare {\em et~al.}, ``{Spectra and ratios
  of identified particles in Au+Au and $d$+Au collisions at $\sqrt{s_{NN}}=200$
  GeV},'' \href{http://dx.doi.org/10.1103/PhysRevC.88.024906}{{\em Phys. Rev.
  C} {\bfseries 88} no.~2, (2013) 024906},
  \href{http://arxiv.org/abs/1304.3410}{{\ttfamily arXiv:1304.3410 [nucl-ex]}}.

\bibitem{Abelev:2009hx}
{\bfseries STAR} Collaboration, B.~I. Abelev {\em et~al.}, ``{Inclusive
  $\pi^0$, $\eta$, and direct photon production at high transverse momentum in
  $p+p$ and $d+$Au collisions at $\sqrt{s_{NN}}=200$ GeV},''
  \href{http://dx.doi.org/10.1103/PhysRevC.81.064904}{{\em Phys. Rev. C}
  {\bfseries 81} (2010) 064904},
  \href{http://arxiv.org/abs/0912.3838}{{\ttfamily arXiv:0912.3838 [hep-ex]}}.

\bibitem{STAR:2006xud}
{\bfseries STAR} Collaboration, J.~Adams {\em et~al.}, ``{Identified hadron
  spectra at large transverse momentum in p+p and d+Au collisions at
  s(NN)**(1/2) = 200-GeV},''
  \href{http://dx.doi.org/10.1016/j.physletb.2006.04.032}{{\em Phys. Lett. B}
  {\bfseries 637} (2006) 161--169},
  \href{http://arxiv.org/abs/nucl-ex/0601033}{{\ttfamily
  arXiv:nucl-ex/0601033}}.

\bibitem{ALICE:2016dei}
{\bfseries ALICE} Collaboration, J.~Adam {\em et~al.}, ``{Multiplicity
  dependence of charged pion, kaon, and (anti)proton production at large
  transverse momentum in p-Pb collisions at $\mathbf{\sqrt{{\textit s}_{\rm
  NN}}}$ = 5.02 TeV},''
  \href{http://dx.doi.org/10.1016/j.physletb.2016.07.050}{{\em Phys. Lett. B}
  {\bfseries 760} (2016) 720--735},
  \href{http://arxiv.org/abs/1601.03658}{{\ttfamily arXiv:1601.03658
  [nucl-ex]}}.

\bibitem{ALICE:2018vhm}
{\bfseries ALICE} Collaboration, S.~Acharya {\em et~al.}, ``{Neutral pion and
  $\eta$ meson production in p-Pb collisions at $\sqrt{s_\mathrm{NN}} = 5.02$
  TeV},'' \href{http://dx.doi.org/10.1140/epjc/s10052-018-6013-8}{{\em Eur.
  Phys. J. C} {\bfseries 78} no.~8, (2018) 624},
  \href{http://arxiv.org/abs/1801.07051}{{\ttfamily arXiv:1801.07051
  [nucl-ex]}}.

\bibitem{ALICE:2021est}
{\bfseries ALICE} Collaboration, S.~Acharya {\em et~al.}, ``{Nuclear
  modification factor of light neutral-meson spectra up to high transverse
  momentum in p-Pb collisions at $\sqrt{s_{NN}}$ = 8.16 TeV},''
  \href{http://arxiv.org/abs/2104.03116}{{\ttfamily arXiv:2104.03116
  [nucl-ex]}}.

\bibitem{Berge:1989hr}
J.~Berge {\em et~al.}, ``{A Measurement of Differential Cross-Sections and
  Nucleon Structure Functions in Charged Current Neutrino Interactions on
  Iron},'' \href{http://dx.doi.org/10.1007/BF01555493}{{\em Z. Phys. C}
  {\bfseries 49} (1991) 187--224}.

\bibitem{CCFRNuTeV:2000qwc}
{\bfseries CCFR/NuTeV} Collaboration, U.-K. Yang {\em et~al.}, ``{Measurements
  of $F_2$ and $xF^{\nu}_3 - x F^{\bar{\nu}}_3$ from CCFR $\nu_\mu-$Fe and
  $\bar{\nu}_\mu-$Fe data in a physics model independent way},''
  \href{http://dx.doi.org/10.1103/PhysRevLett.86.2742}{{\em Phys. Rev. Lett.}
  {\bfseries 86} (2001) 2742--2745},
  \href{http://arxiv.org/abs/hep-ex/0009041}{{\ttfamily arXiv:hep-ex/0009041}}.

\bibitem{Yang:2001rm}
U.-K. Yang, ``{A Measurement of Differential Cross Sections in Charged Current
  Neutrino Interactions on Iron and a Global Structure Functions Analysis},''.

\bibitem{Segarra:2020gtj}
E.~P. Segarra {\em et~al.}, ``{Extending nuclear PDF analyses into the high-$x$
  , low-$Q^2$ region},''
  \href{http://dx.doi.org/10.1103/PhysRevD.103.114015}{{\em Phys. Rev. D}
  {\bfseries 103} no.~11, (2021) 114015},
  \href{http://arxiv.org/abs/2012.11566}{{\ttfamily arXiv:2012.11566
  [hep-ph]}}.

\bibitem{Owens:2007kp}
J.~F. Owens, J.~Huston, C.~E. Keppel, S.~Kuhlmann, J.~G. Morfin, F.~Olness,
  J.~Pumplin, and D.~Stump, ``{The Impact of new neutrino DIS and Drell-Yan
  data on large-x parton distributions},''
  \href{http://dx.doi.org/10.1103/PhysRevD.75.054030}{{\em Phys. Rev.}
  {\bfseries D75} (2007) 054030},
\href{http://arxiv.org/abs/hep-ph/0702159}{{\ttfamily arXiv:hep-ph/0702159
  [HEP-PH]}}.

\bibitem{Stump:2003yu}
D.~Stump, J.~Huston, J.~Pumplin, W.-K. Tung, H.~L. Lai, S.~Kuhlmann, and J.~F.
  Owens, ``{Inclusive jet production, parton distributions, and the search for
  new physics},'' \href{http://dx.doi.org/10.1088/1126-6708/2003/10/046}{{\em
  JHEP} {\bfseries 10} (2003) 046},
  \href{http://arxiv.org/abs/hep-ph/0303013}{{\ttfamily arXiv:hep-ph/0303013}}.

\bibitem{NOMAD:2007krq}
{\bfseries NOMAD} Collaboration, Q.~Wu {\em et~al.}, ``{A Precise measurement
  of the muon neutrino-nucleon inclusive charged current cross-section off an
  isoscalar target in the energy range 2.5 \ensuremath{<} E(nu) \ensuremath{<}
  40-GeV by NOMAD},''
  \href{http://dx.doi.org/10.1016/j.physletb.2007.12.027}{{\em Phys. Lett. B}
  {\bfseries 660} (2008) 19--25},
  \href{http://arxiv.org/abs/0711.1183}{{\ttfamily arXiv:0711.1183 [hep-ex]}}.

\bibitem{Petti:2006tu}
{\bfseries NOMAD} Collaboration, R.~Petti, ``{Cross-section measurements in the
  NOMAD experiment},''
  \href{http://dx.doi.org/10.1016/j.nuclphysbps.2006.08.026}{{\em Nucl. Phys. B
  Proc. Suppl.} {\bfseries 159} (2006) 56--62},
  \href{http://arxiv.org/abs/hep-ex/0602022}{{\ttfamily arXiv:hep-ex/0602022}}.

\bibitem{IceCube:2017roe}
{\bfseries IceCube} Collaboration, M.~G. Aartsen {\em et~al.}, ``{Measurement
  of the multi-TeV neutrino cross section with IceCube using Earth
  absorption},'' \href{http://dx.doi.org/10.1038/nature24459}{{\em Nature}
  {\bfseries 551} (2017) 596--600},
  \href{http://arxiv.org/abs/1711.08119}{{\ttfamily arXiv:1711.08119
  [hep-ex]}}.

\bibitem{PhysRevD.93.071101}
{\bfseries The MINER\ensuremath{\nu}A Collaboration} Collaboration, J.~Mousseau
  {\em et~al.}, ``Measurement of partonic nuclear effects in deep-inelastic
  neutrino scattering using minerva,''
  \href{http://dx.doi.org/10.1103/PhysRevD.93.071101}{{\em Phys. Rev. D}
  {\bfseries 93} (Apr, 2016) 071101}.
  \url{https://link.aps.org/doi/10.1103/PhysRevD.93.071101}.

\bibitem{Abramowicz:1982zr}
H.~Abramowicz {\em et~al.}, ``{Experimental Study of Opposite Sign Dimuons
  Produced in Neutrino and anti-neutrinos Interactions},''
  \href{http://dx.doi.org/10.1007/BF01573422}{{\em Z. Phys. C} {\bfseries 15}
  (1982) 19}.

\bibitem{CHORUS:2008vjb}
{\bfseries CHORUS} Collaboration, A.~Kayis-Topaksu {\em et~al.}, ``{Leading
  order analysis of neutrino induced dimuon events in the CHORUS experiment},''
  \href{http://dx.doi.org/10.1016/j.nuclphysb.2008.02.013}{{\em Nucl. Phys. B}
  {\bfseries 798} (2008) 1--16},
  \href{http://arxiv.org/abs/0804.1869}{{\ttfamily arXiv:0804.1869 [hep-ex]}}.

\bibitem{NOMAD:2013hbk}
{\bfseries NOMAD} Collaboration, O.~Samoylov {\em et~al.}, ``{A Precision
  Measurement of Charm Dimuon Production in Neutrino Interactions from the
  NOMAD Experiment},''
  \href{http://dx.doi.org/10.1016/j.nuclphysb.2013.08.021}{{\em Nucl. Phys. B}
  {\bfseries 876} (2013) 339--375},
  \href{http://arxiv.org/abs/1308.4750}{{\ttfamily arXiv:1308.4750 [hep-ex]}}.

\bibitem{Accardi:2021ysh}
A.~Accardi, T.~J. Hobbs, X.~Jing, and P.~M. Nadolsky, ``{Deuterium scattering
  experiments in CTEQ global QCD analyses: a comparative investigation},''
  \href{http://dx.doi.org/10.1140/epjc/s10052-021-09318-y}{{\em Eur. Phys. J.
  C} {\bfseries 81} no.~7, (2021) 603},
  \href{http://arxiv.org/abs/2102.01107}{{\ttfamily arXiv:2102.01107
  [hep-ph]}}.

\bibitem{Kopeliovich:2012kw}
B.~Z. Kopeliovich, J.~G. Morfin, and I.~Schmidt, ``{Nuclear Shadowing in
  Electro-Weak Interactions},''
  \href{http://dx.doi.org/10.1016/j.ppnp.2012.09.004}{{\em Prog. Part. Nucl.
  Phys.} {\bfseries 68} (2013) 314--372},
  \href{http://arxiv.org/abs/1208.6541}{{\ttfamily arXiv:1208.6541 [hep-ph]}}.

\bibitem{NuSOnG:2009rcm}
{\bfseries NuSOnG} Collaboration, T.~Adams {\em et~al.}, ``{QCD Precision
  Measurements and Structure Function Extraction at a High Statistics, High
  Energy Neutrino Scattering Experiment: NuSOnG},''
  \href{http://dx.doi.org/10.1142/S0217751X10047828}{{\em Int. J. Mod. Phys. A}
  {\bfseries 25} (2010) 909--949},
  \href{http://arxiv.org/abs/0906.3563}{{\ttfamily arXiv:0906.3563 [hep-ex]}}.

\bibitem{Anchordoqui:2021ghd}
L.~A. Anchordoqui {\em et~al.}, ``{The Forward Physics Facility: Sites,
  Experiments, and Physics Potential},''
  \href{http://arxiv.org/abs/2109.10905}{{\ttfamily arXiv:2109.10905
  [hep-ph]}}.

\bibitem{Accardi:2012qut}
A.~Accardi {\em et~al.}, ``{Electron Ion Collider: The Next QCD Frontier}:
  {Understanding the glue that binds us all},''
  \href{http://dx.doi.org/10.1140/epja/i2016-16268-9}{{\em Eur. Phys. J. A}
  {\bfseries 52} no.~9, (2016) 268},
  \href{http://arxiv.org/abs/1212.1701}{{\ttfamily arXiv:1212.1701 [nucl-ex]}}.

\bibitem{AbdulKhalek:2021gbh}
R.~Abdul~Khalek {\em et~al.}, ``{Science Requirements and Detector Concepts for
  the Electron-Ion Collider: EIC Yellow Report},''
  \href{http://arxiv.org/abs/2103.05419}{{\ttfamily arXiv:2103.05419
  [physics.ins-det]}}.

\bibitem{DAgostini:1993arp}
G.~D'Agostini, ``{On the use of the covariance matrix to fit correlated
  data},'' \href{http://dx.doi.org/10.1016/0168-9002(94)90719-6}{{\em Nucl.
  Instrum. Meth. A} {\bfseries 346} (1994) 306--311}.

\bibitem{Stump:2001gu}
D.~Stump, J.~Pumplin, R.~Brock, D.~Casey, J.~Huston, J.~Kalk, H.~Lai, and
  W.~Tung, ``{Uncertainties of predictions from parton distribution functions.
  1. The Lagrange multiplier method},''
  \href{http://dx.doi.org/10.1103/PhysRevD.65.014012}{{\em Phys. Rev. D}
  {\bfseries 65} (2001) 014012},
  \href{http://arxiv.org/abs/hep-ph/0101051}{{\ttfamily arXiv:hep-ph/0101051}}.

\bibitem{10.1214/aoms/1177729893}
J.~Sherman and W.~J. Morrison, ``{Adjustment of an Inverse Matrix Corresponding
  to a Change in One Element of a Given Matrix},''
  \href{http://dx.doi.org/10.1214/aoms/1177729893}{{\em The Annals of
  Mathematical Statistics} {\bfseries 21} no.~1, (1950) 124 -- 127}.
  \url{https://doi.org/10.1214/aoms/1177729893}.

\bibitem{cox_2006}
D.~R. Cox, \href{http://dx.doi.org/10.1017/CBO9780511813559}{{\em Principles of
  Statistical Inference}}.
\newblock Cambridge University Press, 2006.

\bibitem{Ball:2009qv}
{\bfseries NNPDF} Collaboration, R.~D. Ball, L.~Del~Debbio, S.~Forte,
  A.~Guffanti, J.~I. Latorre, J.~Rojo, and M.~Ubiali, ``{Fitting Parton
  Distribution Data with Multiplicative Normalization Uncertainties},''
  \href{http://dx.doi.org/10.1007/JHEP05(2010)075}{{\em JHEP} {\bfseries 05}
  (2010) 075}, \href{http://arxiv.org/abs/0912.2276}{{\ttfamily arXiv:0912.2276
  [hep-ph]}}.

\end{thebibliography}\endgroup
\end{document}